\documentclass[11pt]{article}
\usepackage[margin=1in]{geometry}
\usepackage{amsmath,amssymb,amsthm,booktabs,array}
\usepackage{graphicx}
\usepackage{setspace}
\usepackage{placeins}
\usepackage[round]{natbib}
\setcitestyle{aysep={}}
\usepackage{xurl}
\usepackage{tikz}
\usetikzlibrary{arrows.meta,positioning,fit}
\usepackage[hidelinks]{hyperref}
\hypersetup{pdftitle={Risk Ceilings and Development Deadlines: Pacing AI under Uncertain Safety Productivity},pdfauthor={Li Gan}}
\newtheorem{proposition}{Proposition}
\theoremstyle{definition}
\newtheorem{definition}{Definition}
\theoremstyle{plain}

\newcommand{\ipComputeDoubling}{6}
\newcommand{\ipLam}{0.5}
\newcommand{\ipRSI}{10}
\newcommand{\ipXzero}{50}
\newcommand{\ipSzero}{5}
\newcommand{\ipIzero}{45}
\newcommand{\ipPsi}{0.5}
\newcommand{\ipAlpha}{0.5}
\newcommand{\ipMu}{10}

\newcommand{\ipShareSigma}{8.3}

\newcommand{\ipPauseEnd}{5.2}

\newcommand{\ipEqPauseSafetyP}{20}

\newcommand{\ipEqFloorsShare}{7.7}

\newcommand{\ipEqLagMonths}{21}
\newcommand{\ipEqCeilingPct}{1.6}
\newcommand{\ipEqPauseBlogStart}{10}
\newcommand{\ipEqPauseSafetyStart}{1.5}

\newcommand{\rbBasePauseBlog}{12.3}
\newcommand{\rbBasePauseSafety}{54.8}
\newcommand{\rbBaseFloors}{58.6}
\newcommand{\rbBaseLag}{6.8}
\newcommand{\rbBaseCeiling}{10.5}
\newcommand{\rbBaseSafetyFloor}{29.1}

\newcommand{\rbYieldQuarterPauseBlog}{6.4}
\newcommand{\rbYieldQuarterPauseSafety}{20.8}
\newcommand{\rbYieldQuarterFloors}{15.3}
\newcommand{\rbYieldQuarterLag}{$-$12.1}
\newcommand{\rbYieldQuarterCeiling}{$-$24.5}
\newcommand{\rbYieldQuarterSafetyFloor}{7.4}

\newcommand{\rbYieldHalfPauseBlog}{8.2}
\newcommand{\rbYieldHalfPauseSafety}{35.0}
\newcommand{\rbYieldHalfFloors}{34.7}
\newcommand{\rbYieldHalfLag}{$-$6.4}
\newcommand{\rbYieldHalfCeiling}{$-$20.8}
\newcommand{\rbYieldHalfSafetyFloor}{14.9}

\newcommand{\rbYieldOneHalfPauseBlog}{16.7}
\newcommand{\rbYieldOneHalfPauseSafety}{65.5}
\newcommand{\rbYieldOneHalfFloors}{68.1}
\newcommand{\rbYieldOneHalfLag}{20.3}
\newcommand{\rbYieldOneHalfCeiling}{60.1}
\newcommand{\rbYieldOneHalfSafetyFloor}{41.0}

\newcommand{\rbRisingPauseBlog}{13.4}
\newcommand{\rbRisingPauseSafety}{55.3}
\newcommand{\rbRisingFloors}{63.2}
\newcommand{\rbRisingLag}{18.9}
\newcommand{\rbRisingCeiling}{65.0}
\newcommand{\rbRisingSafetyFloor}{35.9}

\newcommand{\rbTransferPauseBlog}{9.5}
\newcommand{\rbTransferPauseSafety}{41.9}
\newcommand{\rbTransferFloors}{47.2}
\newcommand{\rbTransferLag}{$-$0.1}
\newcommand{\rbTransferCeiling}{$-$1.9}
\newcommand{\rbTransferSafetyFloor}{21.3}

\newcommand{\rbComputePauseBlog}{16.6}
\newcommand{\rbComputePauseSafety}{67.8}
\newcommand{\rbComputeFloors}{70.9}
\newcommand{\rbComputeLag}{36.0}
\newcommand{\rbComputeCeiling}{73.7}
\newcommand{\rbComputeSafetyFloor}{43.7}

\newcommand{\rbSpillPauseBlog}{12.3}
\newcommand{\rbSpillPauseSafety}{48.7}
\newcommand{\rbSpillFloors}{58.6}
\newcommand{\rbSpillLag}{6.8}
\newcommand{\rbSpillCeiling}{$-$23.9}
\newcommand{\rbSpillSafetyFloor}{27.7}

\newcommand{\rbTailSmallPauseBlog}{11.2}
\newcommand{\rbTailSmallPauseSafety}{54.3}
\newcommand{\rbTailSmallFloors}{60.3}
\newcommand{\rbTailSmallLag}{12.4}
\newcommand{\rbTailSmallCeiling}{19.8}
\newcommand{\rbTailSmallSafetyFloor}{30.2}

\newcommand{\rbTailPauseBlog}{11.8}
\newcommand{\rbTailPauseSafety}{55.5}
\newcommand{\rbTailFloors}{64.3}
\newcommand{\rbTailLag}{29.8}
\newcommand{\rbTailCeiling}{52.3}
\newcommand{\rbTailSafetyFloor}{32.9}

\newcommand{\rbPauseSafetyMin}{21}

\newcommand{\rbPauseSafetyMax}{68}
\newcommand{\rbFloorsMin}{15}

\newcommand{\rbFloorsMax}{71}

\newcommand{\rbLagMinAbs}{12}
\newcommand{\rbLagMax}{36}

\newcommand{\rbCeilingMinAbs}{24}
\newcommand{\rbCeilingMax}{74}

\newcommand{\irFixedHalfBase}{93.5}
\newcommand{\irFixedHalfCeil}{88.1}
\newcommand{\irFixedHalfFall}{5.4}
\newcommand{\irFixedHalfLead}{32.8}
\newcommand{\irWeakOneHundredth}{79.6}

\newcommand{\irHybridHazard}{0.421}
\newcommand{\irHybridLead}{11.500}
\newcommand{\irHybridLossTwo}{45.16}
\newcommand{\irHybridLossFive}{5.18}

\newcommand{\irFlatKnowledge}{1.55009}
\newcommand{\irRisingKnowledge}{2.12972}
\newcommand{\irServiceRatio}{13.65}
\newcommand{\irConstructWait}{3.70}
\newcommand{\irConstructHazard}{0.500}
\newcommand{\irConstructLossTwo}{13.36}
\newcommand{\irConstructLossFive}{0.98}
\newcommand{\irSigmaB}{0.086116}
\newcommand{\irTailEta}{0.0665009}
\newcommand{\irJTwo}{72.125}
\newcommand{\irJFive}{984.785}
\newcommand{\irPauseSafetyLead}{11.98}
\newcommand{\irPauseBlogLead}{11.97}
\newcommand{\irHybridX}{0.0827}
\newcommand{\irHybridS}{0.50}
\newcommand{\irHybridI}{0.4173}
\newcommand{\irRisingExtra}{37.4}
\newcommand{\insCap}{0.0492}
\newcommand{\insStateEnd}{0.60}

\newcommand{\insYbar}{2.74}
\newcommand{\insScaleMin}{0.1}
\newcommand{\insScaleMax}{2}
\newcommand{\insCeilLeadHalf}{25}

\newcommand{\insCeilLeadQuarter}{52}

\newcommand{\insFloorsFallQuarter}{15.3}
\newcommand{\insPauseFallQuarter}{20.5}
\newcommand{\insCeilRiseQuarter}{20.1}

\newcommand{\insFrNecAtCap}{0.57}
\newcommand{\insFrSzeroAtCap}{1.08}
\newcommand{\insFrSpareAtCap}{0.69}
\newcommand{\insFrPureFactor}{1.033}
\newcommand{\insFrPureGapPct}{3.3}

\newcommand{\insFrHalfSpareMax}{72}

\newcommand{\insFrReplayShare}{12.5}
\newcommand{\insFrYspare}{2.59}
\newcommand{\insFrLossSzeroFive}{4.7}
\newcommand{\insFrLossSpareFive}{13}
\newcommand{\insFrLossSzeroTwo}{51}

\newcommand{\insFrOptMonths}{7.5}
\newcommand{\insFrOptFive}{4.1}
\newcommand{\insFrOptTwo}{52}
\newcommand{\insFrFixedFive}{5.0}
\newcommand{\insFrFixedTwo}{53}
\newcommand{\insFrOptResearch}{21.0}
\newcommand{\insFrOptSafety}{40.1}
\newcommand{\insFrOptCustomers}{38.9}
\newcommand{\insFrRiseRatio}{2.5}
\newcommand{\insFrTransferRatio}{1.8}
\newcommand{\insFrRiseRatioHalf}{2.9}
\newcommand{\insFrTransferRatioHalf}{2.2}
\newcommand{\insZeroFloors}{5.4}
\newcommand{\insZeroLag}{16.8}
\newcommand{\insZeroPauseSafety}{4.9}

\newcommand{\insZeroLatePause}{7.7}
\newcommand{\insLatePauseStart}{15}
\newcommand{\insLifetimeBound}{77.2}
\newcommand{\inrRatio}{57}
\newcommand{\inrNeed}{4.05}

\newcommand{\inrFloorM}{2.5}
\newcommand{\insRiskFull}{75}
\newcommand{\insRiskSpareFit}{2}
\newcommand{\insRiskSpareBound}{8}
\newcommand{\insRiskCheapFit}{8}
\newcommand{\insRiskUsualFit}{9}
\newcommand{\insFrScenGap}{21}

\newcommand{\insDoubleTwelve}{7.0}
\newcommand{\insDoubleTwentyFour}{10.6}
\newcommand{\ovHzero}{0.00235}
\newcommand{\ovHoneNorm}{0.598}
\newcommand{\ovFfifteen}{0.493}
\newcommand{\ovWaitZero}{0.0282}
\newcommand{\ovSavedZero}{0.201}
\newcommand{\ovDeltaHZero}{-0.173}
\newcommand{\ovPointsZero}{4.72}
\newcommand{\ovPointsScen}{4.87}
\newcommand{\ovHoneFixed}{2.82}
\newcommand{\ovFullPFixed}{99.6}
\newcommand{\ovPointsZeroFixed}{0.46}
\newcommand{\lfComputeYear}{4}
\newcommand{\lfComputeEnd}{8}
\newcommand{\lfSpeedRatio}{28}
\newcommand{\lfHazStart}{0.0047}
\newcommand{\lfHazEnd}{1.20}
\newcommand{\lfHazZeroEnd}{5.6}
\newcommand{\lfHazEndProb}{70}
\newcommand{\lfCapProb}{4.8}
\newcommand{\lfCrossMonth}{13.1}
\newcommand{\lfHalfMonth}{15.1}
\newcommand{\lfMonthsAfterHalf}{3}
\newcommand{\lfHfull}{1.386}
\newcommand{\lfEffMonths}{7.48}
\newcommand{\lfDisc}{0.356}
\newcommand{\lfHafter}{0.493}
\newcommand{\lfExpo}{0.018}
\newcommand{\lfHzero}{0.00235}
\newcommand{\lfHone}{2.82}
\newcommand{\lfVzeroUsual}{0.117}
\newcommand{\lfFitSaving}{0.875}
\newcommand{\lfFitTotal}{0.511}
\newcommand{\lfVzeroSafety}{0.255}
\newcommand{\lfTzeroUsual}{15.9}
\newcommand{\lfTzeroSafety}{16.7}
\newcommand{\lfTzeroInference}{14.8}
\newcommand{\lfMultSafety}{2.16}
\newcommand{\lfRate}{0.236}
\newcommand{\lfMonths}{17.2}

\newcommand{\lfHalvings}{5.8}
\newcommand{\lfHalvingMonths}{2.9}
\newcommand{\lfExtra}{0.17}
\newcommand{\lfExtraPct}{17}
\newcommand{\lfGapPred}{1.21}

\newcommand{\lfWorstExtraPct}{67}
\newcommand{\lfWorstLnreq}{1.04}
\newcommand{\lfFactorTwelve}{1.070}
\newcommand{\lfFactorTwentyFour}{1.106}
\newcommand{\lfFactorNone}{1.168}
\newcommand{\lfFactorNonePct}{17}
\newcommand{\lfShareTwelve}{25.0}
\newcommand{\lfShareTwentyFour}{35.4}
\newcommand{\lfShareNone}{50}
\newcommand{\lfYTwelve}{2.44}
\newcommand{\lfYTwentyFour}{2.30}
\newcommand{\lfYNone}{2.08}
\newcommand{\lfKnowLearn}{1.96}
\newcommand{\lfKnowReplay}{2.78}
\newcommand{\lfKnowEnd}{4.74}
\newcommand{\lfFullCrossBound}{12.0}
\newcommand{\lfPauseCross}{12.8}
\newcommand{\lfFloorsEnd}{30.0}
\newcommand{\tsPauseMult}{2.16}
\newcommand{\tsBasePoints}{60}
\newcommand{\tsBaseProb}{15}

\newcommand{\tsTransferStart}{2.1}
\newcommand{\tsTransferGain}{0.03}
\newcommand{\tsBothStart}{9.2}
\newcommand{\tsBothFirst}{44.8}
\newcommand{\tsBothBest}{47.9}

\newcommand{\tsBothProbFirst}{30}
\newcommand{\tsBothProbBest}{27}
\newcommand{\tsScenMinInitial}{35.5}
\newcommand{\tsBaseMinInitial}{17.2}
\newcommand{\tsCapLoose}{0.15}

\newcommand{\tsCapLooseBest}{13.5}
\newcommand{\tsCapLooseLo}{13.5}
\newcommand{\tsCapLooseHi}{15.8}
\newcommand{\tsCapLoosePoints}{43.2}
\newcommand{\tsCapLooser}{0.30}

\newcommand{\tsCapLooserLo}{8.5}

\newcommand{\tsCapLoosest}{0.50}

\newcommand{\tsCapLoosestHi}{17.2}

\newcommand{\tkTakeoffEighteen}{18}
\newcommand{\tkComputeEndEighteen}{8}
\newcommand{\tkSigmaBEighteen}{0.086116}
\newcommand{\tkHoneEighteen}{2.82}
\newcommand{\tkCapEighteen}{0.0492}
\newcommand{\tkCapPctEighteen}{1.6}
\newcommand{\tkFloorsShareEighteen}{7.7}
\newcommand{\tkLagMonthsEighteen}{21}
\newcommand{\tkPauseSafetyStartEighteen}{1.5}
\newcommand{\tkPauseBlogStartEighteen}{10}
\newcommand{\tkNecBoundEighteen}{0.57}
\newcommand{\tkSpareBoundEighteen}{0.69}
\newcommand{\tkUsualBoundEighteen}{1.08}
\newcommand{\tkScenGapEighteen}{21}
\newcommand{\tkRiskSpareBoundEighteen}{7.8}
\newcommand{\tkRiskSpareFitEighteen}{2.0}
\newcommand{\tkLossSpareFiveEighteen}{13}
\newcommand{\tkOptMonthsEighteen}{7.5}
\newcommand{\tkOptFiveEighteen}{4.1}
\newcommand{\tkCeilLeadHalfEighteen}{25}
\newcommand{\tkCeilLeadQuarterEighteen}{52}
\newcommand{\tkCeilRiseHalfEighteen}{6.6}
\newcommand{\tkCeilRiseQuarterEighteen}{20.1}
\newcommand{\tkTimingConstStartEighteen}{0.0}
\newcommand{\tkTimingRiseStartEighteen}{0.0}
\newcommand{\tkTimingTransferStartEighteen}{2.1}
\newcommand{\tkTimingBothStartEighteen}{9.2}
\newcommand{\tkTimingTransferGainEighteen}{0.03}
\newcommand{\tkTimingBothFirstEighteen}{44.8}
\newcommand{\tkTimingBothBestEighteen}{47.9}
\newcommand{\tkZeroFloorsEighteen}{5.4}
\newcommand{\tkZeroLagEighteen}{16.8}
\newcommand{\tkZeroPauseEighteen}{4.72}
\newcommand{\tkPauseCrossEighteen}{12.8}

\newcommand{\tkTakeoffThirtySix}{36}
\newcommand{\tkComputeEndThirtySix}{64}
\newcommand{\tkSigmaBThirtySix}{0.083958}
\newcommand{\tkHoneThirtySix}{9.08}
\newcommand{\tkCapThirtySix}{0.0358}
\newcommand{\tkCapPctThirtySix}{1.2}
\newcommand{\tkFloorsShareThirtySix}{10.9}
\newcommand{\tkLagMonthsThirtySix}{22}
\newcommand{\tkPauseSafetyStartThirtySix}{17.8}
\newcommand{\tkPauseBlogStartThirtySix}{27.2}
\newcommand{\tkNecBoundThirtySix}{0.50}
\newcommand{\tkSpareBoundThirtySix}{0.59}
\newcommand{\tkUsualBoundThirtySix}{1.08}
\newcommand{\tkScenGapThirtySix}{17}
\newcommand{\tkRiskSpareBoundThirtySix}{6.9}
\newcommand{\tkRiskSpareFitThirtySix}{0.5}
\newcommand{\tkLossSpareFiveThirtySix}{43}
\newcommand{\tkOptMonthsThirtySix}{4.7}
\newcommand{\tkOptFiveThirtySix}{6.1}
\newcommand{\tkCeilLeadHalfThirtySix}{26}
\newcommand{\tkCeilLeadQuarterThirtySix}{53}
\newcommand{\tkCeilRiseHalfThirtySix}{$-$0.3}
\newcommand{\tkCeilRiseQuarterThirtySix}{15.6}
\newcommand{\tkTimingConstStartThirtySix}{0.0}
\newcommand{\tkTimingRiseStartThirtySix}{0.0}
\newcommand{\tkTimingTransferStartThirtySix}{21.3}
\newcommand{\tkTimingBothStartThirtySix}{26.1}
\newcommand{\tkTimingTransferGainThirtySix}{1.08}
\newcommand{\tkTimingBothFirstThirtySix}{43.8}
\newcommand{\tkTimingBothBestThirtySix}{50.0}
\newcommand{\tkZeroFloorsThirtySix}{1.7}
\newcommand{\tkZeroLagThirtySix}{14.5}
\newcommand{\tkZeroPauseThirtySix}{2.30}
\newcommand{\tkPauseCrossThirtySix}{30.0}

\title{Risk Ceilings and Development Deadlines:\\ Pacing AI under Uncertain Safety Productivity}
\author{Li Gan\thanks{Department of Economics, Texas A\&M University. Email: ganli@tamu.edu. I thank Eric Gan for helpful discussions. AI software (Anthropic's Claude) was used under the author's direction to derive and check the algebra, write the calibration and simulation code, and edit the draft. All output was checked by the author, who is responsible for the content. An interactive companion page, which runs the paper's model in the browser, is at \url{https://ligan98.github.io/frontier-pacing/en/ceilings.html}. JEL codes: D81, L51, O33, O38. Keywords: artificial intelligence, catastrophic risk, regulation, safety research, instrument choice.}}
\date{November 2026}
\newcommand{\hlNec}{0.57}
\newcommand{\hlSpare}{0.69}
\newcommand{\hlCapHi}{0.0492}
\newcommand{\hlCapTwentyFive}{0.0102}
\newcommand{\hlCapTen}{0.0037}
\newcommand{\hlCapFive}{0.0018}

\begin{document}
\maketitle
\begin{abstract}
\noindent Proposals to pace frontier AI promise a ceiling on monthly catastrophe hazard and a deadline on development delay. Both rest on an unknown: how productive safety research is. A lab divides growing compute among capability research, safety research, and customers. A ceiling requires a minimum stock of safety knowledge before the deadline, so a joint guarantee requires a lower bound on productivity. In a calibrated scenario, learning first and then replaying development needs a bound within \insFrScenGap{} percent of the minimum any rule needs. At that bound it leaves \insRiskSpareBound{} percent catastrophe risk, against \insRiskFull{} at full speed, and costs \insFrLossSpareFive{} percent of five-year services.
\end{abstract}

\section{Introduction}\label{sec:intro}

Proposals to pace frontier artificial intelligence make two kinds of promises. The first limits danger. The AI Futures Project proposes a cap on the existential risk each frontier company may incur per month, for example 1 percent, as judged by third-party assessors \citep{aifutures2026}. The second limits delay. Pacing must not hand the lead to a rival, and the same proposal estimates that a rival would need about a year to catch up if the leading frontier halted. \citet{amodei2026} argues for pacing in the same spirit: slowing capability gains could buy a year or two for alignment, interpretability, and testing.

Both promises rest on one number that nobody knows: how productive safety research is. Extra time is valuable only if safety research turns it into protection. If research is productive, a little delay buys a lot of safety, and a strict limit on danger costs little time. If research is weak, the same limit on danger can require years of delay. The regulator must write the rules before it learns which case holds.

The first promise is a \emph{risk ceiling}: a cap on the monthly hazard of catastrophe. The second is a \emph{development deadline}: a cap on how many months late any capability milestone may arrive. The deadline stands for the lead over a rival; the rival itself is outside the model. The paper asks whether both promises are feasible, when safety research should occur, and what customers give up.

In the numerical scenario of Section~\ref{sec:scenario}, the tension is easy to see. The final model's unprotected hazard is about \inrRatio{} times the ceiling, so safety knowledge must cut it by that factor, about \lfHalvings{} halvings. At the fitted productivity, with all compute in safety research, one halving takes \lfHalvingMonths{} months, so compliance takes at least \lfMonths{} months. Development may last at most 30 months: 18 at full speed plus the 12 the deadline allows. Keeping both promises requires a productivity of at least \insFrNecAtCap{} of the fitted value. This is a necessary condition; a feasible policy must also allocate compute to development.

The paper builds a model of a frontier lab that divides growing compute among customers, capability research, and safety research. Capability research advances the frontier, and the lab's own models speed up that research. Safety research builds a stock of knowledge that scales down the hazard of every later model. Hazard comes from operating the current model, called \emph{state hazard}, and from advancing to a stronger one, called \emph{step hazard}. The regulator commits to a rule for allocating compute. A guarantee must hold at every productivity the regulator cannot rule out; the lowest such productivity is the \emph{productivity bound}. The analysis asks the three questions of feasibility, timing and customer cost in turn, then compares the proposals that fix allocation instead of hazard.

First, a tight ceiling requires a minimum stock of safety knowledge before the deadline (Section~\ref{sec:keeping}). A lower bound on the required productivity applies to every policy. A constructive upper bound comes from \emph{learn-then-replay}: the lab first devotes all compute to safety, then reproduces the full-speed development path. Growing compute lets it replay with a smaller research share, leaving more compute for safety. The two bounds bracket the productivity needed for a joint guarantee. In the scenario, the bracket is \insFrScenGap{} percent wide. The bracket widens when early research is less productive or its knowledge loses value.

Second, feasibility does not settle when additional learning should occur (Section~\ref{sec:month}). Research done later may be more productive, and its knowledge has less development over which to lose value. But waiting leaves earlier hazards unprotected. An exact comparison weighs these forces for pauses of the same length and safety share. The ceiling and deadline then decide which pauses qualify: a late pause cannot repair a violation that occurs before it begins.

Third, customers pay for safety through reduced compute and delayed improvements in service quality (Section~\ref{sec:price}). The spare-compute replay, which puts all compute the replay does not need into safety research, gives up \insFrLossSpareFive{} percent of five-year service value and all of two-year value. Its catastrophe probability is \insRiskSpareBound{} percent at its productivity bound, against \insRiskFull{} percent in the fitted full-speed benchmark. Stronger evidence about productivity lowers the customer cost. At fitted productivity, the cheapest learn-then-replay policy learns for \insFrOptMonths{} months and gives up \insFrOptFive{} percent of five-year value. Service value and risk are reported separately; the paper does not convert one into the other.

Fourth, rules that fix compute allocation offer a different kind of promise (Section~\ref{sec:fixed}). They fix development dates and let risk vary with safety productivity; a ceiling fixes hazard and lets delay adjust. The scenario gives all four comparisons numerical scale, using the AI Futures Project's judgments rather than estimated parameters.

The paper connects AI-risk regulation to several literatures in economics. Environmental economics models catastrophes whose probability depends on a stock the planner controls \citep{cropper1976,clarke1994,tsur1996,polasky2011}. Here a second stock, safety knowledge, scales the hazard down, and it is built with the same compute that advances the frontier. \citet{jones2024,jones2025} ask how much existential risk to accept for AI's growth and how much to spend against it; research builds protective technology in \citet{jones2016} and \citet{aschenbrenner2019}. \citet{trammell2026} separate risk that accrues with time from risk that accrues with progress, the accounting Section~\ref{sec:fixed} applies; the state and step distinction is \citeauthor{bostrom2014}'s \citeyearpar{bostrom2014}. \citet{acemoglu2024} and \citet{gans2025} study adoption while society learns about harms. \citet{koh2026} show that an adaptive speed limit gives the best worst-case guarantee across ways of learning about a technology's dangers. Here the uncertainty is about how productive safety research is. The value-of-a-month condition in Section~\ref{sec:month} is the safety-versus-exposure condition of \citet{binder2021} and \citet{bostrom2026}, with the pause's own safety allocation and an unknown productivity. Bostrom's multiphase model, in which a later pause can be worth more, motivates the rising yield of Section~\ref{sec:timing}. Racing models make the rival strategic \citep{armstrong2016,buenodemesquita2026,fudenberg2026}; here the deadline is a fixed limit on delay. As in prices versus quantities \citep{weitzman1974}, each rule fixes one outcome and lets another bear the uncertainty. Hard limits on an instrument already matter in \citet{herweg2022}, and intensity standards can induce output expansion \citep{holland2009}. The rules act on compute, which is detectable \citep{sastry2024}, monitorable \citep{shavit2023}, and separable into training and inference \citep{heim2024}.

The contribution links feasibility, timing, and customer cost under uncertain safety productivity. A knowledge requirement and a feasible construction bracket the productivity needed to keep a ceiling and deadline together. An exact timing comparison then operates within a set of pauses that keep both promises, allowing rising productivity and imperfect knowledge transfer. Service costs show what specified guarantees require customers to give up.

Section~\ref{sec:background} introduces the objects behind pacing proposals, Section~\ref{sec:model} builds the model, and Section~\ref{sec:scenario} sets out the scenario. Sections~\ref{sec:keeping}, \ref{sec:month}, and~\ref{sec:price} address feasibility, timing, and customer cost. Section~\ref{sec:fixed} compares fixed allocation rules, and Section~\ref{sec:conclusion} draws the policy lessons.

\section{Pacing Proposals and the Objects behind Them}\label{sec:background}

Pacing proposals concern three uses of compute, two sources of danger, and two promises. This section introduces those objects in words. Table~\ref{tab:rules} previews notation that Section~\ref{sec:model} defines.

\subsection{Compute and the Takeoff Period}

Frontier AI is built with \emph{compute}: specialized chips, and the data centers and electricity that run them. Compute is a frontier lab's main input, and it grows fast. Since the early 2010s, the compute used to train machine-learning models has doubled about every six months \citep{sevilla2022}.

A lab uses its compute in three ways. It serves customers, running the lab's models to answer questions, write code, and do other paid work; this use is called \emph{inference}. It does capability research, training stronger models and running the experiments that find better methods. And it does safety research, which makes models less likely to cause harm. Safety research includes alignment (making models pursue the goals their developers intend), interpretability (understanding what happens inside a model), and testing and evaluation \citep{amodei2026}. The three uses compete for the same chips. Compute used for safety is compute not used to serve customers or to advance the frontier.

One feature makes this allocation unusual: AI models can do AI research. As models improve, the research that produces the next model speeds up. The AI Futures Project's milestones are built on this feedback. Its first milestone, the \emph{Automated Coder}, is an AI that fully automates software engineering. Much of AI research is software engineering, so such a model speeds up the research that produces its successors.

The AI Futures Project describes a \emph{takeoff period} from the Automated Coder to \emph{top-expert-dominating AI}: AI that dominates top human experts at virtually all cognitive tasks. It estimates that at full speed the takeoff period takes 1.5 years. This paper models that period. Capability is an index that runs from 0 at the Automated Coder to 1 at top-expert-dominating AI. Any capability level in between is a \emph{milestone}. \emph{Full speed} is the path a lab follows with its usual allocation of compute and no regulation.

\subsection{Catastrophe and Safety Knowledge}

The catastrophe in this paper is an irreversible loss of human control over AI. The paper reads this danger as misalignment. A more capable model is more likely to pursue goals its developers did not intend and better able to escape control, so the danger rises with capability. Leading AI researchers list this outcome among the extreme risks of advanced AI and warn that safety research is lagging \citep{bengio2024}. The AI Futures Project's worked example puts its probability at 75 percent over a full-speed takeoff period.

The paper treats catastrophe as a \emph{hazard}, as models of environmental catastrophes do. The hazard rate is the chance per month that the catastrophe happens now, given that it has not happened yet. Hazards add up over time. If the hazard is 0.01 a month for a year, the cumulative hazard is 0.12, and the probability of catastrophe within the year is $1-e^{-0.12}$, about 11 percent. In general, a cumulative hazard $H$ gives a probability $1-e^{-H}$. So a probability of 75 percent corresponds to a cumulative hazard of $\ln 4\approx\lfHfull$.

Danger comes from two sources \citep{bostrom2014}. \emph{State hazard} is the danger of having a model of a given capability in the world. A deployed model can act against its developers' intentions, and the longer it is in use, the more chances it has. So state hazard accrues per month spent at a capability level. \emph{Step hazard} is the danger of advancing the frontier: of training a stronger model whose properties nobody yet knows. It accrues per unit of capability gained. It depends on how far the frontier moves, not on how fast.

The distinction matters for pacing. Slowing down spreads the same capability gains over more months. It adds months of state hazard and leaves step hazard unchanged. A pause removes step hazard while it lasts, but state hazard continues.

Safety research builds knowledge: methods to align models, to see inside them, and to test them. The paper treats this knowledge as a stock. Knowledge learned now keeps protecting later, stronger models, so a month of research lowers the hazard of everything that follows. Section~\ref{sec:replay} shows how much the results depend on this transfer.

How much protection a month of research buys is the paper's central unknown. Nobody has measured the productivity of safety research in units of hazard. \citet{amodei2026} argues that today's models make safety research far more productive than it was a few years ago. That is a reason for optimism, but not a number a regulator can rely on.

\subsection{The Proposals and the Two Promises}

The AI Futures Project lists four options for pacing a frontier lab (Table~\ref{tab:rules}). The first is a temporary pause on improving frontier capabilities, with all compute serving customers. The second is a set of \emph{compute floors}: minimum shares of compute for customers and for published safety research, 70 and 25 percent in its example. The third is a lag on the models used for AI research, nine months in its example. The fourth is a maximum risk threshold enforced by third-party assessors, 1 percent existential risk per company per month in its example.

Amodei's case for using the extra time for safety work suggests a variant of the pause in which freed research compute goes to safety research. A \emph{safety floor} isolates one half of the compute floors: it moves compute from customers to safety research and leaves capability research alone. Table~\ref{tab:rules} lists the six rules this paper studies. Each is written as a constraint on how compute is used, with what a regulator would have to observe to enforce it.

\begin{table}[htbp]
\centering
\caption{Six Rules for Pacing a Frontier Lab}
\label{tab:rules}
\small
\begin{tabular}{@{}>{\raggedright\arraybackslash}p{2.9cm}>{\raggedright\arraybackslash}p{7.7cm}>{\raggedright\arraybackslash}p{4.6cm}@{}}
\toprule
Rule & Constraint while it applies & What must be observed\\
\midrule
Compute floors & Minimum shares for customers and for safety research; capability research gets the rest. & How compute is classified; published safety research.\\
Pause, all inference & No compute devoted to capability or safety research: $x=s=0$, $i=1$. & That no compute trains or improves models.\\
Pause with safety & No dedicated capability research; freed compute goes to safety research: $x=0$, $s=s_0+x_0$, $i=i_0$. & That dedicated capability research stops and safety compute does safety research.\\
R\&D-model lag & Capability research uses the model of $\ell$ months earlier. & Which model capability research uses.\\
Safety floor & A safety share drawn from customer compute; capability research unchanged. & How compute is classified; published safety research.\\
Ceiling & The hazard may not exceed $\bar c$ a month; when it binds, the lab slows down just enough to stay at the ceiling and moves the freed compute into safety research. & The hazard itself, as assessed by third parties.\\
\bottomrule
\end{tabular}
\par\smallskip
\begin{minipage}{\textwidth}\footnotesize
\emph{Notes:} The shares $i$, $x$, and $s$ are the fractions of compute used for customers, capability research, and safety research; $(i_0,x_0,s_0)$ are their full-speed values. $F$ is the capability of the frontier. Section~\ref{sec:model} defines each symbol.
\end{minipage}
\end{table}

The ceiling is the fourth option. It caps the monthly hazard of catastrophe. The scenario's ceiling is close to the AI Futures Project's example cap of 1 percent a month per company.

The deadline comes from a rival. The AI Futures Project puts the leading country's lead over its closest rival at four to eight months. If the leader's frontier halted, the rival would need about a year to catch up. So the leader's regulator can afford about a year of delay. The deadline in this paper applies to every milestone, not only to the last one: a rival could overtake the leading lab at any capability level. So no milestone may arrive more than the allowed number of months after its full-speed date. Call that number of months the \emph{lead}.

The regulator must commit before it knows how productive safety research is. A promise is a \emph{guarantee} if it holds at every productivity the regulator cannot rule out. The paper asks when a ceiling and a deadline can be guaranteed together, by which rule, and at what cost.

\subsection{The Climb}\label{sec:climb}

The model maps onto a climbing expedition with two rules: a safety rule (the ceiling) and a weather window (the deadline). Appendix~\ref{supp:exhibits} tabulates the correspondence. The analogy is a reading aid, not part of the argument. It makes the paper's trade-offs concrete. The two kinds of hazard have natural counterparts: the danger of each day spent high up, and the danger of each stretch climbed. One feature of the model has none. In the model, compute keeps growing while the lab waits. So a late start can follow the standard schedule with a small share of compute for capability research, and the rest can keep building protection. The paper states that mechanism in its own terms.

\section{The Model}\label{sec:model}

\begin{figure}[htbp]
\centering
\begin{tikzpicture}[
  box/.style={draw=black!70, rounded corners=2pt, align=center, minimum height=1.0cm, text width=3.0cm, font=\small},
  arr/.style={-{Stealth[length=2.2mm]}, thick, black!70},
  lab/.style={font=\footnotesize, align=center}]
\node[box] (K) {Compute $K(t)$\\ grows at rate $G$};
\node[box, right=1.4cm of K, yshift=1.5cm] (x) {Capability research\\ share $x$};
\node[box, right=1.4cm of K] (s) {Safety research\\ share $s$};
\node[box, right=1.4cm of K, yshift=-1.5cm] (i) {Customers\\ share $i$};
\node[box, right=1.4cm of x] (F) {Capability $F$};
\node[box, right=1.4cm of s] (S) {Safety knowledge $S$};
\node[box, right=1.4cm of i] (J) {Services};
\node[box, right=0.9cm of S, yshift=0.75cm, text width=2.4cm] (lam) {Hazard $\lambda$};
\draw[arr] (K) -- (x.west);
\draw[arr] (K) -- (s.west);
\draw[arr] (K) -- (i.west);
\draw[arr] (x) -- (F);
\draw[arr] (s) -- (S);
\draw[arr] (i) -- (J);
\draw[arr] (F) -- (lam);
\draw[arr] (S) -- (lam);
\draw[arr] (F.north) to[out=150, in=30] node[lab, above] {recursive improvement} (x.north);
\node[lab, below=0.15cm of lam, text width=2.6cm] {ceiling: $\lambda\le\bar c$};
\node[lab, above=0.15cm of F, xshift=1.6cm, text width=3.0cm] {deadline: delay $\le L$};
\end{tikzpicture}
\caption{The Structure of the Model}
\label{fig:schematic}
\par\smallskip
\begin{minipage}{\textwidth}\footnotesize
\emph{Notes:} Arrows show which quantity drives which. The regulator's rule acts on the shares $i$, $x$, and $s$. Safety knowledge lowers the hazard of every later model; the model used for research speeds up capability research. Capability also raises the quality of the services customers receive.
\end{minipage}
\end{figure}
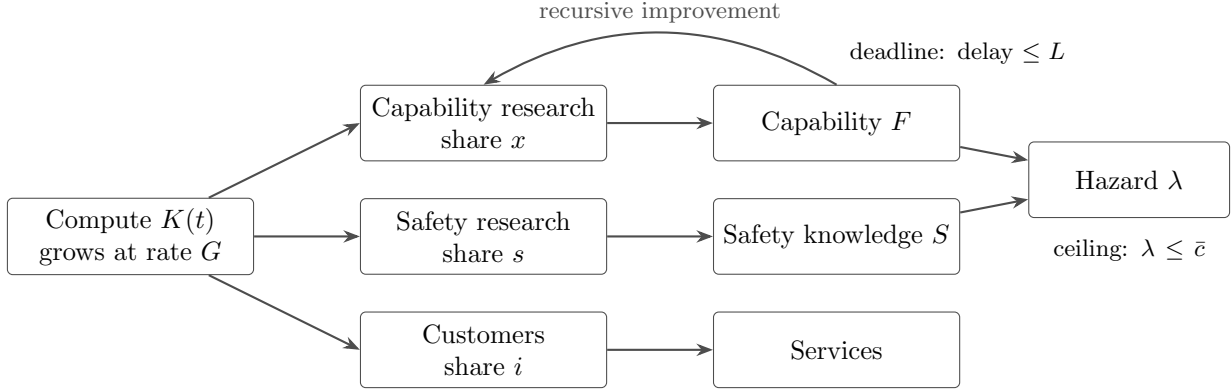

The model has four blocks: compute and capability, safety knowledge, hazard, and customer services. Figure~\ref{fig:schematic} shows how they connect. The lab divides compute among three uses. Capability research raises the frontier, safety research raises knowledge, and inference serves customers. Capability and knowledge together set the hazard. The regulator's rule acts on the division of compute; its two promises constrain the hazard and the dates.

\subsection{Compute, Capability, and Services}\label{sec:capability}\label{sec:services}

Time $t$ is measured in months from the Automated Coder milestone. Capability $F(t)$ rises from $F_0=0$ to $F_1=1$. The takeoff period ends at $T$, the first date at which $F=1$.

Compute grows exogenously at rate $G>0$:
\begin{equation}
K(t)=K_0e^{Gt}.
\label{eq:compute}
\end{equation}
In the scenario, compute doubles every \ipComputeDoubling{} months, so it grows \lfComputeYear-fold in a year and \lfComputeEnd-fold over an 18-month takeoff period. The lab divides compute among customers, capability research, and safety research in nonnegative shares $i$, $x$, and $s$, with $i+x+s=1$.

Capability research advances the frontier at the rate
\begin{equation}
\dot F=\phi(F)\,(xK)^\varepsilon .
\label{eq:research}
\end{equation}
The term $xK$ is the compute used for capability research. The elasticity $\varepsilon>0$ measures how research speed responds to research compute. With $\varepsilon=1/2$, quadrupling research compute doubles research speed. The function $\phi(F)>0$ is the research productivity of a model of capability $F$. In the scenario, it captures recursive improvement: a more capable model does research faster. The scenario of Section~\ref{sec:scenario} uses $\phi(F)=\theta e^{\rho F}$, under which research at top-expert-dominating AI is ten times as fast as at the Automated Coder, for the same compute. The results of Sections~\ref{sec:keeping} to~\ref{sec:fixed} hold for any continuous positive $\phi$.

At full speed, the lab uses the shares $(i_0,x_0,s_0)$, with $x_0,s_0>0$. In the scenario, half of compute goes to capability research, \ipSzero{} percent to safety research, and the rest to customers. The resulting \emph{full-speed path} $F^0(t)$ reaches the final milestone at $T_0$. Write $t_0(F)$ for the date at which milestone $F$ arrives at full speed, and $f^0(F)$ for the full-speed pace $\dot F$ at that milestone.

Research capacity $(x_0K)^\varepsilon$ grows at rate $\gamma=\varepsilon G$, and equation~\eqref{eq:research} separates: $dF/\phi(F)=(x_0K_0)^\varepsilon e^{\gamma t}dt$. So the full-speed path is fixed by $\phi$, the elasticity, and compute growth, and it reaches $F=1$ at $T_0$. When $\phi$ rises with capability, the path rises slowly at first and fast at the end. Compute keeps growing, and each capability gain speeds up the research that produces the next one. Section~\ref{sec:forms} gives the scenario's closed form.

Growing compute has a consequence that the main policy of the paper uses. A lab that starts the full-speed path $\tau$ months late can follow it exactly with a smaller research share:
\begin{equation}
x_0e^{-G\tau}\,K(\tau+u)=x_0K(u)\qquad\text{for all }u\ge0.
\label{eq:replayidentity}
\end{equation}
The left side is the research compute of a lab that starts $\tau$ months late and uses the share $x_0e^{-G\tau}$. The right side is the research compute of the full-speed lab at the same stage $u$ of its path. They are equal, so by equation~\eqref{eq:research} the late lab, researching with its own best model, reaches every milestone exactly $\tau$ months after the full-speed lab. Call equation~\eqref{eq:replayidentity} the \emph{replay identity}. In the scenario compute quadruples in a year. So a lab that starts a year late can reproduce the full-speed research path with one eighth of its compute instead of one half. The rest is free for other uses.

The share $i$ of compute serves customers. They receive services worth $\sqrt{e^{\mu F}iK}$ per month. Inference compute $iK$ produces services whose quality is multiplied by $e^{\mu F}$. The parameter $\mu$ is the increase in log quality per unit of capability. The square root gives diminishing monthly value, so customers value smooth delivery over time. Section~\ref{sec:price} turns this flow into the service value of a policy and measures the price of the guarantee in it.

\subsection{Safety Knowledge and Hazard}\label{sec:knowledge}\label{sec:hazard}

Safety research builds knowledge $S(t)$ from $S(0)=0$ at the rate
\begin{equation}
\dot S=\sigma(F)\,y(s).
\label{eq:safety}
\end{equation}
Knowledge is measured by the protection it provides. The hazard rate below makes this precise: adding $\ln 2$ to $S$ halves the hazard at a given capability and pace.

The equation has two parts. The \emph{safety yield} $\sigma(F)\ge0$ is the monthly growth of knowledge at capability $F$ when the safety share is at its full-speed value $s_0$. The \emph{learning multiplier} $y(s)$ scales that growth with the share of compute in safety research. It is continuous and increasing in $s$, with $y(s_0)=1$, and because $s$ cannot exceed one it is at most $\bar y=y(1)$, the fastest learning possible at a given capability. The scenario gives it a concave form (Section~\ref{sec:forms}); the results need only that it rises with the share.

The productivity of safety research is unknown. Write the yield as
\begin{equation}
\sigma(F)=a\,\bar\sigma(F),
\label{eq:productivity}
\end{equation}
where the profile $\bar\sigma(F)\ge0$ is known and the \emph{productivity} $a\ge0$ is not. The profile says how research opportunities vary with capability; the productivity says how much protection they produce. Let $\bar\sigma_{\max}=\max_F\bar\sigma(F)>0$. In the scenario the profile is constant and fitted to the AI Futures Project's judgments (Section~\ref{sec:scenario}). At the fitted productivity $a=1$, a month of research at the usual share removes \ipShareSigma{} percent of every later hazard.

Conditional on no catastrophe so far, catastrophe arrives at the hazard rate
\begin{equation}
\lambda(t)=[h(F)+k(F)\dot F]\,e^{-S}.
\label{eq:hazard}
\end{equation}
The first term in brackets is state hazard: $h(F)\ge0$ is the hazard per month of operating a model of capability $F$. The second is step hazard: $k(F)\ge0$ is the hazard per unit of capability gained, so per month it is $k(F)\dot F$. Both are the hazards before any safety knowledge, and both are continuous in $F$. The factor $e^{-S}$ is the protection knowledge provides. It scales both kinds of hazard down in the same proportion. Adding $\ln2$ to $S$ halves the hazard; adding \inrNeed{} divides it by about \inrRatio.

Cumulative hazard over the takeoff period is $H=\int_0^T\lambda(t)\,dt$, and the probability of catastrophe is $P=1-e^{-H}$. The rate $\lambda(t)$ is a hazard, not a monthly probability, but over a short interval $dt$ the probability is about $\lambda(t)\,dt$. The baseline counts hazard only during the takeoff period, as the AI Futures Project's 75 percent does. So it assumes that the danger ends at completion, for example because a top-expert-dominating AI still under control can help align its successors \citep{aifutures2026}. Appendix~\ref{app:calibration} considers danger that continues afterward.

A simple accounting shows how speed enters. Take a path without pauses and without safety knowledge. Because $dt=dF/\dot F$, cumulative hazard can be written as a sum over capability levels rather than over months:
\begin{equation}
H=\int_0^1 h(F)\,t'(F)\,dF+\int_0^1 k(F)\,dF,
\label{eq:accounting}
\end{equation}
where $t'(F)=1/\dot F$ is the time spent per unit of capability at milestone $F$. The first integral is state hazard: it grows with the time spent at each level. The second is step hazard: it does not depend on speed at all. A slower path raises the first term and leaves the second unchanged.

\subsection{Rules, Promises, and the Guarantee}\label{sec:rules}\label{sec:promises}

A \emph{rule} specifies the shares, and the model used for research, over time. It may depend on the calendar and on the state of the takeoff period. A rule is \emph{admissible} if its shares are nonnegative and sum to one, and the laws of motion~\eqref{eq:research} and~\eqref{eq:safety} then have a unique solution.\footnote{Throughout, $\sigma$, $h$, and $k$ are continuous and nonnegative, $\phi$ is continuous and positive, and $K_0,x_0,s_0>0$. Prescribed capability dynamics, including their initial history when research is lagged, have unique solutions.} The six rules of Table~\ref{tab:rules} take the following form.
\begin{itemize}
\item \emph{Compute floors} fix the safety and customer shares and leave the rest for capability research.
\item A \emph{safety floor} only moves compute from customers to safety, keeping $x=x_0$.
\item An \emph{all-inference pause} sets $x=s=0$ and $i=1$ for a stated number of months.
\item A \emph{pause with safety} moves the freed research share to safety: $x=0$, $s=s_0+x_0$, $i=i_0$.
\item An \emph{R\&D-model lag} makes research use the model of $\ell$ months earlier: in equation~\eqref{eq:research}, $\phi(F)$ is replaced by $\phi(F(t-\ell))$. Before $\ell$ months have passed, the research model follows the full-speed path extended back before the takeoff period, $F^0(t-\ell)<0$: the lab's research models before the Automated Coder were weaker than it.
\item A \emph{ceiling} requires $\lambda(t)\le\bar c$ at every date before completion. The simulations enforce it by slowing development and moving the freed research compute to safety research (Section~\ref{sec:conventions}).
\end{itemize}
Each rule is one design of a proposal. Other designs exist; for example, a lab could meet a ceiling by adding safety compute rather than by slowing, and a fixed allocation can also satisfy a ceiling.

The deadline concerns delay at every milestone. The \emph{lead used} by a path is its largest milestone delay,
\begin{equation}
D=\sup_{0<F\le1}\,[t(F)-t_0(F)],
\label{eq:lead}
\end{equation}
where $t(F)$ is the date at which the path reaches milestone $F$. The \emph{deadline} requires $D\le L$: no milestone may arrive more than $L$ months late. It implies $T\le T_0+L$, but it is stronger than a deadline for the final milestone alone. A path that falls more than $L$ months behind early on and then catches up still breaks the deadline, because the rival could have overtaken it in between.

The regulator knows the profile $\bar\sigma$, the hazard functions $h$ and $k$, and the capability technology, but not the productivity $a$. Hold $h$, $k$, and the technology fixed as $a$ varies, so a lower $a$ means the same research produces less knowledge, not that the models have become more dangerous. The regulator can rule out productivities below some level, the \emph{productivity bound} $a_{\min}$. It is the lowest productivity the regulator cannot rule out: a lower bound on what is possible, not an average or a forecast.

\begin{definition}\label{def:guarantee}
Let $b>0$. An admissible rule \emph{guarantees} the ceiling $\bar c$ and the deadline $L$ at productivity bound $b$ if, for every $a\ge b$, its path satisfies $\lambda(t)\le\bar c$ for all $t<T$ and $D\le L$. The \emph{threshold} $a^*$ is the infimum of the bounds $b>0$ at which some single admissible rule provides this guarantee.
\end{definition}

The threshold $a^*$ is the paper's target. Below it, no rule keeps both promises at every productivity the regulator must allow for. The threshold need not be attained, and the paper does not compute it exactly. Instead, it brackets the threshold from below, with a bound no rule can beat, and from above, with a simple rule that works.

The timing is simple. The regulator commits to a rule. Nature draws the productivity $a$, which the regulator does not observe. The lab follows the rule, and capability, knowledge, and hazard evolve. Catastrophe arrives at rate $\lambda(t)$.

The model leaves three things out. The lab follows the rule: the paper asks what a rule can achieve if it is followed, not whether the lab will comply. Compute grows exogenously, whatever the rule. And the rival is not modeled; the lead $L$ stands for it. Section~\ref{sec:policy} returns to each.

\section{A Calibrated Scenario}\label{sec:scenario}

The scenario gives every result an example with numbers. It follows the AI Futures Project's worked example, so its numbers show what that example's judgments imply. They are not estimates.

\subsection{Targets, Parameters, and Functional Forms}\label{sec:forms}

The AI Futures Project supplies five judgments. At full speed, the takeoff period from the Automated Coder to top-expert-dominating AI takes 18 months and carries a 75 percent probability of catastrophe. A safety-case regime, in which third-party assessors estimate each company's risk and the government caps it, cuts that probability to about 40 percent while staying ahead of the rival. There are three frontier companies. And the lead over the rival allows about a year of delay. Appendix~\ref{app:calibration} repeats the main results with a slower, 36-month takeoff period; all but one comparison keeps its direction (Table~\ref{tab:takeoff36}).

Table~\ref{tab:calib} lists these targets with the functional forms and the parameters they fix.

\begin{table}[htbp]
\centering
\caption{The Scenario: Source Judgments, Functional Forms, and Fitted Parameters}
\label{tab:calib}
\small
\setlength{\tabcolsep}{4pt}
\begin{tabular}{@{}>{\raggedright\arraybackslash}p{6.3cm}>{\raggedright\arraybackslash}p{9.4cm}@{}}
\toprule
Element & Value and role\\
\midrule
\multicolumn{2}{@{}l}{\emph{Judgments taken from the AI Futures Project's worked example}}\\
Full-speed takeoff period, Automated Coder to top-expert-dominating AI & 18 months; 75 percent probability of catastrophe\\
Safety-case regime & 40 percent probability of catastrophe; mapped here to a 12-month pause at the start with compute standing still (a modeling assumption)\\
Deadline & 12 months of maximum milestone delay, the estimated time a rival would need to catch up\\
Number of frontier companies & 3, for stating the ceiling per company\\
\midrule
\multicolumn{2}{@{}l}{\emph{Functional forms}}\\
Hazard & $[h(F)+k(F)\dot F]e^{-S}$; full-speed hazard per unit of capability grows exponentially, three quarters in the second half; half state hazard\\
Safety research & knowledge scales every later hazard by $e^{-S}$; yield constant over the takeoff period; the compute-using part depends on the safety share\\
Capability research & speed $\phi(F)(xK)^{\varepsilon}$ with $\phi(F)=\theta e^{\rho F}$\\
Services & $Q=e^{\mu F}iK$; value proportional to $\sqrt{Q}$; discounted at 3 percent a year\\
\midrule
\multicolumn{2}{@{}l}{\emph{Illustrative parameters}}\\
Compute; research elasticity; self-improvement & doubles every \ipComputeDoubling{} months; $\varepsilon=\ipLam$; research \ipRSI{} times faster at the endpoint\\
Compute shares at full speed & research \ipXzero{} percent, safety \ipSzero{} percent, customers \ipIzero{} percent\\
Safety research without compute; returns & $\psi=\ipPsi$; $\alpha=\ipAlpha$\\
Services & quality per unit of inference rises by a factor of \ipMu{} at the endpoint\\
\midrule
\multicolumn{2}{@{}l}{\emph{Fitted to the targets}}\\
Safety yield & \ipShareSigma{} percent of later hazard removed per month, to the 40 percent outcome\\
Takeoff speed $\theta$ & to the 18-month takeoff period\\
Hazard scale & to the 75 percent probability at full speed\\
\bottomrule
\end{tabular}
\end{table}

The scenario now specifies capability research and the learning multiplier. Research productivity is $\phi(F)=\theta e^{\rho F}$ with $\rho=\ln10$: research at top-expert-dominating AI is ten times as fast as at the Automated Coder, for the same compute. The constant $\theta>0$ sets the overall speed and is fitted to the 18-month takeoff period. Appendix~\ref{app:proofs} gives the closed form of the path $F^0(t)$ and of the arrival dates $t_0(F)$. In the scenario, full-speed research speed rises \lfSpeedRatio-fold over the takeoff period. The frontier covers the upper half of its capability range in the last \lfMonthsAfterHalf{} months (Figure~\ref{fig:scenario}a).

The learning multiplier is
\begin{equation}
y(s)=\psi+(1-\psi)\Bigl(\frac{s}{s_0}\Bigr)^\alpha ,
\label{eq:multiplier}
\end{equation}
with $\psi=\alpha=1/2$. A share $\psi$ of knowledge growth needs no dedicated compute, for example what the lab learns from studying its deployed models. The rest uses compute, with diminishing returns, $\alpha<1$.

Diminishing returns make dedicated compute a limited lever. In the scenario, $\psi=\alpha=1/2$ and $s_0=0.05$. Moving all compute into safety research multiplies the safety share by 20, but multiplies learning only by $\bar y=0.5+0.5\sqrt{20}\approx\insYbar$.

A pause that moves the research share into safety research, so that $s=s_0+x_0$, gives $y=\lfMultSafety$. A pause that sends all compute to customers gives $y=\psi=0.5$: learning continues at half its usual speed.

\subsection{The Full-Speed Path and the Fitted Productivity}\label{sec:fit}

\begin{figure}[htbp]
\centering
\includegraphics[width=0.92\textwidth]{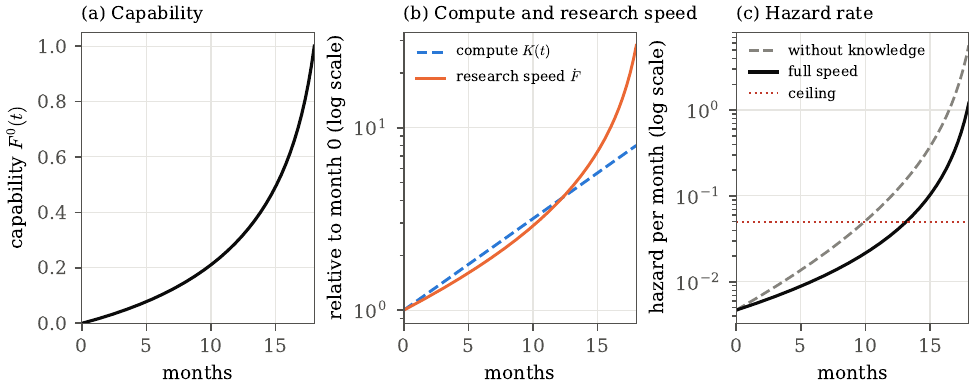}
\caption{The Full-Speed Scenario}
\label{fig:scenario}
\par\smallskip
\begin{minipage}{\textwidth}\footnotesize
\emph{Notes:} Panel (a) shows capability on the full-speed path, equation~\eqref{eq:fullspeed}. Panel (b) shows compute and research speed relative to their values at the Automated Coder. Panel (c) shows the hazard rate per month on a log scale: with the knowledge the lab builds at full speed (solid), and without any safety knowledge (dashed). The dotted line is the scenario's ceiling of \insCap{} a month. At full speed, state and step hazard are equal at every date.
\end{minipage}
\end{figure}

Figure~\ref{fig:scenario} shows the full-speed scenario. Capability creeps up for a year and then accelerates (panel a). Compute grows eightfold over the 18 months, and research speed grows faster still, because recursive improvement compounds compute growth (panel b).

The hazard rises steeply with capability (panel c). Full-speed hazard per unit of capability is $b_He^{\kappa F}$, with $\kappa=2\ln3$, so three quarters of the full-speed cumulative hazard comes after capability passes one half. That happens at month \lfHalfMonth, so three quarters of the danger is concentrated in the last \lfMonthsAfterHalf{} months. Even with safety knowledge, the full-speed hazard rises from \lfHazStart{} a month at the start to \lfHazEnd{} a month at the end. A constant hazard at that final rate would imply about \lfHazEndProb{} percent catastrophe risk within one month. At the ceiling of \insCap{} a month, the corresponding probability is \lfCapProb{} percent. Without any knowledge, the final hazard would be \lfHazZeroEnd{} a month, half of it state hazard ($h(1)=\lfHone$) and half step hazard.

The hazard functions $h$ and $k$ are chosen to deliver this density. Each undoes the knowledge the full-speed lab has built by milestone $F$. Net of that knowledge, the hazard per unit of capability is exactly $b_He^{\kappa F}$, with $b_H=-\ln(1-0.75)/\int_0^1e^{\kappa F}dF$ fitted to the 75 percent probability. A share $\omega$ of the density comes from the model's state and $1-\omega$ from each step; the scenario sets $\omega=1/2$. The explicit forms are in Appendix~\ref{app:calibration}.

The safety yield is fitted to the 40 percent outcome of the safety-case regime. The paper maps that regime to a 12-month pause at the start of the takeoff period, with the safety share at $s_0$ and the compute clock stopped. The later path then repeats the full-speed path. This mapping is a modeling assumption. The fitted constant yield $\sigma_b$ solves
\begin{equation}
\frac{h(0)(1-e^{-12\sigma_b})}{\sigma_b}
+e^{-12\sigma_b}\,[-\ln(1-0.75)]=-\ln(1-0.40).
\label{eq:fit}
\end{equation}
The first term is cumulative exposure during the 12-month pause. Its starting hazard, $h(0)=\omega b_Hf^0(0)=\lfHzero$ a month, does not depend on $\sigma_b$. Knowledge then reduces that hazard throughout the pause. The second term is the full-speed cumulative hazard, $\ln4\approx\lfHfull$, scaled down by the knowledge the pause adds. The right side is the cumulative hazard of a 40 percent probability.

The solution is $\sigma_b=\irSigmaB$ a month: a month of research at the usual share removes \ipShareSigma{} percent of every later hazard. With it, the pause's knowledge multiplies all later hazard by $e^{-12\sigma_b}=\lfDisc$, cutting the full-speed cumulative hazard from \lfHfull{} to \lfHafter. Waiting adds \lfExpo{} in cumulative hazard. This cost is small because the initial model has a low hazard. The total, about \lfFitTotal, is a 40 percent probability. Section~\ref{sec:month} shows that this is the value of a month at work.

Productivity is measured relative to this fit: $a=1$ is the \emph{fitted productivity}, and $a=0.5$ means research produces half as much knowledge per month.

\subsection{The Rules and Two Hazard Conventions}\label{sec:conventions}

The deadline is 12 months of maximum milestone delay. The ceiling is set so that the ceiling controller uses exactly that year of lead at the fitted productivity. It is about \ipEqCeilingPct{} percent hazard per month per company for three companies, an aggregate ceiling of $\bar c=\insCap$ a month. That is close to the AI Futures Project's example of 1 percent a month per company. At full speed, the hazard exceeds this ceiling from month \lfCrossMonth{} on (Figure~\ref{fig:scenario}c).

The ceiling controller works as follows. In the mixed-hazard simulations, $k(F)>0$, and the controller moves freed research compute to safety research. Provided state hazard $h(F)e^{-S}$ does not exceed $\bar c$, it sets the pace $\dot F=[\bar ce^{S}-h(F)]/k(F)$, or the available full-research pace if lower. If state hazard exceeds the ceiling, no nonnegative pace can satisfy it at that date. The case $k(F)=0$ requires a separate rule (Appendix~\ref{app:calibration}).

The compute floors, pauses, and R\&D-model lag also use a year of lead at the fitted productivity. Compute floors fix research at \ipEqFloorsShare{} percent and safety at 25 percent. The two pauses last 12 months and start at months \ipEqPauseSafetyStart{} and \ipEqPauseBlogStart. The R\&D-model lag is \ipEqLagMonths{} months (reaching into the later months of the slowed takeoff period), and the safety floor uses no lead. These are fixed settings for comparison, not each rule's best design.

Two exercises vary productivity, and they hold different things fixed. Under \emph{fixed hazards}, $h$ and $k$ stay at their scenario values while productivity varies. This is the convention of the propositions: a lower productivity means only that research produces less knowledge. Then the full-speed probability of catastrophe changes with productivity.

Under \emph{normalized hazards}, $h$ and $k$ are rescaled at each productivity so that full speed keeps its 18-month takeoff period and its 75 percent probability. This convention matches a common risk judgment across versions of the safety technology, and it is the one Section~\ref{sec:fixed} uses to compare rules. Each figure and table states its convention. The 75 percent is the AI Futures Project's value. Others put it lower: \citet{samotsvety2022} give 25 percent for misaligned AI takeover by 2100, and surveys of AI researchers report medians of 5 to 10 percent \citep{grace2025}. Because the ceiling is set to use the lead, a lower full-speed probability scales the hazard and the matching ceiling together and leaves every bound unchanged (Appendix~\ref{supp:checks}, which reruns the scenario at 25, 10 and 5 percent).

\section{Feasibility: Keeping Both Promises}\label{sec:keeping}

Feasibility comes first. A tight ceiling requires safety knowledge, and the deadline limits the time available to produce it. A necessary condition rules out some productivity bounds. Learning first and then replaying development supplies a sufficient condition. The gap between them records what these results leave unresolved.

\subsection{What a Ceiling Requires}\label{sec:requirement}

Four steps bound the time any path needs to respect a ceiling $\bar c$ below the final model's unprotected state hazard $h(1)$. Throughout, productivity enters through the yield $\sigma(F)=a\bar\sigma(F)$, and the hazard functions stay fixed as $a$ varies.

\emph{Step 1: the final model must be under the ceiling.} Step hazard is never negative, so the hazard is at least the state hazard: $\lambda(t)\ge h(F(t))e^{-S(t)}$. Just before the takeoff period ends, capability is close to one. So the ceiling requires $h(1)e^{-S(T)}\le\bar c$, or
\[
S(T)\ge\ln\frac{h(1)}{\bar c}.
\]
The right side is the knowledge the final model needs: the log of the factor by which its state hazard must fall.

\emph{Step 2: knowledge grows at a bounded rate.} The yield is at most $a\bar\sigma_{\max}$, and the learning multiplier is at most $\bar y$. So knowledge grows by at most $a\bar\sigma_{\max}\bar y$ a month, and $S(T)\le a\bar\sigma_{\max}\bar yT$.

\emph{Step 3: combine.} Steps 1 and 2 give a minimum duration for the takeoff period:
\[
T\ge\frac{\ln[h(1)/\bar c]}{a\bar\sigma_{\max}\bar y}.
\]

\emph{Step 4: compare with the deadline.} The final milestone is one of the milestones the deadline covers, so the lead used is at least the delay at completion, $D\ge T-T_0$. So the deadline requires $T\le T_0+L$. If productivity is so low that the minimum duration exceeds $T_0+L$, no path keeps both promises.

The four steps prove part (b) of the next proposition. Part (a) is the mirror image for rules that ignore safety knowledge, which Section~\ref{sec:fixed} develops.

\begin{proposition}\label{prop:fix}
(a) Prescribed allocation rules fixed independently of $a,S,\lambda$ have arrival dates independent of $a$.

\noindent(b) For a fixed ceiling $0<\bar c<h(1)$ and productivity $a>0$, every path that respects the ceiling satisfies
\begin{equation}
T\ge\frac{\ln[h(1)/\bar c]}{a\bar\sigma_{\max}\bar y}.
\label{eq:bound}
\end{equation}
It violates the deadline if
$a<\ln[h(1)/\bar c]/[\bar\sigma_{\max}\bar y(T_0+L)]$.
\end{proposition}

The proof is in Appendix~\ref{app:proof-fix}. In equation~\eqref{eq:bound}, the numerator $\ln[h(1)/\bar c]$ is the knowledge required. If the unprotected final hazard is twice the ceiling, at least $\ln 2$ of knowledge is needed to halve it. The denominator $a\bar\sigma_{\max}\bar y$ is an upper bound on knowledge produced per month, even with all compute assigned to safety at the most productive capability. So the ratio is a minimum learning time that no development path can evade. Slowing down, pausing, or changing the order of research does not help: the knowledge must be produced, and it cannot be produced faster than the bound allows.

The scenario makes the bound concrete. The final model's unprotected state hazard is $h(1)=\lfHone$ a month, about \inrRatio{} times the ceiling of \insCap. So the lab needs $S(T)\ge\ln[h(1)/\bar c]\approx\inrNeed$, about \lfHalvings{} halvings of the final hazard. At the fitted productivity, the fastest learning is $\sigma_b\bar y\approx\lfRate$ a month, so one halving takes \lfHalvingMonths{} months and compliance takes at least \lfMonths{} months. That fits within $T_0+L=30$ months. But it does not fit once productivity falls below $\lfMonths/30\approx\insFrNecAtCap$ of the fitted value. A regulator who cannot rule out productivity below \insFrNecAtCap{} cannot promise this ceiling and this deadline together, whatever rule it chooses.

The requirement can be read as a limit on the ceiling. Let $a_{\min}$ be the productivity bound. A necessary condition for guaranteeing the deadline is then
\begin{equation}
\bar c\ge h(1)e^{-a_{\min}\bar\sigma_{\max}\bar y(T_0+L)}.
\label{eq:necessary}
\end{equation}
The right side is the final model's state hazard after $T_0+L$ months of maximal learning at productivity $a_{\min}$. No ceiling below it can be guaranteed. The smallest productivity bound that satisfies condition~\eqref{eq:necessary} is the \emph{necessary bound}. It rises as the ceiling tightens, because more knowledge is required. It falls as the lead lengthens, because there is more time to learn.

Two implications follow. First, if productivity can be arbitrarily small, every fixed ceiling below $h(1)$ fails in some state of the world: a ceiling can be guaranteed only together with a productivity bound. Second, the condition is generous. It lets learning run at the maximal rate for the whole time. Attaining $\bar y$ uses all compute for safety research, which leaves none to advance capability, and the most productive capability may be unavailable early on. So the condition is necessary but not sufficient. The next section asks how far from sufficient it is.

The same argument applies at every milestone, not only the last. Appendix~\ref{supp:extensions} states the milestone version, which can bind at an interior milestone when the deadline leaves little time to learn before it.

\subsection{Learn, Then Replay}\label{sec:replay}

The policy has two phases. In the first, the lab learns: for $\tau$ months, all compute goes to safety research, and capability stays at the Automated Coder. In the second, the lab replays: it follows the full-speed development path, $\tau$ months late. Call this policy \emph{learn-then-replay}.

Growing compute is what makes the replay cheap. By the replay identity~\eqref{eq:replayidentity}, the lab needs only the research share $x_0e^{-G\tau}$ to follow the full-speed path. The rest of the compute is free. The regulator chooses how much of it goes to safety research, the \emph{replay safety share} $s_r$; customers get the remainder, $1-x_0e^{-G\tau}-s_r$. Every milestone arrives exactly $\tau$ months late, so the replay uses exactly $\tau$ months of lead.

\begin{figure}[htbp]
\centering
\includegraphics[width=0.92\textwidth]{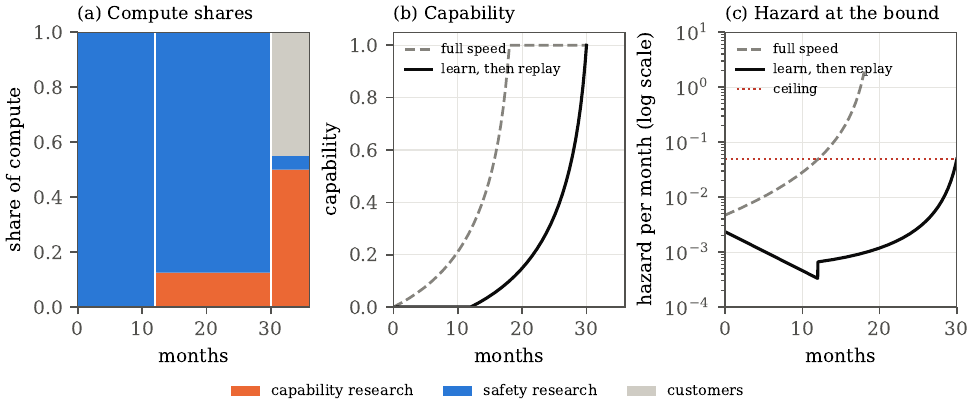}
\caption{Learn-then-Replay at Its Productivity Bound}
\label{fig:replay}
\par\smallskip
\begin{minipage}{\textwidth}\footnotesize
\emph{Notes:} The spare-compute replay with 12 months of learning, at the scenario's ceiling, evaluated at its productivity bound of \insFrSpareAtCap{} times the fitted productivity, with the scenario's fixed hazards. Panel (a): compute shares; after the takeoff period ends at month 30, full-speed shares resume. Panel (b): capability, against the full-speed path. Panel (c): hazard per month on a log scale, against the full-speed path at the same productivity and the ceiling (dotted).
\end{minipage}
\end{figure}

Figure~\ref{fig:replay} shows the policy in the scenario with a year of learning and all spare compute in safety research. During the learning year, capability stays at zero and the hazard is the initial model's state hazard, falling as knowledge accumulates (panel c). At month 12, development resumes. Research takes \insFrReplayShare{} percent of compute, one eighth instead of one half, and safety research takes the rest; customers get nothing until the takeoff period ends at month 30 (panel a). Capability follows the full-speed path exactly a year late (panel b).

The hazard jumps when development resumes, because step hazard starts again; at full pace it equals state hazard, so the hazard doubles. It then rises with capability and touches the ceiling exactly at the last milestone.

The figure is drawn at the policy's productivity bound, \insFrSpareAtCap{} of the fitted value, the lowest productivity at which it keeps the ceiling. At that productivity, the year of learning builds \lfKnowLearn{} of knowledge and the replay adds \lfKnowReplay. The total, \lfKnowEnd, is exactly what the last milestone needs: its hazard before knowledge, state plus step, divided by the ceiling, is $e^{\lfKnowEnd}$. Full speed at the same productivity would breach the ceiling from month \lfFullCrossBound{} on. At any higher productivity, the policy's hazard stays below the ceiling throughout.

\subsection{When Learn-then-Replay Works}

Three steps show when the policy keeps both promises at every productivity $a\ge a_{\min}$. Two quantities along the full-speed path enter. The first is the hazard at zero safety knowledge, $\lambda^0(u)=h(F^0(u))+k(F^0(u))\dot F^0(u)$. The second is the knowledge accumulated per unit of productivity at the usual safety share, $\bar S^0(u)=\int_0^u\bar\sigma(F^0(v))\,dv$.

\emph{Step 1: the learning phase.} While the lab learns, capability stays at zero, so the hazard is $h(0)$ times a falling knowledge factor. It never exceeds $h(0)$. So the learning phase respects the ceiling if and only if the initial model does: $\bar c\ge h(0)$.

\emph{Step 2: the replay.} At stage $u$ of the replay, the lab holds $a\bar\sigma(0)\bar y\tau$ of knowledge from the learning phase and $ay(s_r)\bar S^0(u)$ from the replay. Its hazard factors into the replay piece and the initial-learning piece, $\lambda(\tau+u)=[\lambda^0(u)e^{-ay(s_r)\bar S^0(u)}]\,e^{-a\bar\sigma(0)\bar y\tau}$. Because hazard falls as $a$ rises, the worst case over $a\ge a_{\min}$ is $a=a_{\min}$. Let
\[
B(s_r)=\max_{0\le u\le T_0}\,\lambda^0(u)\,e^{-a_{\min}y(s_r)\bar S^0(u)}
\]
be the largest replay hazard at $a_{\min}$ before counting the knowledge from the learning phase. The replay respects the ceiling if $B(s_r)e^{-a_{\min}\bar\sigma(0)\bar y\tau}\le\bar c$.

\emph{Step 3: learn as long as the lead allows.} A longer learning phase only helps: it adds knowledge, and it lowers the research share the replay needs. So if any learning time up to $L$ works, $\tau=L$ works.

\begin{proposition}\label{prop:construct}
Suppose $\bar\sigma(0)>0$. Fix a positive ceiling $\bar c$ and a replay safety share $s_r$ with $s_0\le s_r\le1-x_0e^{-GL}$. Consider precommitted policies that devote all compute to safety research for $\tau$ months. They then replay the reference capability path with research share $x_0e^{-G\tau}$, safety share $s_r$, and the remainder for customers. Let
\[
B(s_r)=\max_{0\le u\le T_0}\,\lambda^0(u)\,e^{-a_{\min}y(s_r)\bar S^0(u)}.
\]
Some policy in this class respects the ceiling and the deadline for every $a\ge a_{\min}>0$ if and only if
\begin{equation}
\bar c\ge h(0)\qquad\text{and}\qquad
B(s_r)e^{-a_{\min}\bar\sigma(0)\bar y L}\le\bar c.
\label{eq:construct}
\end{equation}
\end{proposition}

The proof is in Appendix~\ref{app:proof-construct}. The two inequalities check the two phases separately. The initial model must already satisfy the ceiling, because knowledge starts at zero. On the replay, $B(s_r)$ is the largest hazard after allowing for learning during development but before adding the initial learning stock. The exponential factor is the protection that $L$ months of initial learning provides. The condition is exact within the class and sufficient among all policies.

For a given replay safety share, the smallest productivity bound that satisfies condition~\eqref{eq:construct} is the \emph{replay bound}. Call the set of these policies, with any learning time $\tau\le L$ and replay safety share $s_0\le s_r\le1-x_0e^{-G\tau}$, the \emph{replay class}.

The replay safety share is the regulator's lever, and two choices mark its range. The \emph{usual-share replay}, $s_r=s_0$, leaves customers most of the compute that development does not need. The \emph{spare-compute replay}, $s_r=1-x_0e^{-GL}$, puts all of it in safety research, as in Figure~\ref{fig:replay}. It raises learning during the replay from $y(s_0)=1$ to $y_r=y(1-x_0e^{-GL})$, which approaches $\bar y$ when compute growth makes the replay research share small.

At the scenario's ceiling, the spare-compute replay needs a productivity bound of \insFrSpareAtCap{} and the usual-share replay needs \insFrSzeroAtCap. The necessary bound is \insFrNecAtCap.

\subsection{How Close the Two Bounds Are}

The necessary bound imagines that learning can use all compute for the entire available time. A feasible policy must also develop capability. Learn-then-replay comes close when later compute is abundant: reproducing the original research path takes a small share, leaving nearly all compute for learning even during development. The next proposition makes this precise when all hazard comes from the frontier's state and the yield is constant.

The proposition needs one restriction on the shape of the hazard. Knowledge keeps accumulating during the replay, so the protected hazard could peak in the middle of the replay rather than at the end. The restriction rules this out: unprotected state hazard must rise along the full-speed path at least as fast as the replay's knowledge grows. Then the protected hazard keeps rising along the replay, as it does in Figure~\ref{fig:replay}c, and the last milestone is the hardest to keep below the ceiling.

\begin{proposition}\label{prop:sharp}
Suppose all hazard is state hazard ($k\equiv0$), the yield is constant ($\bar\sigma(F)=\bar\sigma>0$), $0<\bar c<h(1)$, $\bar c\ge h(0)$, and $h(F^0(u))>0$ throughout the reference path. Let $y_r=y(1-x_0e^{-GL})$ and
\[
\bar a=\frac{\ln[h(1)/\bar c]}{\bar\sigma(\bar yL+y_rT_0)},
\]
and suppose $\ln h(F^0(u))-\ln h(F^0(v))\ge\bar a\bar\sigma y_r(u-v)$ for all $0\le v\le u\le T_0$. Define $a^*$ as the infimum over $b>0$ for which a single admissible policy guarantees the ceiling and the deadline for every $a\ge b$; the infimum need not be attained. Then
\begin{equation}
\frac{\ln[h(1)/\bar c]}{\bar\sigma\bar y(T_0+L)}\ \le\ a^*\ \le\ \bar a.
\label{eq:sharp}
\end{equation}
\end{proposition}

\begin{figure}[htbp]
\centering
\includegraphics[width=0.92\textwidth]{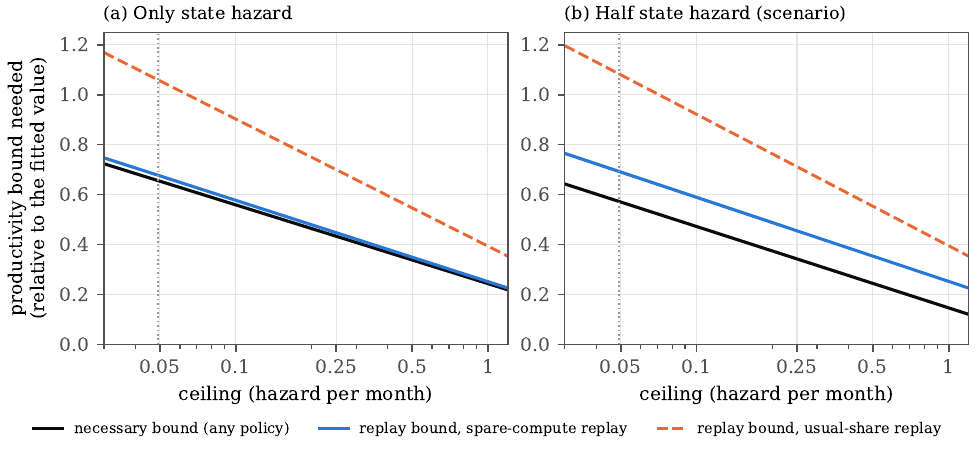}
\caption{What Can Be Promised: The Productivity Needed for a Ceiling and a One-Year Deadline}
\label{fig:bounds}
\par\smallskip
\begin{minipage}{\textwidth}\footnotesize
\emph{Notes:} Productivity bounds are relative to each version's fitted productivity; hazard functions stay fixed as productivity varies. The black line is the necessary bound for any policy (Proposition~\ref{prop:fix}); the other lines are replay bounds with 12 months of initial learning (Proposition~\ref{prop:construct}). Panel (a) refits the scenario with only state hazard; its spare-compute line is the replay bound $\bar a$ of Proposition~\ref{prop:sharp}. Panel (b) uses the scenario's even split between state and step hazard. The dotted line marks the scenario's ceiling. Below the black line, no policy keeps both promises; on or above a replay line, that replay keeps both.
\end{minipage}
\end{figure}

Proposition~\ref{prop:sharp} brackets the productivity needed for a joint promise between the necessary bound and the replay bound $\bar a$. The lower bound is Proposition~\ref{prop:fix}. The upper bound is Proposition~\ref{prop:construct} applied to the spare-compute replay. With the shape restriction, the largest replay hazard is at the last milestone, and condition~\eqref{eq:construct} holds with equality at $\bar a$. Appendix~\ref{app:proof-sharp} gives the proof. The scenario satisfies the restriction at every ceiling shown in Figure~\ref{fig:bounds}a.

The two bounds differ only because the replay must spend some compute on research, so it learns more slowly than the necessary bound assumes. Their ratio is
\[
\frac{\bar a}{\text{necessary bound}}=\frac{\bar y(T_0+L)}{\bar yL+y_rT_0}.
\]
The numerator is the learning the necessary bound allows: the maximal rate $\bar y$ for the whole lead-extended takeoff period. The denominator is the learning the policy achieves: the maximal rate for $L$ months, then the replay rate $y_r$ for $T_0$ months. The factor does not depend on the ceiling, although the shape restriction limits the ceilings to which Proposition~\ref{prop:sharp} applies. It falls to one as compute growth makes the replay's research share vanish, as long as the shape restriction still holds.

In the state-hazard version of the scenario, a year of compute growth cuts the replay research share to \insFrReplayShare{} percent. The replay then learns at $y_r=\insFrYspare$ against $\bar y=\insYbar$. The factor is $\insYbar\times30/(\insYbar\times12+\insFrYspare\times18)\approx\insFrPureFactor$, so the bounds differ by \insFrPureGapPct{} percent. In Figure~\ref{fig:bounds}a the two solid lines nearly coincide.

Compute growth drives the result. The factor depends on compute growth only through the replay research share. With a 12-month doubling time the gap is \insDoubleTwelve{} percent, with 24 months \insDoubleTwentyFour{} percent, and with no compute growth at all the formula gives \lfFactorNonePct{} percent (Appendix~\ref{supp:doubling}).

\subsection{When the Bracket Widens}\label{sec:widens}

The near coincidence of the bounds rests on three conditions: all hazard is state hazard, knowledge learned early keeps its value, and compute grows fast. Relaxing the first two widens the bracket.

\emph{Step hazard.} The necessary bound counts only the final model's state hazard. A replay at the reference pace also carries the hazard of each capability step. In the scenario, state and step hazard are equal along the full-speed path. So at the last milestone, replayed at full pace, the hazard before knowledge is twice the state hazard: $\lambda^0(T_0)=2h(1)$. The replay must hold one more halving of knowledge, $\ln 2$, than the necessary bound counts. When the last milestone binds, the ratio of the two bounds is
\[
\frac{\text{replay bound}}{\text{necessary bound}}=\Bigl(1+\frac{\ln 2}{\ln[h(1)/\bar c]}\Bigr)\times\frac{\bar y(T_0+L)}{\bar yL+y_rT_0}.
\]
At the scenario's ceiling, the necessary bound requires $\ln[h(1)/\bar c]\approx\inrNeed$, so the extra halving adds \lfExtraPct{} percent to the knowledge requirement. The ratio of bounds is then $(1+\lfExtra)\times\insFrPureFactor\approx\lfGapPred$, so the replay bound is \insFrScenGap{} percent above the necessary bound.

At looser ceilings the requirement is smaller, so the same extra halving weighs more. At a ceiling of about 1 a month, the requirement is only about \lfWorstLnreq, so the extra halving adds \lfWorstExtraPct{} percent, and the gap is \insFrHalfSpareMax{} percent. This holds for every ceiling between the scenario's and 1 a month (Figure~\ref{fig:bounds}b). A necessary bound that also charged for step hazard could narrow this band.

Rising yields or knowledge loss widen the bracket. The policy works because knowledge learned on the initial model keeps protecting stronger ones. If the yield instead triples between the first and last milestones, the necessary bound lets learning run at the final yield, while the policy learns at the initial one. With only state hazard, the factor at the scenario's ceiling is then \insFrRiseRatio. If half of knowledge is lost as capability rises, the factor is \insFrTransferRatio, and the replay needs more than the fitted productivity itself. With the scenario's even split between state and step hazard, the two factors are \insFrRiseRatioHalf{} and \insFrTransferRatioHalf{} (Appendix~\ref{supp:exhibits}).

The necessary bound is loose in both cases, so these factors bound the policy's shortfall rather than measure it. Learning first is cheap only when early knowledge protects later models.

The construction establishes feasibility when its condition holds. It does not show that learning first is the best timing when stronger models make safety research more productive \citep{bostrom2026}. A later pause may produce more useful knowledge, but it must also respect the ceiling before learning begins. The next section compares timings and tests which pauses meet both promises.

\section{Timing: When to Learn}\label{sec:month}

An extra month of research protects later development but adds exposure while the lab learns. Developing further can improve productivity and reduce transfer loss, while leaving earlier hazards unprotected. We first value delay on a prescribed path, then compare finite pauses and restrict choices to those that meet the ceiling and deadline.

\subsection{A Benchmark That Isolates Time}

A benchmark isolates the two effects. Suppose that after a pause the lab resumes exactly the path it would have followed, as a function of capability. It keeps the same pace at each milestone and the same safety share. Call this the \emph{repeated-path benchmark}. In the benchmark, delay changes when the lab reaches each milestone but not how it moves afterward. Growing compute breaks the benchmark, because a lab that waits has more compute when it resumes; Section~\ref{sec:pause} returns to this.

Two quantities measure the trade-off. The first is the hazard still ahead,
\[
A(t)=\int_t^T\lambda(u)\,du,
\]
the cumulative hazard, state and step, that remains on the reference path after date $t$. The second is the hazard of standing still, $w(t)=h(F(t))e^{-S(t)}$. During a pause $\dot F=0$, so step hazard stops and state hazard is the only danger: $w(t)$ is the exposure cost of waiting. Let $s_p$ be the safety share during the pause, so knowledge grows by $m=\sigma(F(t))y(s_p)$ a month.

A short pause of length $d\delta$ adds about $w(t)\,d\delta$ of exposure. It also adds $m\,d\delta$ of knowledge. That knowledge reduces every later hazard by about the same fraction, so it saves about $mA(t)\,d\delta$ of the hazard still ahead. The net saving per month of pause is
\begin{equation}
v_{s_p}(t)=mA(t)-w(t).
\label{eq:month}
\end{equation}
Equation~\eqref{eq:month} values time in cumulative hazard saved. The term $mA(t)$ is the benefit of learning: the knowledge a month adds, times the hazard it protects. The term $w(t)$ is the cost of exposure. A month of delay is worth taking when the first exceeds the second. The exact effect of a finite pause is given next.

For positive learning, three steps give the exact effect (Appendix~\ref{app:proofs}).

\emph{Step 1: exposure during the pause.} During the pause, the exposure rate falls as knowledge accumulates, from $w(t)$ to $w(t)e^{-mu}$ after $u$ months, so the pause adds $w(t)(1-e^{-m\delta})/m$.

\emph{Step 2: protection after the pause.} After the pause, the lab repeats the reference path with $m\delta$ more knowledge, so every remaining hazard is multiplied by $e^{-m\delta}$, which subtracts $(1-e^{-m\delta})A(t)$.

\emph{Step 3: net change.} Adding the two pieces gives the formula in the proposition.

\begin{proposition}\label{prop:month}
Under the repeated-path benchmark, a pause of $\delta$ months changes cumulative hazard by
$\Delta H=-[(1-e^{-m\delta})/m]v_{s_p}(t)$.
For $m>0$ it lowers catastrophe probability if and only if $v_{s_p}(t)>0$; at $m=0$, $\Delta H=\delta w(t)$.
\end{proposition}

The proof is in Appendix~\ref{app:proof-month}. The factor $(1-e^{-m\delta})/m$ measures the pause in effective months. Because exposure shrinks as knowledge accumulates, a pause of $\delta$ months costs as much exposure as $(1-e^{-m\delta})/m$ months at the initial rate. It is close to $\delta$ for a short pause and smaller for a long one.

A pause that lowers cumulative hazard also lowers the probability of catastrophe, because $P=1-e^{-H}$ rises with $H$. In probability units, the marginal value of a month is $e^{-H}v_{s_p}(t)$. A larger safety share during the pause raises $m$, and so the benefit of waiting, as long as dedicated compute contributes to learning (the multiplier $y$ rises with $s$). When research does nothing ($m=0$), a pause only adds exposure: $\Delta H=\delta w(t)$.

The fit of the safety yield in Section~\ref{sec:scenario} is Proposition~\ref{prop:month} at work: the safety-case regime is a 12-month pause at the start with compute standing still. At the start, $A(0)=\ln4\approx\lfHfull$ and $w(0)=\lfHzero$ a month, so with $m=\sigma_b$ the value of a month is $\irSigmaB\times\lfHfull-\lfHzero\approx\lfVzeroUsual$. The pause counts as $(1-e^{-12\sigma_b})/\sigma_b=\lfEffMonths$ effective months, so it lowers cumulative hazard by about $\lfEffMonths\times\lfVzeroUsual\approx\lfFitSaving$, from \lfHfull{} to \lfFitTotal: a 40 percent probability.

\subsection{Earlier Months Are Worth More}\label{sec:earlier}

Under the benchmark, an earlier month is worth more than a later one. Along the scenario's full-speed path, the value of a month falls with the start date for every safety share during the pause (Appendix~\ref{supp:exhibits} plots it). At the usual share, a month at the start saves \lfVzeroUsual{} of cumulative hazard. The value falls over the takeoff period and turns negative at month \lfTzeroUsual. A pause that moves research compute into safety research learns \lfMultSafety{} times as fast, so its value starts higher, at \lfVzeroSafety, and stays positive until month \lfTzeroSafety. An all-inference pause learns at half the usual speed and stops paying at month \lfTzeroInference. Late in the takeoff period, the exposure cost overwhelms any learning: the current model is already dangerous, and little hazard is left for new knowledge to protect.

Two forces make later delay less valuable. Advancing the frontier raises the cost of standing still. And a later pause arrives after some of the step hazard that new knowledge could have prevented. A rising yield can reverse this: stronger models can make a later pause's safety research more productive \citep{bostrom2026}. The next subsection gives the exact comparison for finite pauses and allows knowledge to lose value during later development.

\subsection{When to Learn First}\label{sec:timing}

Learning first need not be best. Stronger models can make safety research more productive, and knowledge acquired later has less development over which to lose its value. Waiting for those gains leaves earlier hazards unprotected. This section gives an exact condition for choosing between the two timings, within a class of pauses followed by the same development path.

For this comparison, allow knowledge to lose value as capability advances:
\begin{equation}
\dot S=\sigma(F)y(s)-d\dot F S,\qquad d\ge0.
\label{eq:timing-knowledge}
\end{equation}
The yield $\sigma(F)$ may rise with capability. The parameter $d$ measures transfer loss: knowledge carried across a capability gain $\Delta F$ retains the fraction $e^{-d\Delta F}$. With $d=\ln2$, half survives a unit gain. This is the transfer law behind the knowledge-loss cases of Section~\ref{sec:keeping}; the baseline has $d=0$.

Fix a reference path $F(t),S(t)$ that finishes at $T$. A candidate policy follows it until date $t<T$, pauses development for $\delta$ months with safety share $s_p$, then repeats the remaining pace and safety shares. The reference path may itself include safety research at its own share. The choice is when to insert an additional period of learning, holding its duration and safety share fixed.\footnote{The comparison excludes research-model lags, safety-to-capability spillovers, and calendar-dependent learning. Hazard ends at completion.}

Three steps give the risk saving. All hazards below are those of the reference path, with the same $h,k$ across candidate timings.

\emph{Step 1: track the added knowledge.} The pause produces $m(t)\delta$ knowledge, where $m(t)=\sigma(F(t))y(s_p)$. At a later reference date $u$, the fraction still useful is
\[
R(t,u)=e^{-d[F(u)-F(t)]}.
\]
The replay holds $m(t)\delta R(t,u)$ more knowledge than the reference. Earlier knowledge protects more of the path, but loses more value before reaching a given later model.

\emph{Step 2: count exposure while learning.} The initial hazard of waiting is $w(t)=h(F(t))e^{-S(t)}$. It falls during the pause as knowledge grows. Total pause exposure is $w(t)g_\delta(m(t))$, where
\[
g_\delta(m)=\int_0^\delta e^{-mv}\,dv
=\frac{1-e^{-m\delta}}{m}\quad(m>0),\qquad g_\delta(0)=\delta.
\]
This is the effective duration in Proposition~\ref{prop:month}.

\emph{Step 3: subtract exposure from protection.} Write $\mathcal V_\delta(t)$ for cumulative hazard saved by the pause. Additional knowledge multiplies each later hazard by $e^{-m(t)\delta R(t,u)}$, giving
\begin{equation}
\mathcal V_\delta(t)=
\int_t^T\lambda(u)\left[1-e^{-m(t)\delta R(t,u)}\right]du
-w(t)g_\delta(m(t)).
\label{eq:timing-saving}
\end{equation}
The first term is future hazard removed; the second is hazard incurred while learning. A pause lowers catastrophe probability exactly when this difference is positive. With full transfer, it equals $g_\delta(m)[mA-w]$, recovering Proposition~\ref{prop:month}.

The comparison between two dates separates the opposing forces.

\begin{proposition}\label{prop:timing}
Suppose knowledge follows equation~\eqref{eq:timing-knowledge}, with continuous nonnegative $\sigma,h,k$, a fixed pause safety share, and the same prescribed continuation after either pause. For start dates $t_1<t_2<T$, let $z_i(u)=m(t_i)\delta R(t_i,u)$ be the extra knowledge retained at reference date $u$, and $E_i=w(t_i)g_\delta(m(t_i))$ the exposure during the pause. Starting the pause at $t_2$ gives lower cumulative hazard than starting at $t_1$ if and only if
\begin{equation}
\int_{t_2}^T\lambda(u)\left[e^{-z_1(u)}-e^{-z_2(u)}\right]du
>
\int_{t_1}^{t_2}\lambda(u)\left[1-e^{-z_1(u)}\right]du+E_2-E_1.
\label{eq:timing-comparison}
\end{equation}
\end{proposition}

The left side is the later pause's extra protection after $t_2$. The right side is the protection forgone before $t_2$ plus the change in exposure during the pause. Waiting pays only when the extra protection later outweighs both. If any pause improves on no pause, a best start exists and maximizes $\mathcal V_\delta(t)$ (Appendix~\ref{app:proof-timing}).

Two forces favor waiting. The yield may rise between $t_1$ and $t_2$, and knowledge learned at $t_2$ has less development over which to lose value. So the later pause's retained knowledge, $z_2$, exceeds $z_1$ by the productivity gain times the retention gain. Neither settles the comparison: waiting can sacrifice too much earlier protection or make the pause itself too hazardous. A marginal version of the condition (Appendix~\ref{supp:extensions}) shows the same forces at work. A rising yield and better retention of new knowledge favor waiting; rising state hazard, step hazard already passed, and erosion of existing knowledge oppose it. A local inequality alone does not rank distant start dates.

Table~\ref{tab:timing} runs the comparison in the scenario. The reference is the full-speed path with its usual safety share. The pause lasts the full 12-month lead and moves research compute to safety, the pause with safety of Table~\ref{tab:rules}, so safety knowledge grows \tsPauseMult{} times as fast as at the usual share. After it, the lab replays the remaining path. Hazards are normalized so that each row's reference keeps the scenario's 75 percent; the best start date does not depend on that scale.

With the scenario's constant yield and full transfer, learning first is best. A pause at the Automated Coder cuts the probability of catastrophe by \tsBasePoints{} points, to \tsBaseProb{} percent, and no later start does better. (The scenario's pause with safety in Section~\ref{sec:fixed} leaves \ipEqPauseSafetyP{} percent. It starts at month \ipEqPauseSafetyStart{} and resumes at the full research share instead of replaying.) A yield that triples over the takeoff period does not change the best start. Half transfer alone moves it to month \tsTransferStart{} and gains \tsTransferGain{} points, a negligible gain. Together they move the best start to month \tsBothStart. The pause then cuts the probability by \tsBothBest{} points instead of \tsBothFirst, from 75 to \tsBothProbBest{} percent rather than \tsBothProbFirst. Waiting pays only when stronger models make research more productive and early knowledge would have faded.

The pause at the Automated Coder is not free. Every milestone arrives a year late, so customers wait a year longer for each improvement in service quality (Section~\ref{sec:price}). And at the scenario's ceiling, the pause still fails the monthly test (Section~\ref{sec:feasible-timing}).

\begin{table}[htbp]
\centering
\caption{When Developing Further Makes a Safety Pause More Valuable}
\label{tab:timing}
\small
\begin{tabular}{@{}llrrr@{}}
\toprule
Yield at endpoint & Knowledge retained & Best start & \multicolumn{2}{c}{Fall in catastrophe probability}\\
 & across the takeoff period & (month) & start now & best start\\
 & & & \multicolumn{2}{c}{(percentage points)}\\
\midrule
1 times initial & 100 percent & 0.0 & 60.2 & 60.2\\
3 times initial & 100 percent & 0.0 & 59.8 & 59.8\\
1 times initial & 50 percent & 2.1 & 44.4 & 44.4\\
3 times initial & 50 percent & 9.2 & 44.8 & 47.9\\
\bottomrule
\end{tabular}
\par\smallskip
\begin{minipage}{\textwidth}\footnotesize
\emph{Notes:} The scenario's full-speed path at the fitted productivity, with the usual safety share $s_0$, is the reference. The pause lasts $\delta=L=12$ months at share $s_0+x_0$ (learning multiplier \tsPauseMult), followed by the replay of Section~\ref{sec:replay} with research compute at its reference level and share $s_0$. The yield is $\sigma(F)=\sigma_be^{gF}$ with $g=0$ or $\ln3$, so the initial yield stays at its fitted value; transfer loss is $d=0$ or $\ln2$. Normalized hazards: $h$ and $k$ are rescaled so that each row's reference keeps the 75 percent probability. Best starts maximize equation~\eqref{eq:timing-saving}; the fall in probability is $e^{-H}(e^{\mathcal V_\delta}-1)$, with $H$ the reference's cumulative hazard. Appendix~\ref{supp:numerics} describes the checks.
\end{minipage}
\end{table}

After the pause, the lab replays the reference at its own pace, as in Section~\ref{sec:replay}. At reference date $u$, research compute is held at its reference level, $x^p(u+\delta)=e^{-G\delta}x(u)$, the safety share is unchanged, and the spare compute serves customers. For a full-speed reference, every later milestone is exactly $\delta$ months late, so $\delta\le L$ meets the deadline. The ceiling must be checked separately. If the reference already meets it, every such pause preserves it: the pause removes step hazard and the replay adds protection. If it does not, the next subsection gives the test.

\subsection{Timing That Keeps Both Promises}\label{sec:feasible-timing}

A useful pause must also keep both promises. Lower cumulative hazard does not by itself keep monthly hazard below the ceiling. The pause must begin before any violation, last long enough to protect resumed development, and finish within the lead.

Take the full-speed capability path, which ends at $T_0$, as the reference, with its prescribed safety share and the transfer law~\eqref{eq:timing-knowledge}; the reference need not meet the ceiling. For a fixed pause share $s_p$ and a positive ceiling $\bar c$, three checks identify every feasible pause in this class.

\emph{Step 1: check the path before learning.} A pause leaves its prior history unchanged. Define $\Lambda_-(t)=\sup_{0\le u<t}\lambda(u)$, with $\Lambda_-(0)=0$. If this exceeds the ceiling, starting at $t$ is already too late.

\emph{Step 2: check the learning period.} Hazard during the pause is $w(t)e^{-m(t)v}$ at elapsed time $v$ and falls as research adds protection, so the whole pause meets the ceiling exactly when $w(t)\le\bar c$.

\emph{Step 3: protect the replay within the lead.} At reference date $u\ge t$, resumed hazard is $\lambda(u)e^{-m(t)\delta R(t,u)}$. For $m(t)>0$, the smallest pause that brings every resumed hazard below the ceiling is
\begin{equation}
\delta_c(t)=\sup_{t\le u<T_0}
\frac{[\ln(\lambda(u)/\bar c)]_+}{m(t)R(t,u)}.
\label{eq:minimum-feasible-pause}
\end{equation}
The numerator is the log-hazard reduction needed at a later milestone; the denominator is the protection retained there from one month of research at the pause date. (If $m(t)=0$, set $\delta_c(t)=0$ when every resumed hazard meets the ceiling and $+\infty$ otherwise.) Every later milestone arrives exactly $\delta$ months late, so the deadline requires $\delta\le L$. The minimum learning time and the available lead bound the same choice.

\begin{proposition}\label{prop:feasible-timing}
Under the assumptions of Proposition~\ref{prop:timing}, take the full-speed path as the reference and fix a positive pause duration $\delta$. The pause starting at $t<T_0$ satisfies the ceiling and the deadline if and only if
\begin{equation}
\Lambda_-(t)\le\bar c,\qquad
w(t)\le\bar c,\qquad
\delta_c(t)\le\delta\le L.
\label{eq:feasible-timing}
\end{equation}
Among pauses passing these tests, the comparison in Proposition~\ref{prop:timing} ranks cumulative hazard. With fixed hazards, $S(0)=0$, and $\sigma=a\bar\sigma$, the tests at $a_{\min}$ guarantee both promises for every $a\ge a_{\min}$.
\end{proposition}

The test works because it checks every phase of the path: past exposure, exposure while learning, and protection before the deadline. When productivity is uncertain, run the tests at the productivity bound $a_{\min}$. Among the pauses that pass, Proposition~\ref{prop:timing} picks the lowest cumulative hazard. No pause is an option only if the reference itself meets both promises. Within this class the best duration is either no pause or the full lead $L$; an intermediate duration is never strictly best (Appendix~\ref{app:proof-feasible}). If no pause passes, this class of policies fails, and the necessary bound of Section~\ref{sec:keeping} says whether any policy can succeed.

Appendix~\ref{supp:exhibits} tabulates the test in the scenario, for the technology of the last row of Table~\ref{tab:timing} and the same 12-month pause with safety. It reads as follows. At the scenario's own ceiling of \insCap{} a month, no start works. Even an immediate pause would need \tsScenMinInitial{} months of learning to hold the replayed hazards below the ceiling, and the lead allows 12. With the scenario's own technology it would need \tsBaseMinInitial{} months, a different quantity from the \lfMonths-month minimum compliance time of Section~\ref{sec:requirement}, which assumes learning at the fastest possible rate throughout. This is the same mechanism that makes the usual-share replay of Section~\ref{sec:replay} fail at the fitted productivity. With the usual safety share after learning, holding this ceiling takes more than 12 months of learning. The spare-compute replay holds it by putting spare compute in safety during the replay, which this class does not do. At a ceiling of \tsCapLoose{} a month, only starts between months \tsCapLooseLo{} and \tsCapLooseHi{} pass. Earlier starts fail the duration test: the yield is still low and the knowledge would fade, so 12 months of learning cannot protect the final models. Later starts fail the first test, because the reference crosses the ceiling before they begin. The best feasible start is month \tsCapLooseBest, later than the unconstrained best of \tsBothStart, and it cuts the probability by \tsCapLoosePoints{} points instead of \tsBothBest. At \tsCapLooser{} a month, starts from month \tsCapLooserLo{} pass, and the unconstrained best is among them. At \tsCapLoosest, every start before month \tsCapLoosestHi{} passes.

The constraint can move the pause later as well as earlier: first identify the dates that keep the promise, then compare the protection they buy.

\FloatBarrier
\section{Customer Cost: What Services Are Lost}\label{sec:price}

Customers give up services when compute moves to safety and when capability improvements arrive later. This section measures those losses for the learn-then-replay policies introduced in Section~\ref{sec:keeping}. It compares their service cost with the risk they leave and chooses the cheapest policy within that replay class.

Service value is measured as follows. Customers receive services worth $\sqrt{e^{\mu F}iK}$ per month (Section~\ref{sec:services}). With $\mu=\ln10$, service quality rises tenfold from the Automated Coder to top-expert-dominating AI. At fixed inference compute $iK$, monthly service value rises by a factor of $\sqrt{10}$. Over $q$ months, discounted service value is
\begin{equation}
J_q=\int_0^q e^{-rt}\sqrt{e^{\mu F(t)}i(t)K(t)}\,dt,
\label{eq:services}
\end{equation}
and the \emph{service loss} of a policy is $100(1-J_q/J_q^0)$ percent, where $J_q^0$ is the full-speed value. After the takeoff period ends, capability stays at one, the full-speed shares resume, and compute keeps growing.

Two channels link pacing to services. Moving compute from customers to safety research lowers current services. Moving it from capability research delays improvements in service quality. Service loss is an opportunity cost: it does not weight services by the chance of survival, and the paper does not convert risk into service value. This section reports the two side by side. The same measure applies to the pauses of Section~\ref{sec:feasible-timing}; the numbers below are for the learn-then-replay class, which varies the learning time and the replay safety share.

The replay safety share trades service value against the productivity bound. At the scenario's ceiling of \insCap{} a month, the necessary bound is \insFrNecAtCap{} of the fitted productivity. With a year of learning, the spare-compute replay needs a productivity bound of \insFrSpareAtCap. It gives up \insFrLossSpareFive{} percent of five-year service value and all of two-year value, because customers receive nothing until the takeoff period ends at month 30. At its productivity bound, it holds the probability of catastrophe to \insRiskSpareBound{} percent, against \insRiskFull{} percent at full speed. At the fitted productivity, its risk is \insRiskSpareFit{} percent.

The usual-share replay is cheaper but needs a higher bound. It gives up \insFrLossSzeroFive{} percent of five-year value and \insFrLossSzeroTwo{} percent of two-year value. But it needs a productivity bound of \insFrSzeroAtCap{}, above the fitted productivity itself. At the fitted productivity its ceiling fails, and its risk is \insRiskUsualFit{} percent.

\begin{figure}[htbp]
\centering
\includegraphics[width=0.74\textwidth]{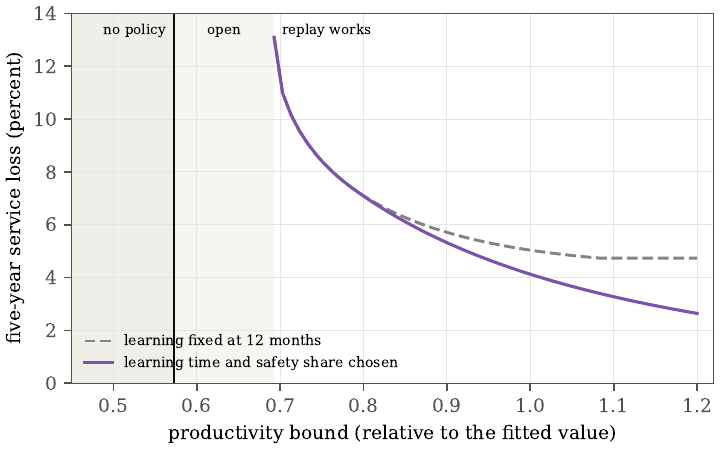}
\caption{The Service Cost of a Guaranteed Ceiling and Deadline}
\label{fig:service}
\par\smallskip
\begin{minipage}{\textwidth}\footnotesize
\emph{Notes:} Five-year service loss of the cheapest learn-then-replay policy that guarantees the scenario's ceiling and the 12-month deadline at each productivity bound, with the scenario's fixed hazards. Purple: learning time and replay safety share both chosen. Gray: learning fixed at 12 months, replay safety share chosen. Left of the black line (the necessary bound), no policy keeps both promises; between the black line and the start of the curves, no replay policy does. Costs are minimized within the replay class.
\end{minipage}
\end{figure}

A shorter learning phase lowers the price. For each productivity bound, Figure~\ref{fig:service} chooses the learning time $\tau\le L$ and the replay safety share that minimize five-year service loss, while respecting the ceiling at the worst productivity.

The choice turns on one constraint. For the scenario's hazards, $\ln\lambda^0(u)$ is convex, so the replay hazard peaks at an endpoint of the replay, and at the scenario's ceiling only the final endpoint binds. That endpoint requires
\begin{equation}
a_{\min}\sigma_b[\bar y\tau+y(s_r)T_0]\ge\ln[\lambda^0(T_0)/\bar c].
\label{eq:endpoint}
\end{equation}
The left side is the knowledge the policy holds at the end of the replay at the worst productivity. It comes from $\tau$ months of maximal learning and $T_0$ months at the replay rate. The right side is the log reduction the final hazard needs.

At each learning time, the smallest safety share that satisfies the constraint leaves the most compute for customers. Waiting longer builds more initial protection and lowers the research share needed for the replay, but it also prolongs the phase with no customer services.

The cost falls steeply as the productivity bound rises (Figure~\ref{fig:service}). A regulator that can rule out productivity below the fitted value pays far less than one that must allow for \insFrSpareAtCap. At the fitted productivity, the cheapest policy learns for \insFrOptMonths{} months. It then gives \insFrOptResearch{} percent of compute to research, \insFrOptSafety{} percent to safety, and \insFrOptCustomers{} percent to customers. It gives up \insFrOptFive{} percent of five-year service value, against \insFrFixedFive{} percent when learning lasts a year. Over two years both give up about half (\insFrOptTwo{} and \insFrFixedTwo{} percent), so the saving is small.

The cheaper policy also leaves more risk. At the fitted productivity, it leaves \insRiskCheapFit{} percent catastrophe risk, against \insRiskSpareFit{} percent for the spare-compute replay. The cheapest policy is designed to meet the ceiling exactly at the fitted productivity, so at that productivity its hazard reaches the ceiling at the last milestone. The spare-compute replay is designed for productivity as low as \insFrSpareAtCap, so at the fitted productivity it has room to spare.

The five-year loss is smaller than the two-year loss only because the policy supplies full-speed services after completion. So service budgets need a stated horizon: the full-speed value over five years is \irServiceRatio{} times that over two, so the same percentage allows \irServiceRatio{} times the absolute loss.

\section{Rules That Fix Compute Allocation}\label{sec:fixed}

Compute floors, pauses, and lags work the other way round from a ceiling. This section shows that they fix development dates and let risk absorb the uncertainty about productivity, and that their effect on risk can have either sign.

A rule that sets the shares and the research model without reference to knowledge, hazard, or productivity produces the same capability path at every productivity. The reason is simple: the law of capability, equation~\eqref{eq:research}, contains neither knowledge nor productivity. This is Proposition~\ref{prop:fix}(a). So the lead such a rule uses is fixed in advance, and only its effect on risk varies with productivity.

The contrast with a ceiling is the contrast between prices and quantities \citep{weitzman1974}. A ceiling fixes the hazard and lets delay absorb the uncertainty about productivity. An allocation rule fixes the dates and lets risk absorb it.

\subsection{Slowing Down When Research Does Nothing}

The sharpest case is research that produces no knowledge at all. Then slowing down cannot buy protection; it can only move time between capability levels. Three steps show what a constant limit on the research share does.

\emph{Step 1: dates.} With a constant research share $x$ and no lag, equation~\eqref{eq:research} separates as for the full-speed path: $dF/\phi(F)=(xK_0)^\varepsilon e^{\gamma t}dt$. The dates are
\[
t_x(F)=\frac1\gamma\ln\bigl[1+M(e^{\gamma t_0(F)}-1)\bigr],\qquad M=\Bigl(\frac{x_0}{x}\Bigr)^\varepsilon>1.
\]
The slowdown factor $M$ is the ratio of full-speed to capped research capacity. The quantity $e^{\gamma t}-1$ measures research capacity accumulated since the start. On that scale, every milestone arrives $M$ times later than at full speed.

\emph{Step 2: time per unit of capability.} Differentiating,
\[
t_x'(F)-t_0'(F)=\frac{(M-1)\,t_0'(F)}{1+M(e^{\gamma t_0(F)}-1)}>0.
\]
Both the numerator and the denominator are positive, because $M>1$. So the slower lab spends more time at every capability level.

\emph{Step 3: hazard.} With no knowledge, the accounting identity~\eqref{eq:accounting} applies, so the step-hazard term $\int_0^1k(F)\,dF$ is the same for both paths, and only the state-hazard term $\int_0^1h(F)t'(F)\,dF$, which is larger for the slower path, differs.

\begin{proposition}\label{prop:exposure}
Suppose $a=0$ and $h$ is positive on a capability interval. A constant research share $0<x<x_0$, without a lag, increases cumulative takeoff hazard relative to full speed. The same strict inequality holds for all small enough positive $a$, with $h,k$ and allocation settings fixed.
\end{proposition}

The proof is in Appendix~\ref{app:proof-exposure}. The extension to small positive productivity follows from continuity: dates do not depend on productivity, and the hazard of each path changes continuously with it. This is the risk accounting of \citet{trammell2026}.

The scenario's compute floors are an example. They fix research at \ipEqFloorsShare{} percent of compute, so $M=(\ipXzero/\ipEqFloorsShare)^{1/2}\approx\inrFloorM$. The takeoff period ends at month \lfFloorsEnd{} instead of 18, so the floors use exactly the year of lead. When safety research does nothing, they raise the probability of catastrophe by \insZeroFloors{} points.

A safety floor that keeps $x=x_0$ is different: it leaves the path and the hazard unchanged at zero productivity, and it uses only customer compute.

The R\&D-model lag also slows research, and most where the frontier has recently moved fastest. Research productivity is $\phi(F(t-\ell))$ instead of $\phi(F(t))$, which in the scenario is lower by the factor $e^{-\rho[F(t)-F(t-\ell)]}$. So the lag adds time late in the takeoff period, where state hazard is highest. When research does nothing, it raises the probability of catastrophe by \insZeroLag{} points.

\subsection{Pauses and Growing Compute}\label{sec:pause}

Growing compute breaks the repeated-path benchmark of Section~\ref{sec:month}: the lab can catch up after a pause. A pause adds time at a single capability level, and compute keeps growing while the lab waits. Without compute growth, Proposition~\ref{prop:month} at $m=0$ gives $\Delta H=\delta w$: a pause that teaches nothing only adds exposure. With compute growth, the lab resumes with more compute than it had when it stopped, and it crosses the later milestones faster than at full speed. The exposure saved there can exceed the exposure added while waiting.

The comparison is exact. Suppose research produces no knowledge. At full-speed date $z$, the lab pauses for $\delta$ months at $F_*=F^0(z)$. It then resumes with research share $x_0$ while compute keeps growing. On the research-capacity scale $e^{\gamma t}-1$, the pause forgoes $C=e^{\gamma(z+\delta)}-e^{\gamma z}$ of capacity. Every later milestone must make up this capacity:
\[
e^{\gamma t_p(F)}=e^{\gamma t_0(F)}+C,\qquad C=e^{\gamma z}(e^{\gamma\delta}-1),\qquad
t_p'(F)=t_0'(F)\frac{e^{\gamma t_0(F)}}{e^{\gamma t_0(F)}+C}<t_0'(F).
\]
The lab crosses each later capability interval faster than at full speed. Earlier milestones are unchanged, and the same capability interval is crossed, so step hazard cancels. The exact change in takeoff hazard is
\begin{equation}
\Delta H=\underbrace{\delta h(F_*)}_{\text{exposure while waiting}}
-\underbrace{\int_{F_*}^1 h(F)t_0'(F)
\frac{C}{e^{\gamma t_0(F)}+C}\,dF}_{\text{exposure saved after resuming}}.
\label{eq:zeropause}
\end{equation}
The first term is the state hazard of the held model, for the length of the pause. The second is the state hazard the lab avoids at every later level, because it crosses each level faster. Risk falls if and only if the second term exceeds the first. Faster crossing alone does not ensure this: with constant positive state hazard, $\Delta H=h\,(T_p-T_0)>0$, where $T_p=t_p(1)$. The pause uses $\delta$ months of lead, because the milestone just above $F_*$ arrives $\delta$ months late, even though the takeoff period ends less than $\delta$ months late.

In the scenario, a 12-month pause at the start adds \ovWaitZero{} of exposure while waiting, because the initial model has low state hazard, and saves \ovSavedZero{} afterward. So $\Delta H=\ovDeltaHZero$, and the probability of catastrophe falls by \ovPointsZero{} points. The saved term changes little with the start date, while the waiting term grows with the hazard of the held model. The two are equal for a pause starting at month \lfPauseCross; a later pause raises risk (Appendix~\ref{supp:exhibits} plots both terms).

The early-pause result assumes that step hazard is proportional to speed. If hazard rose more than proportionally with speed, faster catch-up after a pause could add risk \citep{trammell2026}. That is the overhang worry. The AI Futures Project warns that compute left unused under an international agreement could lead to a very fast takeoff period once the agreement ends. \citet{bostrom2026} lists a compute overhang among the ways a pause can backfire.

\subsection{Fixed Rules across Productivity and Technologies}

Appendix~\ref{supp:fixed} varies productivity from a tenth to twice the fitted value, holding each rule at its one-year-lead setting. It also retunes each rule in ten versions of the safety technology (Appendix~\ref{app:calibration}). The allocation rules keep their dates as productivity changes; the ceiling uses \insCeilLeadHalf{} months of lead at half the fitted productivity and \insCeilLeadQuarter{} at a quarter. Rules that move compute into safety research early lower risk most consistently across the ten versions. Floors lower the probability of catastrophe by \rbFloorsMin--\rbFloorsMax{} points across the versions, and the pause with safety by \rbPauseSafetyMin--\rbPauseSafetyMax. Rules that delay development late do not. The ceiling's effect ranges from a \rbCeilingMinAbs-point rise to a \rbCeilingMax-point fall, and the lag's from a \rbLagMinAbs-point rise to a \rbLagMax-point fall. Late delay pays when safety research grows more productive with capability or with compute, or when danger continues after the takeoff period. Knowledge added at the end then protects the whole tail.

\section{Conclusion: What a Regulator Can Promise}\label{sec:conclusion}\label{sec:policy}

A regulator can proceed in three steps: establish feasibility, choose compliant timing, and compare customer cost. First, the bounds test whether a ceiling and deadline can be promised together. This test has three inputs: the log-hazard reduction the ceiling requires, $\ln[h(1)/\bar c]$; the fastest attainable learning, $\bar\sigma_{\max}\bar y$ per unit of productivity; and the time the lead allows, $T_0+L$. It has three outcomes.
\begin{itemize}
\item If the necessary condition~\eqref{eq:necessary} fails at the productivity bound, no compliance strategy keeps both promises. One of them must be relaxed, the safety technology improved, or the final model made less hazardous.
\item If the replay condition~\eqref{eq:construct} holds, learning first and then replaying keeps both, at the service cost in Figure~\ref{fig:service}.
\item Between the two, the question is open: the paper's bounds neither rule out a policy nor provide one.
\end{itemize}
The band between the two conditions is narrow when hazard comes from the model's state and early knowledge stays useful. It widens with step hazard, rising yields, or lost knowledge.

Second, screen the possible pause dates with Proposition~\ref{prop:feasible-timing} and rank the survivors with Proposition~\ref{prop:timing}. Later research can be more productive, but it cannot erase a ceiling violation before learning begins. Third, evaluate customer services on the compliant paths; the policy with the lowest risk need not have the lowest service cost.

In the scenario, the test reads as follows. A regulator who cannot rule out productivity of half the fitted value cannot promise the scenario's ceiling and deadline together, by any rule. One who can rule out productivity below \insFrSpareAtCap{} can promise both with the spare-compute replay, at a cost of \insFrLossSpareFive{} percent of five-year services. One who can rule out productivity below the fitted value pays only \insFrOptFive{} percent. Between \insFrNecAtCap{} and \insFrSpareAtCap, the answer is open.

The binding input is the productivity bound. Evidence that rules out productivity below the replay bound turns an impossible universal promise into a feasible conditional one. Evidence that raises the bound lowers the price of the promise (Figure~\ref{fig:service}). So measuring how much safety research reduces hazard, and how well that knowledge carries to stronger models, is what makes a joint promise affordable. Rules that fix compute shares offer a different commitment: they fix dates and let risk vary with productivity.

A monthly ceiling bounds the probability of catastrophe only loosely: with the deadline, it implies $P\le1-e^{-\bar c(T_0+L)}$, which is \insLifetimeBound{} percent at the scenario's ceiling. The simulated risk is far lower: \insRiskSpareBound{} percent for the spare-compute replay at its productivity bound, and \insRiskSpareFit{} percent at the fitted productivity. The reason is that the ceiling caps the hazard rather than setting it. A cumulative-hazard budget would bound the probability directly and faces the same deadline arithmetic.

The analysis is limited in six respects. Compute grows exogenously. A lab held to a research share could invest to restore its speed, as intensity standards can induce output expansion \citep{holland2009}. The overhang result assumes step hazard proportional to speed \citep{trammell2026}. Enforcement and the measurement of hazard are idealized: compute is detectable \citep{sastry2024} and training can be told from inference \citep{heim2024}, but safety research is harder to tell from capability research \citep{aifutures2026}. The deadline is a fixed limit on delay rather than the outcome of a race. Propositions~\ref{prop:construct} and~\ref{prop:sharp} let early knowledge protect later models fully; Section~\ref{sec:replay} shows the cost when it does not. And the scenario's magnitudes are judgments, not estimates; the analytical results do not depend on them. Three companion papers build on the model: \citet{gan2026allocation} asks whether safety compute should come from customers or from capability research, \citet{gan2026racing} makes the pace the outcome of a race among labs, and \citet{gan2026agreements} studies agreements between countries to pause.

\begin{singlespace}
\end{singlespace}

\appendix
\renewcommand{\thetable}{\Alph{section}\arabic{table}}
\renewcommand{\theequation}{\Alph{section}\arabic{equation}}
\makeatletter
\@addtoreset{table}{section}
\@addtoreset{equation}{section}
\makeatother

\section{Proofs}\label{app:proofs}

Throughout, $\sigma$, $h$, and $k$ are continuous and nonnegative, $\phi$ is continuous and positive, and $K_0,x_0,s_0>0$. Prescribed capability dynamics, including their initial history when research is lagged, have unique solutions.

\subsection{Dates on the Full-Speed Path and under a Constant Research Share}

Write $\Phi(F)=\int_0^F dv/\phi(v)$ and $\gamma=\varepsilon G$. Without a lag and with a constant research share $x$, equation~\eqref{eq:research} separates, and integrating from $F(0)=0$ gives
\begin{equation}
\Phi(F)=(xK_0)^\varepsilon\,\frac{e^{\gamma t}-1}{\gamma}.
\label{eq:app-integrated}
\end{equation}
Compare equation~\eqref{eq:app-integrated} at the same milestone $F$ under shares $x$ and $x_0$. This gives $e^{\gamma t_x(F)}-1=M(e^{\gamma t_0(F)}-1)$ with $M=(x_0/x)^\varepsilon$, the formula of Section~\ref{sec:fixed}. It holds for any continuous positive $\phi$.

In the scenario, $\phi(F)=\theta e^{\rho F}$, so $\Phi(F)=(1-e^{-\rho F})/(\rho\theta)$. At full speed, $x=x_0$ and $F(T_0)=1$, so $\theta(x_0K_0)^\varepsilon=b_0$, where
\[
b_0=\frac{\gamma(1-e^{-\rho})}{\rho(e^{\gamma T_0}-1)}.
\]
Substituting $b_0$ and solving for $F$ or for $t$ gives
\begin{equation}
F^0(t)=-\frac1\rho\ln\!\left[1-(1-e^{-\rho})\frac{e^{\gamma t}-1}{e^{\gamma T_0}-1}\right],
\qquad
t_0(F)=\frac1\gamma\ln\!\left[1+(e^{\gamma T_0}-1)\frac{1-e^{-\rho F}}{1-e^{-\rho}}\right].
\label{eq:fullspeed}
\end{equation} The full-speed pace at milestone $F$ is $f^0(F)=b_0e^{\gamma t_0(F)+\rho F}$.

\subsection{Proof of Proposition~\ref{prop:fix}}\label{app:proof-fix}

(a) The prescribed capability dynamics and their initial history contain neither $S$ nor $a$. Uniqueness then gives the same capability path and dates at every productivity.

(b) Fix a path that respects the ceiling, so $\lambda(t)\le\bar c$ for every $t<T$. Since $k\ge0$ and $\dot F\ge0$, $\lambda(t)\ge h(F(t))e^{-S(t)}$, so $h(F(t))e^{-S(t)}\le\bar c$ for $t<T$. As $t\uparrow T$, $F(t)\to1$ and $S(t)\to S(T)$; continuity of $h$ gives $h(1)e^{-S(T)}\le\bar c$, that is, $S(T)\ge\ln[h(1)/\bar c]>0$. By the safety law, $\dot S=a\bar\sigma(F)y(s)\le a\bar\sigma_{\max}\bar y$, so $S(T)\le a\bar\sigma_{\max}\bar yT$. Combining the two inequalities gives equation~\eqref{eq:bound}. Finally, $D\ge t(1)-t_0(1)=T-T_0$, so the deadline requires $T\le T_0+L$, which equation~\eqref{eq:bound} rules out when $a<\ln[h(1)/\bar c]/[\bar\sigma_{\max}\bar y(T_0+L)]$.

\subsection{Proof of Proposition~\ref{prop:construct}}\label{app:proof-construct}

Fix $\tau\in[0,L]$ and $a\ge a_{\min}$. While the lab learns, $x=0$, so $\dot F=0$ and $F=0$; with $s=1$, knowledge is $a\bar\sigma(0)\bar yt$ and the hazard is $h(0)e^{-a\bar\sigma(0)\bar yt}\le h(0)$, with equality at $t=0$.

On replay at date $\tau+u$, research compute is $x_0e^{-G\tau}K(\tau+u)=x_0K(u)$. With $F(\tau)=0=F^0(0)$, equation~\eqref{eq:research} and uniqueness give $F(\tau+u)=F^0(u)$ and $\dot F(\tau+u)=\dot F^0(u)$ for $0\le u\le T_0$. Knowledge is
\[
S(\tau+u)=a\bar\sigma(0)\bar y\tau+a\,y(s_r)\int_0^u\bar\sigma(F^0(v))\,dv=a[\bar\sigma(0)\bar y\tau+y(s_r)\bar S^0(u)],
\]
so the hazard is $\lambda^0(u)e^{-a[\bar\sigma(0)\bar y\tau+y(s_r)\bar S^0(u)]}$. It is nonincreasing in $a$, so the largest replay hazard over $a\ge a_{\min}$ is attained at $a_{\min}$ and equals $B(s_r)e^{-a_{\min}\bar\sigma(0)\bar y\tau}$. The shares are feasible when $x_0e^{-G\tau}+s_r\le1$. Every positive milestone is reached exactly $\tau$ months late, so $D=\tau$, and the deadline holds if and only if $\tau\le L$.

So a policy with learning time $\tau$ works for every $a\ge a_{\min}$ if and only if $\bar c\ge h(0)$, $x_0e^{-G\tau}+s_r\le1$, $B(s_r)e^{-a_{\min}\bar\sigma(0)\bar y\tau}\le\bar c$, and $\tau\le L$. The second and third conditions relax as $\tau$ rises. So some $\tau\le L$ works if and only if $\tau=L$ works. At $\tau=L$, the share condition holds by the assumption $s_r\le1-x_0e^{-GL}$, which leaves condition~\eqref{eq:construct}.

\subsection{Proof of Proposition~\ref{prop:sharp}}\label{app:proof-sharp}

\emph{Lower bound.} Suppose a single admissible policy guarantees the ceiling and the deadline for every $a\ge b$, with $b>0$. Then it does so at $a=b$, and Proposition~\ref{prop:fix}(b) with $\bar\sigma_{\max}=\bar\sigma$ gives $b\ge\ln[h(1)/\bar c]/[\bar\sigma\bar y(T_0+L)]$. Taking the infimum over such $b$ gives the lower bound on $a^*$.

\emph{Upper bound.} Apply Proposition~\ref{prop:construct} with $s_r=1-x_0e^{-GL}$, which satisfies $s_r\ge s_0$ because $x_0e^{-GL}\le x_0\le1-s_0$, and with $a_{\min}=\bar a$. With $k\equiv0$, $\lambda^0(u)=h(F^0(u))$, and with a constant yield, $\bar S^0(u)=\bar\sigma u$. The shape restriction says that $\ln h(F^0(u))-\bar a\bar\sigma y_ru$ is nondecreasing in $u$, so $\lambda^0(u)e^{-\bar ay_r\bar\sigma u}$ is maximized at $u=T_0$ and $B=h(1)e^{-\bar ay_r\bar\sigma T_0}$. Condition~\eqref{eq:construct} then reads $\ln[h(1)/\bar c]\le\bar a\bar\sigma(\bar yL+y_rT_0)$, which holds with equality by the definition of $\bar a$; the condition $\bar c\ge h(0)$ is assumed. So the spare-compute replay with $\tau=L$ guarantees both promises at bound $\bar a$, and $a^*\le\bar a$.

\subsection{Proof of Proposition~\ref{prop:month}}\label{app:proof-month}

Under the repeated-path benchmark, the pause holds capability at $F(t)$ for $\delta$ months with safety share $s_p$. Knowledge grows linearly: $S(t+u)=S(t)+mu$ for $0\le u\le\delta$. Since $\dot F=0$, hazard during the pause is $h(F(t))e^{-S(t)-mu}=w(t)e^{-mu}$. The pause adds
\[
\int_0^\delta w(t)e^{-mu}\,du=w(t)\frac{1-e^{-m\delta}}{m}.
\]
After the pause, the lab repeats the reference path from $F(t)$ with the same pace and safety controls as functions of capability. So knowledge at each later milestone exceeds its reference value by exactly $m\delta$, and every remaining hazard is multiplied by $e^{-m\delta}$. Remaining cumulative hazard falls from $A(t)$ to $A(t)e^{-m\delta}$. Adding the two pieces,
\[
\Delta H=w(t)\frac{1-e^{-m\delta}}{m}-(1-e^{-m\delta})A(t)=-\frac{1-e^{-m\delta}}{m}\,[mA(t)-w(t)].
\]
For $m>0$, the factor $(1-e^{-m\delta})/m$ is positive, so $\Delta H<0$ if and only if $v_{s_p}(t)>0$. Since $P=1-e^{-H}$ increases with $H$, the same criterion holds for probability. As $m\to0$, $(1-e^{-m\delta})/m\to\delta$, so $\Delta H=\delta w(t)$.

\emph{Timing.} Along the reference path with $s=s_p=s_0$, constant $\sigma$, and $y=1$, $\dot A=-\lambda=-(h+k\dot F)e^{-S}$ and $\dot w=h'(F)\dot Fe^{-S}-\sigma w$, because $\dot S=\sigma$. So
\[
\dot v=\sigma\dot A-\dot w=-\sigma(h+k\dot F)e^{-S}-h'(F)\dot Fe^{-S}+\sigma he^{-S}=-[h'(F)+\sigma k(F)]\dot Fe^{-S}.
\]
Appendix~\ref{supp:extensions} gives the general case.

\subsection{Proof of Proposition~\ref{prop:timing}}\label{app:proof-timing}

Take a strictly increasing reference path, with safety law~\eqref{eq:timing-knowledge}, fixed hazard functions, and a fixed candidate share $s_p$. The reference and pause share the same initial history. During the pause capability is fixed, so transfer loss is zero and knowledge increases by $m(t)\delta$. On the replay, the reference and candidate have identical gross learning at each matched date. Their knowledge difference solves $\dot{\Delta S}(u)=-d\dot F(u)\Delta S(u)$, with initial difference $m(t)\delta$. Its solution is $\Delta S(u)=m(t)\delta R(t,u)$. Integrating hazard during the pause and on the replay gives equation~\eqref{eq:timing-saving}.

Subtract $\mathcal V_\delta(t_1)$ from $\mathcal V_\delta(t_2)$. Split the first pause's protected-hazard integral at $t_2$. The resulting difference is the left side of equation~\eqref{eq:timing-comparison} minus its right side. Since $H_\delta(t)=H_0-\mathcal V_\delta(t)$ and catastrophe probability rises with cumulative hazard, the inequality is necessary and sufficient. When $m(t_1)>0$, the ratio of retained knowledge makes the gain from waiting explicit:
\[
\frac{z_2(u)}{z_1(u)}=
\frac{\sigma(F(t_2))}{\sigma(F(t_1))}
e^{d[F(t_2)-F(t_1)]},\qquad u\ge t_2.
\]
The first factor is the productivity gain; the second is improved retention.

The terminal date is a limit, not an admissible pause start. Continuity gives $\lim_{t\uparrow T}\mathcal V_\delta(t)=-w(T)g_\delta(m(T))\le0$. If any start gives positive savings, a positive maximum is attained before completion. Otherwise no pause weakly dominates every candidate. The same comparison applies on any specified closed interval of eligible dates before $T$.

\subsection{Proof of Proposition~\ref{prop:feasible-timing}}\label{app:proof-feasible}

Before the pause, hazard equals the reference hazard, so the first test in~\eqref{eq:feasible-timing} is necessary and sufficient there. During the pause, hazard is $w(t)e^{-m(t)v}$ for $0\le v<\delta$. Its supremum is $w(t)$, giving the second test. On the replay, the knowledge difference derived in Appendix~\ref{app:proof-timing} gives hazard $\lambda(u)e^{-m(t)\delta R(t,u)}$ for $t\le u<T_0$. For positive $m(t)$ and $\lambda(u)$, the ceiling is equivalent to $m(t)\delta R(t,u)\ge\ln(\lambda(u)/\bar c)$. Taking the positive part and the supremum gives $\delta\ge\delta_c(t)$. The stated conventions cover zero hazard and zero pause productivity. The terminal date is excluded because hazard ends at completion, but its left limit enters the supremum.

The held milestone has already been reached. Every strictly higher milestone is reached $\delta$ months after its full-speed date, so the largest milestone delay is exactly $\delta$. The replay is implementable using $e^{-G\delta}x(u)$ for capability research, with the prescribed safety share unchanged. These shares leave nonnegative compute for customers. The four restrictions are jointly necessary and sufficient within the insertion class.

For the guarantee, fix the capability path and all safety shares independently of $a$, and write $\sigma=a\bar\sigma$. The linear knowledge law with $S(0)=0$ gives $S_a=aS_1$ on every candidate path. Every hazard is nonincreasing in $a$, while milestone dates are unchanged. Checking the ceiling at $a_{\min}$ is necessary and sufficient for the same candidate to meet it at every $a\ge a_{\min}$. Its worst cumulative hazard also occurs at that bound, so the timing comparison there ranks worst-case cumulative hazard among feasible candidates. Call the pairs $(t,\delta)$ passing~\eqref{eq:feasible-timing} the \emph{feasible pause set}, $\mathcal A(\bar c,L)$; it is evaluated at $a_{\min}$ when productivity is uncertain.

\emph{Duration.} With a full-speed reference, the deadline permits any pause duration up to $L$, and duration has a simple implication when only cumulative hazard matters. At a fixed start with $m>0$,
\[
e^{m\delta}\partial_\delta\mathcal V_\delta
=m\int_t^T\lambda(u)R(t,u)e^{m\delta[1-R(t,u)]}du-w(t).
\]
Because $0<R\le1$, this expression is nondecreasing in $\delta$. The derivative can change sign only from negative to positive, so an interior duration cannot be a strict maximum. At $m=0$, the saving is $-w\delta$. For $L>0$, this gives
\[
\max_{0\le\delta\le L}\mathcal V_\delta(t)
=\max\{0,\mathcal V_L(t)\}.
\]
The hazard-minimizing choice within this class is no pause or a full-$L$ pause at a start maximizing $\mathcal V_L$. This statement imposes no service budget or additional ceiling constraint. It does not extend to free choice of the resumed development pace.

\subsection{Proof of Proposition~\ref{prop:exposure}}\label{app:proof-exposure}

Set $\gamma=\varepsilon G$ and $M=(x_0/x)^\varepsilon>1$. By Appendix~A.1, $t_x(F)=\gamma^{-1}\ln[1+M(e^{\gamma t_0(F)}-1)]$. Differentiating,
\[
t_x'(F)=\frac{Me^{\gamma t_0(F)}t_0'(F)}{1+M(e^{\gamma t_0(F)}-1)},\qquad
t_x'(F)-t_0'(F)=\frac{(M-1)\,t_0'(F)}{1+M(e^{\gamma t_0(F)}-1)}>0.
\]
At $a=0$, $S\equiv0$. Both paths are strictly increasing in $F$, so changing variables from $t$ to $F$ gives $H_x=\int_0^1h(F)t_x'(F)\,dF+\int_0^1k(F)\,dF$, and likewise at full speed. Only the state-hazard term differs, by $\int_0^1h(F)[t_x'(F)-t_0'(F)]\,dF$, which is strictly positive because $h$ is positive on an interval. For $a>0$, the dates are unchanged by Proposition~\ref{prop:fix}(a), and cumulative hazard over the finite takeoff period is continuous in $a$. So the strict inequality persists for all small enough positive $a$.

\subsection{The Exact Pause Comparison at Zero Productivity}

Suppose $a=0$, $S(0)=0$, no lag, and $h,k$ fixed. The lab pauses at reference date $z<T_0$ for $\delta>0$ months, holding $F_*=F^0(z)$, then resumes research share $x_0$ with the compute clock running. Integrating $dF/\phi(F)=(x_0K_0)^\varepsilon e^{\gamma t}dt$ from the date of resumption $z+\delta$, where $F=F_*$, with $\Phi'=1/\phi$,
\[
\Phi(F)-\Phi(F_*)=\frac{(x_0K_0)^\varepsilon}{\gamma}\bigl[e^{\gamma t_p(F)}-e^{\gamma(z+\delta)}\bigr].
\]
On the full-speed path the same integral from $z$ gives $[(x_0K_0)^\varepsilon/\gamma][e^{\gamma t_0(F)}-e^{\gamma z}]$. Equating the two, $e^{\gamma t_p(F)}=e^{\gamma t_0(F)}+C$ with $C=e^{\gamma z}(e^{\gamma\delta}-1)$, and differentiating gives $t_p'(F)=t_0'(F)e^{\gamma t_0(F)}/[e^{\gamma t_0(F)}+C]$ for $F>F_*$. Milestones below $F_*$ are unchanged. The pause adds $\delta h(F_*)$; the step-hazard integral over capability is unchanged; and the state-hazard integral above $F_*$ changes by $\int_{F_*}^1h(F)[t_p'(F)-t_0'(F)]\,dF$. That gives equation~\eqref{eq:zeropause}. With constant positive state hazard $h$, the two terms combine to $h\,(T_p-T_0)>0$. A pause of length $\delta$ at the start has endpoint delay
\[
D_{\rm end}(\delta)=\frac1\gamma\ln\bigl[1+(e^{\gamma\delta}-1)e^{-\gamma T_0}\bigr]<\delta,
\]
which is shorter than the pause because compute keeps growing while the lab waits. The largest milestone delay remains $\delta$. Equation~\eqref{eq:zeropause} concerns takeoff-only risk; a continuing tail adds its own change in cumulative hazard.

\section{Calibration Details}\label{app:calibration}

This appendix completes Section~\ref{sec:scenario}. Set $F_0=0$, $F_1=1$, $K_0=1$, and $K(t)=e^{Gt}$, and use the closed forms of Appendix~A.1. The baseline 12-month pauses start at months \ipEqPauseSafetyStart{} (with safety) and \ipEqPauseBlogStart{} (all inference); their maximum milestone delays are \irPauseSafetyLead{} and \irPauseBlogLead{} months. A pause at the start has endpoint delay \ipPauseEnd{} months.

\emph{Hazard normalization.} Let $\kappa=2\ln3$, $I_\kappa=\int_0^1e^{\kappa F}dF$, and $b_H=H_{\rm full}/I_\kappa$ in the baseline, where $H_{\rm full}=-\ln(1-P_{\rm full})$ and $P_{\rm full}=0.75$. Full-speed hazard density per unit of capability is $b_He^{\kappa F}$, so three quarters of full-speed cumulative hazard occur after $F=1/2$. With state share $\omega=1/2$, the zero-knowledge functions are
\begin{equation}
h(F)=\omega b_He^{\kappa F}f^0(F)e^{S^0(F)},\qquad
k(F)=(1-\omega)b_He^{\kappa F}e^{S^0(F)},
\label{eq:hk}
\end{equation}
where $S^0(F)$ is knowledge accumulated along full speed in that version of the technology; in the baseline, $S^0(F)=\sigma_bt_0(F)$, which gives the scenario's functions. So $(h/f^0+k)e^{-S^0}=b_He^{\kappa F}$. Changing the safety law with the full-speed hazard density fixed changes $S^0$ and so $h,k$ (normalized hazards); the fixed-hazards exercises instead change $\sigma$ while retaining the baseline $h,k$, so the full-speed probability changes.

\emph{The general safety law.} The simulations use
\[
\dot S=\sigma(F)\left[\psi+(1-\psi)
\left(\frac{s}{s_0}\right)^\alpha e^{\beta Gt}\right]-d\,\dot F S .
\]
The baseline has $\beta=d=0$ and $\sigma(F)=\sigma_b$. The fit of equation~\eqref{eq:fit} gives $\sigma_b=\irSigmaB$ per month, a log-hazard reduction rate, or $1-e^{-\sigma_b}=\ipShareSigma$ percent of hazard removed in a benchmark month. Appendix~\ref{supp:versions} defines the ten versions of the safety technology. They vary the yield's level and growth, transfer loss, compute-scaled learning, a spillover from safety to capability research, and hazard after the takeoff period.

\emph{Services.} For horizon $q$ and discount rate $r=0.03/12$, $J_q$ is given by equation~\eqref{eq:services}. For $t\ge T$, capability stays at one and the shares return to $(i_0,x_0,s_0)$; growing compute supplies the common continuation, in all versions. The full-speed values are $J_{24}^0=\irJTwo$ and $J_{60}^0=\irJFive$, a ratio of \irServiceRatio. After completion the integrand grows at $G/2-r>0$, so service value is reported over finite horizons.

\emph{The ceiling controller.} In the mixed-hazard simulations, $k(F)>0$ at every capability. The controller chooses the pace
\[
f=\min\left\{f_{\rm available},
\max\left[0,\frac{\bar c e^S-h(F)}{k(F)}\right]\right\}.
\]
Without spillovers, $x=x_0(f/f_{\rm available})^{1/\varepsilon}$, the freed research share goes to safety, and $i=i_0$. With spillovers, the controller solves $x+\chi s=(x_0+\chi s_0)(f/f_{\rm available})^{1/\varepsilon}$ and $x+s=x_0+s_0$ if the solution has $x\ge0$; otherwise it sets $x=0$, uses the necessary $s$, and sends the remainder to customers. This is a controller rather than cost-minimizing compliance.

When $k(F)=0$, the pace formula does not apply: the ceiling then requires $h(F)e^{-S}\le\bar c$ whatever the pace, and a rule must keep that state constraint satisfied as capability and knowledge evolve.

With a 36-month full-speed takeoff period, all but one comparison keeps its direction (Table~\ref{tab:takeoff36}). Thirty-six months is a slow case. In the AI Futures Model's published forecasts for this stretch, the medians run from 9 to 18 months and the 90th percentiles from 3.6 to 15.5 years \citep{lifland2026}. The other targets stay fixed, and each rule is re-solved to use the 12-month lead. The ceiling tightens to \tkCapThirtySix{} a month. The necessary bound falls to \tkNecBoundThirtySix{} and the replay bound to \tkSpareBoundThirtySix, both below the fitted productivity. The necessary bound stays just above one half, so a regulator who cannot rule out half the fitted productivity still cannot keep both promises. Learning first stays best with a constant yield; when half of the knowledge is lost across the takeoff period, the best start moves from month \tkTimingTransferStartEighteen{} to month \tkTimingTransferStartThirtySix. When research does nothing, floors and the lag still raise the probability of catastrophe. The exception is the ceiling at half the fitted productivity: it now changes that probability by \tkCeilRiseHalfThirtySix{} points instead of raising it by \tkCeilRiseHalfEighteen. The spare-compute replay costs more, \tkLossSpareFiveThirtySix{} percent of five-year services rather than \tkLossSpareFiveEighteen, because customers now get nothing until month 48 of the 60.

\begin{table}[htbp]
\centering
\caption{Headline Results with a 36-Month Full-Speed Takeoff Period}
\label{tab:takeoff36}
\small
\begin{tabular}{@{}>{\raggedright\arraybackslash}p{0.7\textwidth}rr@{}}
\toprule
 & 18 months & 36 months\\
 & (baseline) & \\
\midrule
\multicolumn{3}{@{}l}{\emph{Refitted scenario, and settings re-solved to use the 12-month lead}}\\
Full-speed takeoff period (months) & \tkTakeoffEighteen & \tkTakeoffThirtySix\\
Compute growth over the takeoff period (fold) & \tkComputeEndEighteen & \tkComputeEndThirtySix\\
Fitted safety yield $\sigma_b$ (per month) & \tkSigmaBEighteen & \tkSigmaBThirtySix\\
Unprotected final state hazard $h(1)$ (per month) & \tkHoneEighteen & \tkHoneThirtySix\\
Ceiling $\bar c$ (per month) & \tkCapEighteen & \tkCapThirtySix\\
Ceiling per company, three companies (percent a month) & \tkCapPctEighteen & \tkCapPctThirtySix\\
Compute floors: research share (percent) & \tkFloorsShareEighteen & \tkFloorsShareThirtySix\\
R\&D-model lag (months) & \tkLagMonthsEighteen & \tkLagMonthsThirtySix\\
Pause with safety: start (month) & \tkPauseSafetyStartEighteen & \tkPauseSafetyStartThirtySix\\
All-inference pause: start (month) & \tkPauseBlogStartEighteen & \tkPauseBlogStartThirtySix\\
\addlinespace[4pt]
\multicolumn{3}{@{}l}{\emph{Feasibility and customer cost at the ceiling (fixed hazards)}}\\
Necessary bound & \tkNecBoundEighteen & \tkNecBoundThirtySix\\
Replay bound, spare-compute replay & \tkSpareBoundEighteen & \tkSpareBoundThirtySix\\
Replay bound, usual-share replay & \tkUsualBoundEighteen & \tkUsualBoundThirtySix\\
Gap between the replay and necessary bounds (percent) & \tkScenGapEighteen & \tkScenGapThirtySix\\
Spare-compute replay: probability at its bound (percent) & \tkRiskSpareBoundEighteen & \tkRiskSpareBoundThirtySix\\
Spare-compute replay: probability at the fitted productivity (percent) & \tkRiskSpareFitEighteen & \tkRiskSpareFitThirtySix\\
Spare-compute replay: five-year service loss (percent) & \tkLossSpareFiveEighteen & \tkLossSpareFiveThirtySix\\
Cheapest replay at the fitted productivity: learning (months) & \tkOptMonthsEighteen & \tkOptMonthsThirtySix\\
Cheapest replay at the fitted productivity: five-year loss (percent) & \tkOptFiveEighteen & \tkOptFiveThirtySix\\
\addlinespace[4pt]
\multicolumn{3}{@{}l}{\emph{The ceiling at lower productivity (normalized hazards)}}\\
Lead used at half the fitted productivity (months) & \tkCeilLeadHalfEighteen & \tkCeilLeadHalfThirtySix\\
Lead used at a quarter of the fitted productivity (months) & \tkCeilLeadQuarterEighteen & \tkCeilLeadQuarterThirtySix\\
Rise in probability at half the fitted productivity (points) & \tkCeilRiseHalfEighteen & \tkCeilRiseHalfThirtySix\\
Rise in probability at a quarter of the fitted productivity (points) & \tkCeilRiseQuarterEighteen & \tkCeilRiseQuarterThirtySix\\
\addlinespace[4pt]
\multicolumn{3}{@{}l}{\emph{Timing of a 12-month pause with safety (normalized hazards)}}\\
Best start, constant yield, all knowledge retained (month) & \tkTimingConstStartEighteen & \tkTimingConstStartThirtySix\\
Best start, yield triples, all knowledge retained (month) & \tkTimingRiseStartEighteen & \tkTimingRiseStartThirtySix\\
Best start, constant yield, half of knowledge retained (month) & \tkTimingTransferStartEighteen & \tkTimingTransferStartThirtySix\\
Best start, yield triples, half of knowledge retained (month) & \tkTimingBothStartEighteen & \tkTimingBothStartThirtySix\\
Gain from the best start, constant yield, half retained (points) & \tkTimingTransferGainEighteen & \tkTimingTransferGainThirtySix\\
Fall in probability, yield triples, half retained, start now (points) & \tkTimingBothFirstEighteen & \tkTimingBothFirstThirtySix\\
Fall in probability, yield triples, half retained, best start (points) & \tkTimingBothBestEighteen & \tkTimingBothBestThirtySix\\
\addlinespace[4pt]
\multicolumn{3}{@{}l}{\emph{Zero productivity (normalized hazards)}}\\
Compute floors: change in probability (points) & \tkZeroFloorsEighteen & \tkZeroFloorsThirtySix\\
R\&D-model lag: change in probability (points) & \tkZeroLagEighteen & \tkZeroLagThirtySix\\
12-month pause at the start: fall in probability (points) & \tkZeroPauseEighteen & \tkZeroPauseThirtySix\\
Latest start at which a 12-month pause lowers the probability (month) & \tkPauseCrossEighteen & \tkPauseCrossThirtySix\\
\bottomrule
\end{tabular}
\par\smallskip
\begin{minipage}{\textwidth}\footnotesize
\emph{Notes:} Held fixed: the 75 percent probability of catastrophe at full speed, the 40 percent probability after a 12-month pause at the start with compute standing still, the 12-month deadline, compute doubling every 6 months, research 10 times faster at the endpoint, the compute shares (50, 5, and 45 percent), the learning multiplier, the 24- and 60-month service horizons, and the 3 percent discount rate. Refitted: the takeoff speed $\theta$, the hazard scale, and the safety yield $\sigma_b$. Each rule's setting is re-solved to use the 12-month lead at the fitted productivity, and later panels use each column's own ceiling. Productivity is relative to each column's fitted yield. Probability means the probability of catastrophe; changes are in percentage points, and a positive change is a rise. The pauses in the first panel resume at the full research share; the pause in the timing panel is followed by a replay, as in Table~\ref{tab:timing}; ``half retained'' means half of the knowledge carries across the takeoff period.
\end{minipage}
\end{table}

\clearpage
\section{Exhibits Summarized in the Paper}\label{supp:exhibits}

This section reports the climbing analogy of Section~\ref{sec:climb} and four exhibits whose numbers the main text gives: the productivity bounds when early knowledge loses value (Section~\ref{sec:widens}), the value of a month along the full-speed path (Section~\ref{sec:earlier}), the feasible pause timing under the ceiling (Section~\ref{sec:feasible-timing}), and the two terms of the zero-productivity pause (Section~\ref{sec:pause}).

\begin{table}[htbp]
\centering
\caption{The Model and the Climb}
\label{tab:climb}
\small
\begin{tabular}{@{}>{\raggedright\arraybackslash}p{4.6cm}>{\raggedright\arraybackslash}p{10.6cm}@{}}
\toprule
In the model & On the mountain\\
\midrule
Capability $F$ & Altitude. The Automated Coder is base camp; top-expert-dominating AI is the summit.\\
Full-speed path & The standard schedule: the date the expedition plans to reach each camp.\\
Safety knowledge $S$ & Adjustment to the altitude. It builds with time on the mountain and protects the climber at every higher altitude.\\
Productivity $a$ & How fast this climber adjusts, unknown before the trip.\\
Ceiling $\bar c$ & A safety rule: never climb above the altitude your body is ready for.\\
Deadline and lead $L$ & A weather window: at no camp may the expedition fall more than $L$ days behind the standard schedule.\\
State hazard $h(F)$ & The danger of each day spent high up: cold and thin air, worst in the ``death zone'' near the summit.\\
Step hazard $k(F)$ & The danger of each stretch climbed: a fall. It depends on the distance climbed, not on the speed.\\
A rule fixing compute shares & A fixed itinerary, such as ``climb 500 metres a day'': the dates are certain, the danger is not.\\
Learn-then-replay & The classic expedition plan: adjust at base camp first, then follow the standard schedule.\\
Lost customer services & Days of the trip spent waiting rather than doing anything else.\\
Growing compute $K(t)$ & No counterpart.\\
\bottomrule
\end{tabular}
\end{table}

\begin{table}[htbp]
\centering
\caption{Productivity Bounds When Early Knowledge Loses Value}
\label{tab:scope}
\small
\begin{tabular}{@{}llccc@{}}
\toprule
Hazard & Safety technology & Necessary & Replay & Ratio\\
\midrule
Only state hazard & constant yield & 0.656 & 0.677 & 1.03\\
 & yield rises threefold & 0.246 & 0.625 & 2.54\\
 & half of knowledge carries over & 0.569 & 1.046 & 1.84\\
\addlinespace
Scenario's hazard split & constant yield & 0.573 & 0.693 & 1.21\\
 & yield rises threefold & 0.218 & 0.637 & 2.92\\
 & half of knowledge carries over & 0.486 & 1.074 & 2.21\\
\bottomrule
\end{tabular}
\par\smallskip
\begin{minipage}{\textwidth}\footnotesize
\emph{Notes:} Smallest productivity bounds, relative to each version's own fitted productivity, for a ceiling of \insCap{} a month and a 12-month deadline. ``Necessary'' is the bound of Proposition~\ref{prop:fix}; ``Replay'' is the replay bound with 12 months of learning and all spare compute in safety. Each variant uses its own normalized hazards and keeps the initial yield at its fitted value. The constant-yield rows reproduce Figure~\ref{fig:bounds} at the scenario's ceiling.
\end{minipage}
\end{table}

\begin{figure}[htbp]
\centering
\includegraphics[width=0.72\textwidth]{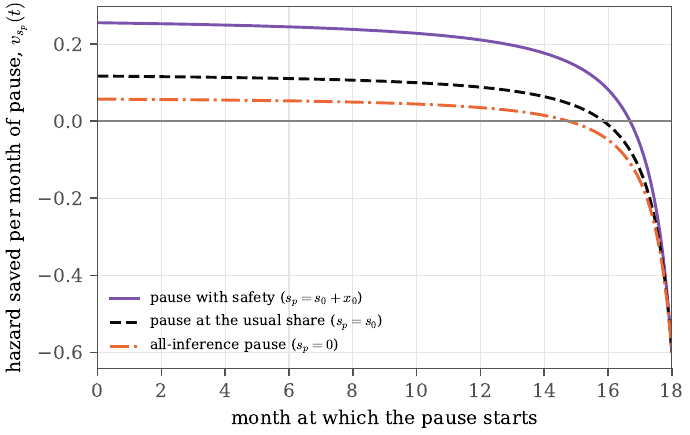}
\caption{The Value of a Month along the Full-Speed Path}
\label{fig:valuemonth}
\par\smallskip
\begin{minipage}{\textwidth}\footnotesize
\emph{Notes:} Net cumulative hazard saved per month of pause, $v_{s_p}(t)=mA(t)-w(t)$ of equation~\eqref{eq:month}, under the repeated-path benchmark, at the fitted productivity and the scenario's hazards. The three lines differ in the safety share during the pause, which sets the monthly growth of knowledge $m=\sigma_b y(s_p)$. A pause lowers cumulative hazard where the line is positive.
\end{minipage}
\end{figure}

\begin{table}[htbp]
\centering
\caption{How the Ceiling Changes the Feasible Pause Timing}
\label{tab:feasible-timing}
\small
\begin{tabular}{@{}lrrrr@{}}
\toprule
Ceiling & Minimum initial & Feasible starts & Best feasible & Fall in probability\\
(per month) & pause (months) & (months) & start (month) & (points)\\
\midrule
0.0492 (scenario) & 35.5 & none & \textemdash & \textemdash\\
0.15 & 23.5 & 13.5 to 15.8 & 13.5 & 43.2\\
0.30 & 16.0 & 8.5 to 16.7 & 9.2 & 47.9\\
0.50 & 10.5 & 0.0 to 17.2 & 9.2 & 47.9\\
\bottomrule
\end{tabular}
\par\smallskip
\begin{minipage}{\textwidth}\footnotesize
\emph{Notes:} The technology of the last row of Table~\ref{tab:timing} (yield tripling over the takeoff period, half transfer, normalized hazards), the same reference, and the same 12-month pause with safety; the lead is $L=12$ months. The minimum initial pause is $\delta_c(0)$ in equation~\eqref{eq:minimum-feasible-pause}. Feasible starts are the dates satisfying~\eqref{eq:feasible-timing}; the best feasible start maximizes the saving among them. The scenario's ceiling is the one of Section~\ref{sec:conventions}; the looser ceilings illustrate the constraint. The reference itself exceeds every ceiling shown before completion, so no pause is not an option.
\end{minipage}
\end{table}

\begin{figure}[htbp]
\centering
\includegraphics[width=\textwidth]{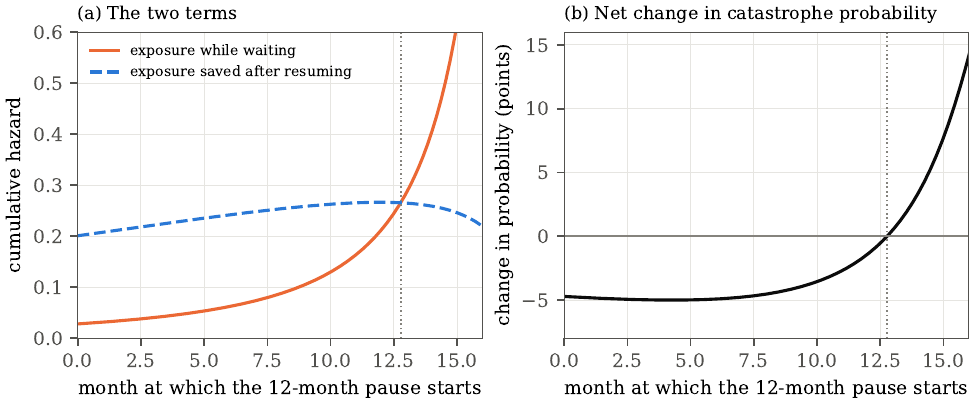}
\caption{A 12-Month Pause When Safety Research Does Nothing}
\label{fig:pausezero}
\par\smallskip
\begin{minipage}{\textwidth}\footnotesize
\emph{Notes:} The two terms of equation~\eqref{eq:zeropause} (panel a) and the resulting change in the probability of catastrophe relative to full speed (panel b), as functions of the month at which a 12-month pause starts. Productivity is zero, and hazards are normalized so that full speed keeps its 75 percent probability. The dotted line marks the start date at which the two terms are equal.
\end{minipage}
\end{figure}

\section{The Ten Versions of the Safety Technology}\label{supp:versions}

This section completes Appendix~\ref{app:calibration} with the definitions behind Table~\ref{tab:robust}.

\emph{Yield levels and growth.} Three versions set $\sigma(F)=a\sigma_b$ for $a=0.25,0.5,1.5$. The rising version sets $\sigma(F)=\sigma_b3^F$: initial productivity is fixed, rather than endpoint knowledge. Full-speed endpoint stocks are \irFlatKnowledge{} in the baseline and \irRisingKnowledge{} in the rising version, \irRisingExtra{} percent higher.

The transfer version sets $d=\ln2$, so knowledge carried through a unit increase in capability retains one half of its value. The compute-scaled version sets $\beta=\alpha=0.5$ and $d=0$, making the compute-using part proportional to absolute safety compute raised to that elasticity.

In the spillover version, with $\chi=0.2$, the research law becomes
\[
\dot F=b_0e^{\gamma t+\rho F}
\left(\frac{x+\chi s}{x_0+\chi s_0}\right)^\varepsilon ,
\]
so dedicated research can be zero during a pause with safety while safety compute still advances capability. Dates remain independent of productivity when $\chi$ and the allocation controls stay fixed.

Two tail versions have continuing state hazard after completion:
\[
\lambda_{\rm tail}(T+u)=\eta h(1)e^{-S(T)-\sigma(1)u},
\qquad H_{\rm tail}=\frac{\eta h(1)e^{-S(T)}}{\sigma(1)}.
\]
Tail hazard starts at the share $\eta$ of the endpoint state hazard. It then decays as knowledge grows at the endpoint yield with safety share $s_0$. One version uses $\eta=1$; the other uses $\eta=\irTailEta$, chosen so that one quarter of full-speed cumulative hazard is in the tail. For these versions,
\[
b_H=\frac{H_{\rm full}}
{I_\kappa+\eta\omega e^\kappa f^0(1)/\sigma(1)},
\]
where the second term in the denominator is the tail's full-speed hazard per unit of $b_H$, so takeoff period and tail together match the full-speed target. Probability is $1-e^{-H_{\rm takeoff}-H_{\rm tail}}$. Initial yield stays at its baseline value outside the three yield-level versions. No alternative technology receives a new 40 percent target.

\section{Fixed Rules across Productivity and Technologies}\label{supp:fixed}

This section reports the exercise summarized in Section~\ref{sec:fixed}.

\begin{figure}[htbp]
\centering
\includegraphics[width=\textwidth]{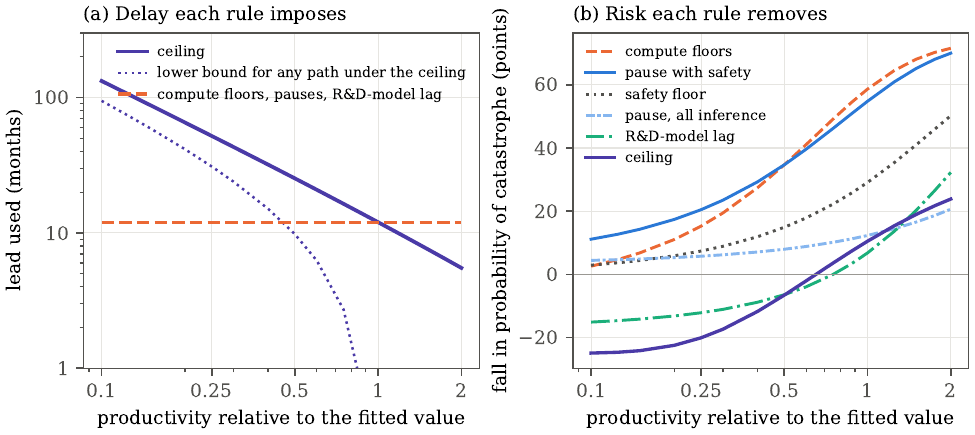}
\caption{Delay and Risk of Fixed Rules as Safety Research Becomes More or Less Productive}
\label{fig:yield}
\par\smallskip
\begin{minipage}{\textwidth}\footnotesize
\emph{Notes:} Each rule keeps the setting that uses a year of lead at the fitted productivity. Zero-knowledge hazard functions are rescaled at each productivity to maintain 75 percent full-speed probability. Panel (a) reports maximum milestone delay; the dotted line is the lower bound on endpoint delay $T-T_0$ implied by Proposition~\ref{prop:fix}(b). Negative bounds are uninformative and lie below the plotted range. Panel (b) reports the fall in probability relative to that version's full-speed path. The ceiling uses the specified controller, rather than optimized compliance.
\end{minipage}
\end{figure}

Figure~\ref{fig:yield} varies productivity from a tenth to twice the fitted value while holding each rule at its one-year-lead setting. It uses normalized hazards, so each version keeps the full-speed takeoff period and its 75 percent probability.

The allocation rules keep their dates as productivity changes (panel a). The ceiling does not. It uses \insCeilLeadHalf{} months of lead at half the fitted productivity and \insCeilLeadQuarter{} at a quarter. Risk tells the opposite story (panel b). At a quarter of the fitted productivity, the ceiling raises catastrophe risk by \insCeilRiseQuarter{} points. The floors instead lower it by \insFloorsFallQuarter{} points, and the pause with safety by \insPauseFallQuarter. With weak research, the controller must slow development more, and a lower hazard rate held for more months can accumulate more danger.

The risk comparison depends on which convention is used. Under fixed hazards, halving the productivity makes the full-speed probability \irFixedHalfBase{} percent. The same ceiling lowers it to \irFixedHalfCeil{} percent, a \irFixedHalfFall-point reduction, while using \irFixedHalfLead{} months of lead. The deadline problem persists even when the risk comparison changes sign.

Under normalized hazards, at one hundredth of the fitted productivity, the floors raise risk to \irWeakOneHundredth{} percent. That exceeds full speed's 75 percent while preserving the year of lead. Section~\ref{supp:checks} reports both checks.

The sign of a rule's effect depends on the safety technology. Table~\ref{tab:robust} retunes each rule to use a year of lead in ten versions of the technology and reports how much it lowers the probability of catastrophe. The versions vary the level of the yield, its growth with capability, and how much knowledge carries over to stronger models. They also vary whether safety research scales with compute or advances capability, and whether danger continues after the takeoff period. Appendix~\ref{app:calibration} defines them.

\begin{table}[htbp]
\centering
\caption{What a Year of Lead Buys under Ten Safety Technologies}
\label{tab:robust}
\small
\setlength{\tabcolsep}{3.5pt}
\begin{tabular}{@{}>{\raggedright\arraybackslash}p{5.2cm}cccccc@{}}
\toprule
& Compute & Pause with & Pause, all & & R\&D-model & Safety\\
& floors & safety & inference & Ceiling & lag & floor\\
\midrule
Scenario & \rbBaseFloors & \rbBasePauseSafety & \rbBasePauseBlog & \rbBaseCeiling & \rbBaseLag & \rbBaseSafetyFloor\\
Yield a quarter of the scenario's & \rbYieldQuarterFloors & \rbYieldQuarterPauseSafety & \rbYieldQuarterPauseBlog & \rbYieldQuarterCeiling & \rbYieldQuarterLag & \rbYieldQuarterSafetyFloor\\
Yield half of the scenario's & \rbYieldHalfFloors & \rbYieldHalfPauseSafety & \rbYieldHalfPauseBlog & \rbYieldHalfCeiling & \rbYieldHalfLag & \rbYieldHalfSafetyFloor\\
Yield 1.5 times the scenario's & \rbYieldOneHalfFloors & \rbYieldOneHalfPauseSafety & \rbYieldOneHalfPauseBlog & \rbYieldOneHalfCeiling & \rbYieldOneHalfLag & \rbYieldOneHalfSafetyFloor\\
Yield rises threefold & \rbRisingFloors & \rbRisingPauseSafety & \rbRisingPauseBlog & \rbRisingCeiling & \rbRisingLag & \rbRisingSafetyFloor\\
Half of knowledge carries over & \rbTransferFloors & \rbTransferPauseSafety & \rbTransferPauseBlog & \rbTransferCeiling & \rbTransferLag & \rbTransferSafetyFloor\\
Safety research scales with compute & \rbComputeFloors & \rbComputePauseSafety & \rbComputePauseBlog & \rbComputeCeiling & \rbComputeLag & \rbComputeSafetyFloor\\
A fifth of safety compute advances capability & \rbSpillFloors & \rbSpillPauseSafety & \rbSpillPauseBlog & \rbSpillCeiling & \rbSpillLag & \rbSpillSafetyFloor\\
A quarter of hazard after the takeoff period & \rbTailSmallFloors & \rbTailSmallPauseSafety & \rbTailSmallPauseBlog & \rbTailSmallCeiling & \rbTailSmallLag & \rbTailSmallSafetyFloor\\
State hazard continues after the takeoff period & \rbTailFloors & \rbTailPauseSafety & \rbTailPauseBlog & \rbTailCeiling & \rbTailLag & \rbTailSafetyFloor\\
\bottomrule
\end{tabular}
\par\smallskip
\begin{minipage}{\textwidth}
\footnotesize
\emph{Notes:} Fall in the probability of catastrophe relative to full speed, in percentage points; a negative entry is a rise. Each rule is retuned separately in each version to use 12 months of maximum milestone delay, and each pause is placed at its best start; the safety floor uses no lead. Every version keeps the full-speed takeoff period, its 75 percent probability of catastrophe, and the even split between state and step hazard during the takeoff period. Zero-knowledge hazard functions are rescaled with the safety technology. Versions with danger after the takeoff period count that danger.
\end{minipage}
\end{table}

\section{How Compute Growth Sets the Gap between the Bounds}\label{supp:doubling}

Compute growth drives the result. The factor depends on compute growth only through the replay research share; the hazard functions and the fitted productivity do not enter. Table~\ref{tab:doubling} shows the factor for slower compute growth, holding everything else fixed. With a 12-month doubling time, the replay needs a quarter of compute for research, and the gap is \insDoubleTwelve{} percent; with a 24-month doubling time, \insDoubleTwentyFour{} percent.

Without any compute growth, the replay would need the full research share of one half, and the factor's formula gives a gap of \lfFactorNonePct{} percent. Diminishing returns to safety compute keep even that gap modest: moving from half of compute to all of it multiplies learning only by $\insYbar/\lfYNone$.

\begin{table}[htbp]
\centering
\caption{The Gap between the Bounds and the Speed of Compute Growth}
\label{tab:doubling}
\footnotesize
\begin{tabular}{@{}lcccc@{}}
\toprule
Compute doubles every & Research share & Learning $y_r$ & Factor & Gap (percent)\\
\midrule
6 months (scenario) & \insFrReplayShare & \insFrYspare & \insFrPureFactor & \insFrPureGapPct\\
12 months & \lfShareTwelve & \lfYTwelve & \lfFactorTwelve & \insDoubleTwelve\\
24 months & \lfShareTwentyFour & \lfYTwentyFour & \lfFactorTwentyFour & \insDoubleTwentyFour\\
No growth (limit of the formula) & \lfShareNone & \lfYNone & \lfFactorNone & \lfFactorNonePct\\
\bottomrule
\end{tabular}
\par\smallskip
\begin{minipage}{\textwidth}\footnotesize
\emph{Notes:} The factor $\bar y(T_0+L)/(\bar yL+y_rT_0)$ of Proposition~\ref{prop:sharp}, with $T_0=18$, $L=12$, $\bar y=\insYbar$, and $y_r=y(1-x_0e^{-GL})$. Research shares are in percent of compute. The factor depends on compute growth only through the replay research share. The last row evaluates the formula with no compute growth; it is not a recalibrated scenario, and it does not check the shape restriction.
\end{minipage}
\end{table}

\section{Extensions of the Value of a Month and the Knowledge Bound}\label{supp:extensions}

The proofs of Propositions~\ref{prop:timing} and~\ref{prop:feasible-timing} are in Appendix~\ref{app:proofs}. This section gives the marginal timing condition and the exact finite-pause derivative. It then gives the test for a reference that already uses lead, the value of a month with a variable yield, and the knowledge bound at every milestone. It ends with the bound under rising yields, depreciation, compute-scaled learning, and normalized hazards.

For the local comparison, let $A_d(t)=\int_t^T\lambda(u)R(t,u)du$ be remaining hazard weighted by how much new knowledge survives, and $v_d(t)=m(t)A_d(t)-w(t)$ the value of a short pause. Put $f(t)=\dot F(t)$ and let $\nu(t)=\sigma(F(t))y(s(t))$ denote gross reference learning. Here $\dot S=\nu-dfS$; gross learning must not be confused with the change in the stock. At differentiable points,
\[
\dot A_d=-\lambda+dfA_d,\qquad
\dot w=fh'e^{-S}-\nu w+dfSw.
\]
The first identity accounts for hazard passed and the shorter distance over which newly acquired knowledge will depreciate. The second accounts for the changing state hazard and the net change in existing knowledge. Expansion of equation~\eqref{eq:timing-saving} at $\delta=0$ gives $\mathcal V_\delta=\delta v_d+O(\delta^2)$. Differentiating $v_d=mA_d-w$ gives
\begin{equation}
\dot v_d=(\dot m+dfm)A_d
-fe^{-S}[h'+mk+dhS]+(\nu-m)w.
\label{eq:timing-general}
\end{equation}
When the reference and pause use $s_0$, $\nu=m=\sigma$ because $y(s_0)=1$, and $\dot m=\sigma'f$. Dividing by positive $f$ shows that, at a developing milestone, postponing a short pause raises its value exactly when
\begin{equation}
[\sigma'(F)+d\sigma(F)]A_d(t)
>
e^{-S(t)}[h'(F)+\sigma(F)k(F)+dS(t)h(F)].
\label{eq:timing-local}
\end{equation}
Here $h$ and $\sigma$ are differentiable, and $\dot F>0$. The left side values more productive research and better retention of new knowledge. The right side counts rising state hazard, step hazard already passed, and erosion of existing knowledge. At the initial state $S=0$, the last cost vanishes. With a higher safety share during the pause, $m>\nu$, the final term is negative: delaying also postpones the extra research effort. Transfer loss has two effects. It favors later production of new knowledge through $dfmA_d$, but raises exposure from erosion of existing knowledge through $-dfSw$.

The exact finite-pause condition need not have the marginal condition's sign. Define two nonnegative weights,
\[
C_\delta(t)=\int_t^T\lambda(u)R(t,u)e^{-m(t)\delta R(t,u)}du,
\qquad J_\delta(m)=\int_0^\delta v e^{-mv}dv.
\]
The first weights hazard that another unit of pause knowledge could still remove. The second measures how a higher yield reduces exposure during the pause. Since $\partial_tR=dfR$ and $\partial_mg_\delta=-J_\delta$, differentiation of equation~\eqref{eq:timing-saving} gives
\begin{equation}
\begin{split}
\partial_t\mathcal V_\delta={}&
\delta(\dot m+dfm)C_\delta+w\dot m J_\delta\\
&-g_\delta\{fe^{-S}[h'+mk+dhS]+(m-\nu)w\}.
\end{split}
\label{eq:timing-finite-derivative}
\end{equation}
The moving lower limit contributes $-\lambda(t)(1-e^{-m\delta})$. Combining it with the derivative of pause exposure gives the terms on the second line. Equation~\eqref{eq:timing-finite-derivative} is exact for any positive duration, including its continuous limit at $m=0$.

This derivative supplies global timing tests within the class. If it is nonpositive throughout the candidate interval, learning first maximizes the saving. With constant yield, full transfer, nondecreasing $h$, and $y(s_p)\ge y(s(t))$, its first line vanishes and its second line is nonpositive. Learning first is then globally best among equal-length pauses. If instead the derivative is positive before one interior date and negative afterward, that date maximizes saving. It dominates no pause when its saving is positive. Without this sign pattern, compare endpoints and all stationary points using equation~\eqref{eq:timing-saving}. In particular, a positive derivative at zero and a nonnegative initial saving imply that some later start beats both immediate learning and no pause.

The replay is feasible under the basic research law. At matched dates, $e^{-G\delta}x(u)K(u+\delta)=x(u)K(u)$, so research speed is identical. For $G\ge0$, holding the safety share fixed leaves nonnegative compute for customers. A full-speed reference gives maximum milestone delay exactly $\delta$; a general reference gives delay no greater than its own maximum plus $\delta$. No calendar-dependent safety yield, research lag, or capability spillover from safety research is used here. If the reference meets a ceiling, pause hazard starts below its reference hazard and falls, while every replay hazard is weakly lower. If it does not, the best cumulative-risk timing need not repair the ceiling violation.

The comparison can also use the main text's uncertainty criterion. Fix $h,k$ and write $\sigma=a\bar\sigma$, with $S(0)=0$ and prescribed capability and allocation paths independent of $a$. For each candidate path, knowledge is proportional to $a$, so cumulative hazard is nonincreasing in $a$. Its worst value over $a\ge a_{\min}$ occurs at $a_{\min}$. Applying the timing comparison there minimizes worst-case cumulative hazard within the same class. This does not turn the comparison into a guarantee of a hazard ceiling.

\emph{A reference that already uses lead.} Let the reference be strictly increasing, and define its milestone delay by $\ell(u)=u-t_0(F(u))$. The pause leaves milestones already reached unchanged and delays every higher one by $\delta$. Define
\[
D_-(t)=\sup_{0<u\le t}\ell(u),\qquad
D_+(t)=\sup_{t<u\le T}\ell(u),\qquad D_-(0)=0.
\]
The first quantity is lead already used; the second is the greatest delay on the remaining reference path. The inserted pause has largest delay $\max\{D_-(t),D_+(t)+\delta\}$. Replace the deadline test in~\eqref{eq:feasible-timing} by $D_-(t)\le L$ and $\delta\le L-D_+(t)$. Replace $T_0$ by $T$ in the hazard tests. This accounts for every milestone, including completion, rather than checking only the final date.

\emph{Choosing duration within the commitments.} At a fixed start passing the prior-exposure and pause-exposure tests, feasible durations form $[\delta_c(t),L]$ for a full-speed reference. The interval is empty if its lower endpoint exceeds its upper endpoint. For a general strictly increasing reference, replace $L$ by $L-D_+(t)$ and require $D_-(t)\le L$. When the interval is nonempty, its maximum saving occurs at an endpoint because the transformed duration derivative is nondecreasing. For the full-speed case with $\delta_c(t)\le L$,
\begin{equation}
\max_{\delta_c(t)\le\delta\le L}\mathcal V_\delta(t)
=\max\{\mathcal V_{\delta_c(t)}(t),\mathcal V_L(t)\}.
\label{eq:feasible-duration}
\end{equation}
The regulator compares the shortest pause that satisfies the ceiling with the longest pause the deadline permits. If $\delta_c(t)=0$, that endpoint is the unpaused reference, which passes the ceiling tests. If $m(t)=0$ and the interval is nonempty, saving is $-w(t)\delta$, and the same conclusion holds. This duration result minimizes cumulative hazard within the feasible pause set; customer services require a separate comparison.

\emph{The value of a month with variable yield.} The repeated-path benchmark of Proposition~\ref{prop:month} repeats future capability pace and safety controls after a pause. Learning depends on capability and the safety share, with no calendar-time dependence or depreciation. On differentiable reference segments, assume locally that $h$ and $\sigma$ are continuously differentiable. For a fixed candidate pause share $s_p$, let $m(t)=\sigma(F(t))y(s_p)$, $r_S(t)=\dot S(t)$, and $v_{s_p}=mA-w$. Since $\dot A=-\lambda$ and $\dot w=h'(F)\dot F e^{-S}-r_Sw$, differentiation gives
\begin{equation}
\dot v_{s_p}=\dot m A-[h'(F)+m k(F)]\dot F e^{-S}+(r_S-m)w.
\label{eq:general-value-derivative}
\end{equation}
The last term matters when the reference safety allocation differs from the candidate pause allocation. With both shares at $s_0$, $y(s_0)=1$ and $r_S=m=\sigma(F)$, so
\begin{equation}
\dot v=\dot F\left\{\sigma'(F)A-[h'(F)+\sigma(F)k(F)]e^{-S}\right\}.
\label{eq:variable-value-derivative}
\end{equation}
A higher yield at later capabilities increases the benefit of waiting there through $\sigma'(F)A$. It can overturn the constant-yield timing conclusion, but only when it exceeds the exposure and step-hazard terms. A rising yield alone does not establish that a late pause is best. These are derivatives of the repeated-path marginal value, not a policy comparison when growing compute changes the resumed path.

\emph{The knowledge bound at every milestone.} The main bound combines $S(T)\ge\ln[h(1)/\bar c]$ with an upper limit on learning, $S(T)\le a\int_0^T\bar\sigma(F(t))y(s(t))\,dt$. Applying it at every milestone, for fixed hazard functions and a completing monotone path under the share-based learning law, the ceiling requires $S(t(F))\ge[\ln(h(F)/\bar c)]_+$ wherever $h(F)>0$. The deadline gives $t(F)\le t_0(F)+L$, while learning is at most $a\bar\sigma_{\max}\bar y$ per month. So a necessary productivity bound is
\begin{equation}
a\ge\sup_{0<F\le1}
\frac{[\ln(h(F)/\bar c)]_+}
{\bar\sigma_{\max}\bar y\,[t_0(F)+L]}.
\label{eq:milestone-bound}
\end{equation}
Here $[z]_+=\max\{z,0\}$, with numerator zero when $h(F)=0$. This includes the endpoint bound of Proposition~\ref{prop:fix} and can detect a more restrictive interior milestone. Replacing $\bar\sigma_{\max}$ in the denominator by $\max_{0\le f\le F}\bar\sigma(f)$ strengthens the bound further: a monotone path cannot use higher-capability yields before reaching that milestone. A zero local maximum makes a milestone with $h(F)>\bar c$ infeasible, and imposes no additional requirement when $h(F)\le\bar c$. The separate initial condition $\bar c\ge h(0)$ still applies. These bounds ignore nonnegative step hazard and relax the joint use of compute for learning and development, so passing them does not establish feasibility.

\emph{Rising yields, depreciation, and compute-scaled learning.} Bounded rising yields and depreciation are covered by $\bar\sigma_{\max}$. Nonnegative depreciation $d\dot F S$ only lowers knowledge, preserving the upper bound in Proposition~\ref{prop:fix}. It generally changes the replay bound of Proposition~\ref{prop:construct}, which relies on knowledge retained during the replay. With compute-scaled learning ($\beta>0$), $y(s,t)\le\bar y(t)=\psi+(1-\psi)s_0^{-\alpha}e^{\beta Gt}$, and the necessary condition is
\[
a\bar\sigma_{\max}\left[\psi T+
\frac{(1-\psi)s_0^{-\alpha}}{\beta G}(e^{\beta GT}-1)\right]
\ge\ln[h(1)/\bar c].
\]
For a fixed deadline the left side tends to zero with $a$, so a ceiling below $h(1)$ eventually fails. Required time diverges as $a\downarrow0$, but exponential growth can make it of order $\ln(1/a)$ rather than $1/a$. The replay result must be rechecked for the modified law; Proposition~\ref{prop:construct} concerns share-based learning without depreciation.

\emph{Normalized hazards.} For the constant-yield sweep under normalized rather than fixed hazards, $h(1;a)=\bar h_1e^{a\sigma_bT_0}$, where $\bar h_1=\insStateEnd$ per month is the state hazard carried by full speed. So
\[
T\ge\frac{\ln(\bar h_1/\bar c)+a\sigma_bT_0}
{a\sigma_b\bar y}.
\]
This bound diverges when $\bar c<\bar h_1$. A ceiling below the baseline zero-knowledge $h(1)$ alone does not imply divergence under this normalization. The scenario's aggregate ceiling \insCap{} satisfies the stronger condition. The knowledge requirement at the endpoint date is unchanged by later tail hazard; total-risk comparisons depend on the tail and continued learning.

\section{Additional Checks}\label{supp:checks}

\emph{Two hazard conventions.} Table~\ref{tab:normalization} applies the same ceiling settings under both conventions. Full-speed risk changes when $h,k$ stay fixed. At half the fitted productivity the ceiling lowers risk by \irFixedHalfFall{} points with fixed hazards, but raises it under the 75 percent normalization. Both half-productivity exercises violate the deadline.

\begin{table}[htbp]
\centering
\caption{A Fixed Ceiling under Two Hazard Conventions}
\label{tab:normalization}
\small
\begin{tabular}{@{}cccccc@{}}
\toprule
Productivity & \multicolumn{2}{c}{Normalized hazards} & \multicolumn{3}{c}{Fixed scenario hazards}\\
& Full-speed $P$ & Ceiling $P$ & Full-speed $P$ & Ceiling $P$ & Ceiling lead\\
\midrule
0.25 & 75.00 & 95.07 & 97.92 & 98.62 & 75.80 \\[0pt]
0.5 & 75.00 & 81.56 & 93.54 & 88.13 & 32.78 \\[0pt]
1 & 75.00 & 64.50 & 75.00 & 64.50 & 12.00 \\[0pt]
\bottomrule
\end{tabular}
\par\smallskip
\begin{minipage}{\textwidth}\footnotesize
\emph{Notes:} Probabilities in percent; lead in months; productivity relative to the fitted value. Normalized hazards rescale $h,k$ to maintain 75 percent full-speed probability; fixed scenario hazards scale only the yield. The ceiling and its controller stay fixed. Step is 0.005 months.
\end{minipage}
\end{table}

\emph{Very weak research.} Table~\ref{tab:weak} extends the compute floors under normalized hazards to productivities far below the range of Figure~\ref{fig:yield}. Fixed dates coexist with a risk increase. These productivities illustrate the scope of a guarantee, not plausible values.

\begin{table}[htbp]
\centering
\caption{Fixed Floors When Safety Research Is Very Weak}
\label{tab:weak}
\small
\begin{tabular}{@{}ccccc@{}}
\toprule
Productivity & Full-speed $P$ & Floor $P$ & Increase in $P$ & Floor lead\\
\midrule
0.01 & 75.00 & 79.60 & 4.60 & 12.000 \\[0pt]
0.001 & 75.00 & 80.28 & 5.28 & 12.000 \\[0pt]
\bottomrule
\end{tabular}
\par\smallskip
\begin{minipage}{\textwidth}\footnotesize
\emph{Notes:} Probabilities in percent, increases in points, lead in months, productivity relative to the fitted value. Full-speed probability is normalized to 75 percent. Settings are the baseline equal-lead floors. A zero-productivity simulation retaining the baseline $h,k$ also verifies the cumulative-hazard increase of Proposition~\ref{prop:exposure}.
\end{minipage}
\end{table}

The scenario's pause with safety starts slightly later, at month \ipEqPauseSafetyStart{}. The held model still has low state hazard, and the post-pause speedup is slightly larger. It lowers the probability by \insZeroPauseSafety{} points even when research does nothing. A pause started at month \insLatePauseStart{} raises it by \insZeroLatePause{} points. Table~\ref{tab:zero} reports both hazard conventions; they give the same signs.

The saved term has no counterpart in the benchmark of Section~\ref{sec:month}: it comes from compute that keeps growing while the lab waits.

\emph{Rules when research does nothing.} Table~\ref{tab:zero} sets productivity to $10^{-9}$ of its fitted value and keeps each rule at its equal-lead setting. At exactly zero learning, a path without pauses has cumulative hazard $\int_0^1h(F)t'(F)\,dF+\int_0^1k(F)\,dF$; pauses also contribute the hazard incurred at the held capability. So a rule changes risk only through the time spent at each capability level. Compute floors slow every step and raise risk, as Proposition~\ref{prop:exposure} implies; the lag also raises it. A pause adds time at one capability level and, because compute keeps growing, removes time at later ones. The scenario's early pauses lower risk, while a pause starting at month \insLatePauseStart{} raises it. Both hazard conventions give the same signs.

\begin{table}[htbp]
\centering
\caption{Fixed Rules When Safety Research Does Nothing}
\label{tab:zero}
\small
\begin{tabular}{@{}lcc@{}}
\toprule
Rule & Normalized hazards & Fixed scenario hazards\\
Full-speed probability (percent) & 75.00 & 99.59\\
\midrule
Compute floors & 5.35 & 0.24\\
Pause with safety, scenario start & $-$4.87 & $-$0.49\\
All-inference pause, scenario start & $-$3.56 & $-$0.48\\
R\&D-model lag & 16.85 & 0.41\\
Safety floor & 0.00 & 0.00\\
12-month pause from month 0 & $-$4.72 & $-$0.46\\
12-month pause from month 15 & 7.73 & 0.28\\
\bottomrule
\end{tabular}
\par\smallskip
\begin{minipage}{\textwidth}\footnotesize
\emph{Notes:} Change in the probability of catastrophe relative to full speed, in percentage points, with productivity at $10^{-9}$ of its fitted value. Normalized hazards rescale $h,k$ so that full speed keeps 75 percent; fixed hazards keep the scenario's $h,k$, so full speed is close to certain catastrophe. Rules keep their scenario equal-lead settings; the scenario's pause with safety starts at month \ipEqPauseSafetyStart{} and its all-inference pause at month \ipEqPauseBlogStart. Step is 0.005 months.
\end{minipage}
\end{table}

In the normalized zero-productivity limit, $h(0)=\ovHzero$ and $h(1)=\ovHoneNorm$ per month, while $F^0(15)=\ovFfifteen$. For a 12-month pause from month zero, the two terms in equation~\eqref{eq:zeropause} are $\ovWaitZero$ and $\ovSavedZero$. So $\Delta H=\ovDeltaHZero$, and $\Delta P=e^{-H_0}(1-e^{-\Delta H})$ is a reduction of $\ovPointsZero$ percentage points, where $H_0$ is unpaused zero-productivity takeoff hazard. The scenario's safety pause instead starts at month $\ipEqPauseSafetyStart$ and reduces probability by $\ovPointsScen$ points in this limit. Its extra benefit comes from timing, because the safety allocation has no effect at zero productivity.

With fixed baseline hazards, $h(1)=\ovHoneFixed$ per month and zero-learning full-speed probability is $\ovFullPFixed$ percent; the month-zero pause reduces probability by $\ovPointsZeroFixed$ points. These calculations reconcile the two timings and hazard conventions in Table~\ref{tab:zero} without treating their probability magnitudes as interchangeable.

\emph{A replay check.} A simulation checks the guarantee of Proposition~\ref{prop:construct} with the usual safety share in the replay, $s_r=s_0$. With fixed baseline $h,k$, take $a_{\min}=1$ and $\bar c=0.50$. The shortest initial learning period that keeps the ceiling is $\tau=\irConstructWait$ months, within the deadline. After it, set $s=s_0$ and $x=x_0e^{-G\tau}$. At the lowest allowed productivity the largest hazard is \irConstructHazard{}, at the ceiling. With $a=1$ and $s_r=s_0$, the replay hazard before the initial period's discount is the reference hazard $b_He^{\kappa F}f^0(F)$, which increases in $F$, so its endpoint value is $B(s_0)$.

Service losses are \irConstructLossTwo{} percent over two years and \irConstructLossFive{} percent over five years. The service path does not depend on $a$, so a five-year 5 percent budget is met for every $a\ge1$ in this example. Proposition~\ref{prop:construct} itself imposes no service budget; Figure~\ref{fig:service} instead chooses timing and safety to minimize five-year service loss within the replay class at the scenario's ceiling.

\emph{The shortest learning period.} Fix $s_0\le s_r<1$, $a_{\min}>0$, $\bar\sigma(0)>0$, and a positive ceiling. If $B(s_r)>0$, the shortest feasible learning period with this replay share is
\begin{equation}
\tau_{\min}(s_r)=\max\left\{0,\,
\frac1G\ln\frac{x_0}{1-s_r},\,
\frac{\ln[B(s_r)/\bar c]}
{a_{\min}\bar\sigma(0)\bar y}\right\}.
\label{eq:replay-wait}
\end{equation}
The second term enforces the compute-share condition $x_0e^{-G\tau}+s_r\le1$; the third supplies enough initial knowledge to meet the ceiling on replay at the worst productivity. If $B(s_r)=0$, omit the third term. A policy with this replay share meets the ceiling and the deadline for every $a\ge a_{\min}$ if and only if $\bar c\ge h(0)$ and $\tau_{\min}(s_r)\le L$. This is an exact result within the fixed-share replay class, not an optimum over unrestricted policies.

\emph{A combined commitment.} Constant shares $x=\irHybridX$, $s=\irHybridS$, $i=\irHybridI$ are simulated in all ten versions of Table~\ref{tab:robust}. The largest sampled takeoff hazard, including the left limit at $T$, is \irHybridHazard{}, below a ceiling of 0.50. The largest lead used is \irHybridLead{} months; the largest service losses are \irHybridLossTwo{} percent over 24 months and \irHybridLossFive{} percent over 60 months. Probability falls in every version. This combination meets the lead limit and the five-year service budgets of 15 and 30 percent. It fails a five-year budget of 5 percent and every two-year budget considered here. It is a tested feasible combination, not an optimal controller or a guarantee for arbitrarily small productivity.

\emph{The full-speed probability.} The scenario takes 75 percent from the AI Futures Project; Section~\ref{sec:conventions} cites lower values. The main text normalizes hazards to the full-speed probability and sets the ceiling so that it limits delay to the 12-month lead, so a lower probability scales the hazard down and the matching ceiling with it. The ratio of the final hazard to the ceiling does not change, and neither do the bounds. Rerunning the scenario at full-speed probabilities of 25, 10 and 5 percent, with the pause judgment cutting hazard in the same proportion as at 75 percent, gives the same necessary bound, \hlNec{}, and the same spare-compute replay bound, \hlSpare{}, at every probability. The matching ceilings are \hlCapHi{} a month at 75 percent, \hlCapTwentyFive{} at 25 percent, \hlCapTen{} at 10 percent and \hlCapFive{} at 5 percent. A ceiling held at the 75 percent value while the probability falls would be a much looser promise; this paper does not study it.

\section{Numerical Methods}\label{supp:numerics}

The timing comparison in Table~\ref{tab:timing} evaluates the exact finite-pause saving in equation~\eqref{eq:timing-saving} on the scenario's full-speed path. The reference knowledge stock solves $\dot S=\sigma(F^0(t))-d\dot F^0S$ with an eighth-order Runge--Kutta scheme at relative tolerance $10^{-12}$. Quadrature evaluates the hazard integrals. A 361-point grid of start dates is refined by bounded minimization; a 721-point grid moves the displayed starts and savings by less than $10^{-5}$. Direct calendar-time integration of the pre-pause, pause, and replay laws agrees with the exact formula to within $10^{-9}$. Because the saving is linear in the hazard scale, the best start is the same under fixed and normalized hazards, to within $10^{-5}$. Finite-difference checks of both timing derivatives in Section~\ref{supp:extensions} have errors below $10^{-8}$ across 36 cases of a stylized technology. The cases cover constant or rising yield, full or imperfect transfer, and equal or unequal safety shares. The full-transfer formula reproduces Proposition~\ref{prop:month}. These checks neither refit the scenario nor alter its numerical inputs.

A fixed-step fourth-order Runge--Kutta scheme integrates capability, knowledge, hazard, and services; the lag uses stored history. The equal-lead settings and the retuned ten-technology comparison in Table~\ref{tab:robust} use a 0.01-month step, root finding for each setting, and a coarse-to-fine search over pause starts. The productivity sweep in Figure~\ref{fig:yield} uses 16 scales from \insScaleMin{} to \insScaleMax{} and a 0.02-month step.

Table~\ref{tab:feasible-timing} evaluates the three tests of Proposition~\ref{prop:feasible-timing} on a 361-point grid of start dates. The supremum in equation~\eqref{eq:minimum-feasible-pause} and the prior peak are taken over 400-point grids and refined by bounded minimization. The best feasible start is refined within the grid cell that attains the largest saving.

The productivity bounds in Figures~\ref{fig:bounds} and~\ref{fig:service} hold the hazard functions fixed as productivity varies. Figure~\ref{fig:bounds}b and Figure~\ref{fig:service} use the scenario's functions. Figure~\ref{fig:bounds}a refits the scenario with only state hazard to the same three targets, and productivity there is relative to that version's fitted yield. The necessary bound is in closed form. For the two fixed-timing replay policies, bisection finds the smallest productivity at which the largest replay hazard on a 4,001-point grid meets the ceiling. This is solved at 41 ceilings from 0.03 to 1.2 a month. In Figure~\ref{fig:bounds}a the ceiling domain and shape restriction of Proposition~\ref{prop:sharp} hold at every displayed ceiling, and the ratio of replay to necessary bounds equals its closed-form factor. This brackets the unrestricted threshold rather than identifying it exactly.

Figure~\ref{fig:service} minimizes five-year service loss numerically within the replay class. At learning time $\tau$, set $x_r=x_0e^{-G\tau}$ and $i_r=1-x_r-s_r$. For the fitted hazards,
\[
\ln \lambda^0(u)=\text{constant}+(\kappa+\rho)F^0(u)+(\gamma+\sigma_b)u
\]
is convex, and subtracting the linear knowledge term preserves convexity. So the replay maximum occurs at an endpoint. At the scenario's ceiling, $\bar c\ge \lambda^0(0)$, so the initial replay endpoint always satisfies it. The final endpoint requires
\[
y(s_r)\ge Y(\tau,a_{\min})
=\frac{\ln[\lambda^0(T_0)/\bar c]/(a_{\min}\sigma_b)-\bar y\tau}{T_0}.
\]
The service-maximizing choice at that timing is $s_r=s_0$ when $Y\le1$; otherwise it is $s_0[(Y-\psi)/(1-\psi)]^{1/\alpha}$, provided $s_r\le1-x_r$.

The maximum feasible safety share rises with the learning time, so root finding identifies the earliest feasible learning time. The search includes this boundary, $\tau=L$, and the point where $s_r$ reaches $s_0$. It also includes every local loss minimum on a 129-point grid of learning times, each refined by bounded scalar optimization. A 513-point grid changes the minimum loss by less than $10^{-7}$ percentage points at all 56 displayed productivity bounds.

Service values use quadrature of the closed-form path, with baseline shares restored at completion; independent integration in calendar time agrees within $10^{-10}$ of full-speed value. Endpoint ceiling checks agree with the 4,001-point replay grid. The gray comparison fixes $\tau=L$ and selects the smallest feasible safety share at the same productivity bound. The objective is five-year services; the reported two-year loss uses that selected policy and a separate denominator.

Table~\ref{tab:scope} uses the same grid and bisection with each version's own normalized hazards. Replay knowledge integrates the transfer-loss law along the reference path by the trapezoid rule. Table~\ref{tab:zero} and the checks of Appendix~\ref{supp:checks} use 0.005-month steps. The checks cover normalized and fixed-hazards cases at a quarter, half, and the fitted productivity; floors at 0.01 and 0.001 of it; and the combined commitment in all ten versions. Ceiling checks sample states and the left limit at $T$; Proposition~\ref{prop:construct} supplies its own analytical guarantee.

The illustrations in Figures~\ref{fig:scenario}, \ref{fig:valuemonth}, \ref{fig:replay}, and~\ref{fig:pausezero} use closed forms. Along the full-speed path, the hazard is $b_He^{\kappa F^0(t)}f^0(F^0(t))$, the hazard still ahead is $A(t)=b_H(e^\kappa-e^{\kappa F^0(t)})/\kappa$, and the hazard of standing still is $w(t)=\omega b_He^{\kappa F^0(t)}f^0(F^0(t))$. The replay hazard in Figure~\ref{fig:replay} is the expression in the proof of Proposition~\ref{prop:construct}, and its maximum equals the ceiling at the replay bound to within $10^{-6}$. The integral in equation~\eqref{eq:zeropause} is evaluated by adaptive quadrature; at the start dates of Table~\ref{tab:zero} it reproduces the simulated changes in probability to the table's two decimals. Section~\ref{sec:replay} decomposes the gap under step hazard. The decomposition reproduces the computed ratio of the two bounds to within 0.002 at every ceiling between the scenario's and 1 a month.

\section{Notation}\label{supp:notation}

Table~\ref{tab:notation} summarizes the notation.

\begin{table}[htbp]
\centering
\caption{Notation}\label{tab:notation}
\small
\begin{tabular}{@{}>{$}l<{$}>{\raggedright\arraybackslash}p{12.2cm}@{}}
\toprule
\multicolumn{1}{@{}l}{Symbol} & Meaning\\
\midrule
F,\ T & Capability, from 0 (Automated Coder) to 1 (top-expert-dominating AI); completion date, the first date $F=1$\\
K(t)=K_0e^{Gt} & Compute, growing at rate $G$\\
i,\ x,\ s & Compute shares for customers, capability research, and safety research ($i+x+s=1$)\\
\phi(F),\ \varepsilon & Research productivity of a model of capability $F$ (scenario: $\theta e^{\rho F}$, with $\rho$ the recursive-improvement rate); research-compute elasticity\\
\ell & Lag of the R\&D-model lag rule: research uses the model of $\ell$ months earlier\\
x_0,\ s_0,\ i_0 & Full-speed shares\\
F^0,\ T_0,\ t_0(F),\ f^0(F) & Full-speed path, completion date, arrival date of milestone $F$, and pace at milestone $F$\\
\gamma=\varepsilon G & Growth rate of research capacity\\
S & Safety knowledge, in units of log-hazard reduction\\
\sigma(F)=a\bar\sigma(F) & Safety yield; $a$ is the unknown productivity, $\bar\sigma(F)$ its known profile\\
y(s),\ \psi,\ \alpha & Learning multiplier of the safety share, increasing with $y(s_0)=1$; in the scenario's form, the share of learning without dedicated compute and the curvature\\
\bar y=y(1),\ y_r & Fastest learning multiplier (all compute in safety); multiplier during the spare-compute replay\\
h(F),\ k(F) & State hazard per month; step hazard per unit of capability (both before any knowledge)\\
\lambda(t),\ H,\ P & Hazard rate; cumulative hazard $\int_0^T\lambda\,dt$; probability of catastrophe $1-e^{-H}$\\
\mu,\ r,\ J_q & Log quality increase per unit of capability; monthly discount rate; discounted service value over $q$ months\\
\bar c & Ceiling on the hazard rate\\
D,\ L & Lead used, the largest milestone delay; the lead, the delay the deadline allows\\
a_{\min} & Productivity bound: the lowest productivity the regulator cannot rule out\\
a^* & Threshold: the infimum of productivity bounds at which some single rule guarantees both promises; it need not be attained\\
b_H,\ \kappa,\ \omega,\ \sigma_b & Hazard scale, hazard growth with capability, state share, fitted yield in the scenario\\
A(t),\ w(t),\ m & Hazard still ahead; hazard of standing still; monthly knowledge growth during a pause\\
v_{s_p}(t) & Net hazard saved per month of pause, $mA(t)-w(t)$\\
\tau,\ s_r & Learning time and replay safety share of learn-then-replay\\
\lambda^0(u),\ \bar S^0(u),\ B(s_r) & Unprotected reference hazard; knowledge per unit productivity; largest replay hazard before initial learning\\
\bar a & Replay bound of Proposition~\ref{prop:sharp}\\
d,\ R(t,u) & Knowledge loss per unit of capability; fraction of pause knowledge retained at reference date $u$\\
g_\delta,\ s_p,\ \mathcal V_\delta & Effective pause duration $\int_0^\delta e^{-mv}dv$; pause safety share; cumulative hazard saved by a finite pause (Section~\ref{sec:timing})\\
A_d,\ v_d & Retention-weighted remaining hazard; value of a short pause, $mA_d-w$ (Section~\ref{supp:extensions})\\
\Lambda_-,\ \delta_c,\ \mathcal A & Largest prior hazard; minimum pause protecting the replay; pause pairs meeting the ceiling and deadline\\
D_-,\ D_+ & Largest reference milestone delays before and after an inserted pause (Section~\ref{supp:extensions})\\
M=(x_0/x)^\varepsilon & Slowdown factor of a constant research share (Proposition~\ref{prop:exposure})\\
\bottomrule
\end{tabular}
\end{table}


\small
\begin{thebibliography}{40}
\bibitem[Acemoglu and Lensman(2024)]{acemoglu2024}
Acemoglu, Daron, and Todd Lensman. 2024. ``Regulating Transformative Technologies.'' \emph{American Economic Review: Insights} 6 (3): 359--76. \url{https://doi.org/10.1257/aeri.20230353}.

\bibitem[AI Futures Project(2026)]{aifutures2026}
AI Futures Project. 2026. ``How to Pace the US Frontier: Tentative Proposals for Domestic AI Regulation.'' August 5. \url{https://blog.aifutures.org/p/how-to-pace-the-us-frontier}.

\bibitem[Amodei(2026)]{amodei2026}
Amodei, Dario. 2026. ``We Must Pace the Frontier.'' September. \url{https://darioamodei.com/post/we-must-pace-the-frontier}.

\bibitem[Armstrong et~al.(2016)]{armstrong2016}
Armstrong, Stuart, Nick Bostrom, and Carl Shulman. 2016. ``Racing to the Precipice: A Model of Artificial Intelligence Development.'' \emph{AI \& Society} 31 (2): 201--06. \url{https://doi.org/10.1007/s00146-015-0590-y}.

\bibitem[Aschenbrenner(2019)]{aschenbrenner2019}
Aschenbrenner, Leopold. 2019. ``Existential Risk and Growth.'' Preliminary draft, version 0.5, Columbia University, September 2. \url{https://leopoldaschenbrenner.github.io/xriskandgrowth/ExistentialRiskAndGrowth050.pdf}.

\bibitem[Bengio et~al.(2024)]{bengio2024}
Bengio, Yoshua, Geoffrey Hinton, Andrew Yao, Dawn Song, Pieter Abbeel, Trevor Darrell, Yuval Noah Harari, et~al. 2024. ``Managing Extreme AI Risks amid Rapid Progress.'' \emph{Science} 384 (6698): 842--45. \url{https://doi.org/10.1126/science.adn0117}.

\bibitem[Binder(2021)]{binder2021}
Binder, Damon. 2021. ``A Simple Model of AGI Deployment Risk.'' Effective Altruism Forum, July 9. \url{https://forum.effectivealtruism.org/posts/aSMexrjGXpNiWpbb5/a-simple-model-of-agi-deployment-risk}.

\bibitem[Bostrom(2014)]{bostrom2014}
Bostrom, Nick. 2014. \emph{Superintelligence: Paths, Dangers, Strategies}. Oxford: Oxford University Press. Chapter 14.

\bibitem[Bostrom(2026)]{bostrom2026}
Bostrom, Nick. 2026. ``Optimal Timing for Superintelligence: Mundane Considerations for Existing People.'' Working paper, version 1.0. \url{https://nickbostrom.com/optimal.pdf}.

\bibitem[Bueno de Mesquita et~al.(2026)]{buenodemesquita2026}
Bueno de Mesquita, Ethan, Wioletta Dziuda, and Mattias Polborn. 2026. ``The AGI Race and Existential Risk.'' NBER Working Paper 35276. \url{https://doi.org/10.3386/w35276}.

\bibitem[Clarke and Reed(1994)]{clarke1994}
Clarke, Harry R., and William J. Reed. 1994. ``Consumption/Pollution Tradeoffs in an Environment Vulnerable to Pollution-Related Catastrophic Collapse.'' \emph{Journal of Economic Dynamics and Control} 18 (5): 991--1010. \url{https://doi.org/10.1016/0165-1889(94)90042-6}.

\bibitem[Cropper(1976)]{cropper1976}
Cropper, Maureen L. 1976. ``Regulating Activities with Catastrophic Environmental Effects.'' \emph{Journal of Environmental Economics and Management} 3 (1): 1--15. \url{https://doi.org/10.1016/0095-0696(76)90009-7}.

\bibitem[Fudenberg and Koh(2026)]{fudenberg2026}
Fudenberg, Drew, and Andrew Koh. 2026. ``Cooperating against Catastrophe.'' arXiv:2609.28291. \url{https://arxiv.org/abs/2609.28291}.

\bibitem[Gan(2026a)]{gan2026allocation}
Gan, Li. 2026a. ``Safety Research and the Allocation of Frontier Compute.'' arXiv:2610.14305. \url{https://arxiv.org/abs/2610.14305}.

\bibitem[Gan(2026b)]{gan2026racing}
Gan, Li. 2026b. ``Prizes, Shared Hazard, and the Speed of Frontier AI.'' arXiv:2611.04266. \url{https://arxiv.org/abs/2611.04266}.

\bibitem[Gan(2026c)]{gan2026agreements}
Gan, Li. 2026c. ``Agreements Have a Clock: Compute Growth and the Verification of AI Restraint.'' arXiv:2611.04272. \url{https://arxiv.org/abs/2611.04272}.

\bibitem[Gans(2025)]{gans2025}
Gans, Joshua S. 2025. ``How Learning about Harms Impacts the Optimal Rate of Artificial Intelligence Adoption.'' \emph{Economic Policy} 40 (121): 199--219. \url{https://doi.org/10.1093/epolic/eiae053}.

\bibitem[Grace et~al.(2025)]{grace2025}
Grace, Katja, Harlan Stewart, Julia Fabienne Sandk{\"u}hler, Stephen Thomas, Ben Weinstein-Raun, Jan Brauner, and Richard C. Korzekwa. 2025. ``Thousands of AI Authors on the Future of AI.'' \emph{Journal of Artificial Intelligence Research} 84: 9. \url{https://arxiv.org/abs/2401.02843}.

\bibitem[Heim et~al.(2024)]{heim2024}
Heim, Lennart, Tim Fist, Janet Egan, Sihao Huang, Stephen Zekany, Robert Trager, Michael A. Osborne, and Noa Zilberman. 2024. ``Governing Through the Cloud: The Intermediary Role of Compute Providers in AI Regulation.'' arXiv:2403.08501. \url{https://arxiv.org/abs/2403.08501}.

\bibitem[Herweg and Schmidt(2022)]{herweg2022}
Herweg, Fabian, and Klaus M. Schmidt. 2022. ``How to Regulate Carbon Emissions with Climate-Conscious Consumers.'' \emph{Economic Journal} 132 (648): 2992--3019. \url{https://doi.org/10.1093/ej/ueac045}.

\bibitem[Holland et~al.(2009)]{holland2009}
Holland, Stephen P., Jonathan E. Hughes, and Christopher R. Knittel. 2009. ``Greenhouse Gas Reductions under Low Carbon Fuel Standards?'' \emph{American Economic Journal: Economic Policy} 1 (1): 106--46. \url{https://doi.org/10.1257/pol.1.1.106}.

\bibitem[Jones(2016)]{jones2016}
Jones, Charles I. 2016. ``Life and Growth.'' \emph{Journal of Political Economy} 124 (2): 539--78. \url{https://doi.org/10.1086/684750}.

\bibitem[Jones(2024)]{jones2024}
Jones, Charles I. 2024. ``The AI Dilemma: Growth versus Existential Risk.'' \emph{American Economic Review: Insights} 6 (4): 575--90. \url{https://doi.org/10.1257/aeri.20230570}.

\bibitem[Jones(2025)]{jones2025}
Jones, Charles I. 2025. ``How Much Should We Spend to Reduce A.I.'s Existential Risk?'' NBER Working Paper 33602. \url{https://doi.org/10.3386/w33602}.

\bibitem[Koh and Sanguanmoo(2026)]{koh2026}
Koh, Andrew, and Sivakorn Sanguanmoo. 2026. ``Technology Speed Limits.'' arXiv:2606.01424. \url{https://arxiv.org/abs/2606.01424}.

\bibitem[Lifland et~al.(2026)]{lifland2026}
Lifland, Eli, Brendan Halstead, Alex Kastner, and Daniel Kokotajlo. 2026. ``AI Futures Model: Timelines \& Takeoff.'' AI Futures Project. Forecasts accessed October 25, 2026. \url{https://www.aifuturesmodel.com/forecast}.

\bibitem[Polasky et~al.(2011)]{polasky2011}
Polasky, Stephen, Aart de Zeeuw, and Florian Wagener. 2011. ``Optimal Management with Potential Regime Shifts.'' \emph{Journal of Environmental Economics and Management} 62 (2): 229--40. \url{https://doi.org/10.1016/j.jeem.2010.09.004}.

\bibitem[Samotsvety Forecasting(2022)]{samotsvety2022}
Samotsvety Forecasting. 2022. ``Samotsvety's AI Risk Forecasts.'' September 9. \url{https://samotsvety.org/blog/2022/09/09/samotsvety-s-ai-risk-forecasts/}.

\bibitem[Sastry et~al.(2024)]{sastry2024}
Sastry, Girish, Lennart Heim, Haydn Belfield, Markus Anderljung, Miles Brundage, Julian Hazell, Cullen O'Keefe, et~al. 2024. ``Computing Power and the Governance of Artificial Intelligence.'' arXiv:2402.08797. \url{https://arxiv.org/abs/2402.08797}.

\bibitem[Sevilla et~al.(2022)]{sevilla2022}
Sevilla, Jaime, Lennart Heim, Anson Ho, Tamay Besiroglu, Marius Hobbhahn, and Pablo Villalobos. 2022. ``Compute Trends Across Three Eras of Machine Learning.'' In \emph{2022 International Joint Conference on Neural Networks}, 1--8. \url{https://doi.org/10.1109/IJCNN55064.2022.9891914}.

\bibitem[Shavit(2023)]{shavit2023}
Shavit, Yonadav. 2023. ``What Does It Take to Catch a Chinchilla? Verifying Rules on Large-Scale Neural Network Training via Compute Monitoring.'' arXiv:2303.11341. \url{https://arxiv.org/abs/2303.11341}.

\bibitem[Trammell and Aschenbrenner(2026)]{trammell2026}
Trammell, Philip, and Leopold Aschenbrenner. 2026. ``Existential Risk and Growth.'' Working paper, January 5 version; an earlier version is Global Priorities Institute Working Paper 13-2024, University of Oxford.

\bibitem[Tsur and Zemel(1996)]{tsur1996}
Tsur, Yacov, and Amos Zemel. 1996. ``Accounting for Global Warming Risks: Resource Management under Event Uncertainty.'' \emph{Journal of Economic Dynamics and Control} 20 (6--7): 1289--1305. \url{https://doi.org/10.1016/0165-1889(95)00900-0}.

\bibitem[Weitzman(1974)]{weitzman1974}
Weitzman, Martin L. 1974. ``Prices vs. Quantities.'' \emph{Review of Economic Studies} 41 (4): 477--91. \url{https://doi.org/10.2307/2296698}.
\end{thebibliography}
\end{document}